\documentclass[aps,prd,showpacs,twocolumn,floatfix,nofootinbib]{revtex4-1} 
\usepackage{graphicx}
\usepackage{amsmath,amssymb,siunitx}
\usepackage{xcolor}
\usepackage{hyperref}
\usepackage{rotating}
\usepackage{ulem}

\usepackage{comment}

\newcommand{\addms}[1]{{\color{black} #1}} 
\newcommand{\addkh}[1]{{\color{black} #1}} 

\begin{document}
\title{General-relativistic neutrino-radiation magnetohydrodynamics simulation of seconds-long black hole-neutron star mergers: Dependence on initial magnetic field strength, configuration, and neutron-star equation of state}
\author{
  Kota Hayashi$^{1}$,
  Kenta Kiuchi$^{2,1}$,
  Koutarou Kyutoku$^{3,1,4}$,
  Yuichiro Sekiguchi$^{5,1}$,  
  Masaru Shibata$^{2,1}$
}
\affiliation{
  $^1$Center for Gravitational Physics, Yukawa Institute for Theoretical Physics,
  Kyoto University, Kyoto 606-8502, Japan\\
  $^2$Max Planck Institute for Gravitational Physics (Albert Einstein Institute),
  Am M{\"u}hlenberg 1, Postdam-Golm 14476, Germany\\
  $^3$Department of Physics, Kyoto University, Kyoto 606-8502, Japan\\
  $^4$Interdisciplinary Theoretical and Mathematical Sciences Program (iTHEMS), RIKEN, Wako, Saitama 351-0198, Japan \\
  $^5$Department of Physics, Toho University, Funabashi, Chiba 274-8510, Japan
}
\date{\today}

\begin{abstract}
\addms{As a follow-up study of our previous work~\cite{hayashi2022jul},} 
numerical-relativity simulations for seconds-long black hole-neutron star mergers are performed for a variety of setups. 
Irrespective of the initial and symmetry conditions, we find qualitatively universal evolution processes: The dynamical mass ejection takes place together with a massive accretion disk formation after the neutron star is tidally disrupted; Subsequently, the magnetic field in the accretion disk is amplified by the magnetic winding, Kelvin-Helmholtz instability, and magnetorotational instability, which establish a turbulent state inducing the dynamo and angular momentum transport; 
The post-merger mass ejection by the effective viscous processes stemming from the magnetohydrodynamics turbulence sets in at $\sim300$--$500$\,ms after the merger and continues for several hundred ms; 
A magnetosphere near the black-hole spin axis is developed and the collimated strong Poynting flux is generated with its lifetime of $\sim0.5$--$2$\,s.
\addms{We have newly found that} the model of no equatorial-plane symmetry shows the reverse of the magnetic-field polarity in the magnetosphere, which is caused by the dynamo associated with the magnetorotational instability in the accretion disk. 
The model with initially toroidal fields shows the tilt of the disk and magnetosphere in the late post-merger stage because of the anisotropic post-merger mass ejection.
These effects could terminate the strong Poynting-luminosity stage within the timescale of $\sim0.5$--$2$\,s.

\end{abstract}

\maketitle

\section{Introduction} \label{sec:intro}

The first direct detection of gravitational waves from a binary black hole merger,  referred to as GW150914~\cite{abbott2016feb}, opened the era of gravitational-wave astronomy. 
To date, advanced LIGO and advanced Virgo have observed $\sim80$ binary black hole merger events~\cite{abbott2021jun,ligoo3b}.
Several neutron-star merger events have also been observed in addition to the binary black hole mergers. A remarkable event is GW170817~\cite{abbott2017oct1}, the first binary neutron star merger event. Associated with this event, a wide variety of electromagnetic counterparts were successfully observed~\cite{abbott2017oct2, abbott2017oct3}, and provided us with invaluable information for understanding the processes of the neutron-star merger and its post-merger evolution. The electromagnetic observations also opened the era of   multi-messenger astronomy including gravitational-wave observation. 

In the latest observational run (O3b), gravitational waves from black hole-neutron star binaries,  referred to as \addkh{(GW200105 and)} GW200115~\cite{abbott2021jun2}, were observed. 
These events surely indicate that black hole-neutron star binaries exist in nature. 
Although no electromagnetic counterpart is observed for them, it is natural to expect that electromagnetic counterparts will be observed in future events, if the binary parameters are suitable for inducing tidal disruption of neutron stars.
A number of numerical-relativity simulations for black hole-neutron star binaries predict that the neutron star could be disrupted by the tidal force of the companion black hole if the black-hole mass is relatively small and/or the black-hole spin is high (e.g., Refs.~\cite{shibata2015textbook,Kyutoku:2021icp}). 
The tidal disruption is accompanied by disk formation and mass ejection, which will result in the $r$-process nucleosynthesis for synthesizing heavy neutron-rich elements~\cite{lattimer1974,eichler1989}.
If the remnant black hole is rapidly spinning and surrounded by a massive magnetized disk or torus, an ultra-relativistic jet could be launched and drive a short-hard gamma-ray burst~\cite{eichler1989,nakar2007apr,berger2014jun}.
Powered by thermal energy generated by the radioactive decay of synthesized heavy neutron-rich elements, the ejecta will shine with high luminosity as kilonovae~\cite{li1998nov,metzger2010jun}. 

The sensitivity of the gravitational-wave detectors is being improved for the forthcoming observational runs (O4 and O5)~\cite{abbott2020sep}. 
Also, large-scale telescopes such as JWST and Vera Rubin telescopes will be in operation during such runs~\cite{jwst2006apr,lsst2019mar}. 
It is quite natural to expect simultaneous detection of gravitational waves and electromagnetic counterparts from black hole-neutron star mergers if the source is within a distance of several hundred Mpc from the earth.
This implies that black hole-neutron star mergers are among the most promising sources for multi-messenger astronomy in the near future.
In view of this situation, it is urgent to theoretically develop the entire evolution scenario from the merger to the post-merger stages, in order to predict  observable signals and to make a reliable model for the interpretation of the forthcoming observational data. 

In the last two decades, a variety of numerical-relativity simulations for black hole-neutron star mergers have been performed~\cite{shibata2006dec,shibata2007may, shibata2008apr,etienne2008apr, duez2008nov, shibata2009feb, etienne2009feb, chawla2010sep, duez2010may, kyutoku2010aug, kyutoku2011sep, foucart2011jan, foucart2012feb, etienne2012mar, etienne2012oct, kyutoku2013aug, kyutoku2015aug, foucart2013apr, lovelace2013jun, deaton2013sep, foucart2014jul, paschalidis2015jun, kawaguchi2015jul, kiuchi2015sep, foucart2017jan, kyutoku2018jan, brege2018sep, ruiz2018dec, foucart2019feb, foucart2019may, hinderer2019sep, hayashi2021feb, foucart2021mar, most2021may, chaurasia2021oct, most2021jul, hayashi2022jul}. 
By improving the input physics and grid resolution, the previous studies have extensively explored the process of the tidal disruption, accretion disk formation, dynamical mass ejection, gravitational-wave emission, and neutrino emission.
However, most of the previous works have focused only on the evolution from the inspiral to early post-merger stages; the evolution was followed at longest for a few hundred ms after the merger. Hence, the long-term seconds-long evolution of the system has not been explored deeply.
In order to compensate for this deficiency, i.e., to explore the entire post-merger evolution processes, many long-term numerical simulations for black hole-accretion disk systems have also been performed, including viscous hydrodynamics or magnetohydrodynamics effects~\cite{fernamdez2013aug,metzger2014may,just2015feb,fernandez2015mar,fernandez2017jul,daniel2018may,fernandez2018oct,agnieszka2019sep,christie2019sep,miller2019jul,fujibayashi2020apr,fujibayashi2020dec,Li2021jun,fernandez2020jul,just2022jan,shibata2021sep}.
These simulations have qualitatively clarified the evolution processes in the post-merger stage such as the post-merger mass ejection and jet launch. For example, it is now widely accepted that the post-merger mass ejection is likely to be driven by the effective viscous processes induced by the magnetohydrodynamics turbulence. 
The jet outflow is also likely to be powered by the Blandford-Znajek mechanism~\cite{blandford1977} by magnetic fields penetrating a rapidly spinning black hole. 
\addkh{However, it is not clear whether the initial conditions given in such simulations are appropriate for exploring the post-merger evolution of the black hole-neutron star merger.
During the merger stage, the neutron-star matter is spread non-axisymmetrically around the remnant black hole, and the non-axisymmetric fall-back matter suppresses the system from being axisymmetric.
Such non-axisymmetric structure of the accretion disk could also significantly affect the magnetic-field amplification process and field profile.}
Therefore, the quantitative details of the post-merger process are not fully understood. 

In order to overcome these deficiencies and acquire the self-consistent evolution scenario of the black hole-neutron star merger starting from the inspiral stage to the late post-merger stage, in a previous paper, we performed seconds-long merger simulations including dynamical general-relativity effect, neutrino-radiation effect, and magnetohydrodynamics effect altogether for the first time~\cite{hayashi2022jul}. 
We confirmed that the post-merger mass ejection is indeed driven by the magnetically-induced viscous effect at several hundred milliseconds after the onset of the merger, and  
this post-merger mass ejection is triggered by the decrease of the temperature and neutrino luminosity in the accretion disk, as previous viscous hydrodynamics studies have clarified~(e.g., Refs.~\cite{fernamdez2013aug,fujibayashi2020apr,just2022jan}). 
We also confirmed that the electron fraction of the post-merger ejecta is not very low, typically with $Y_\mathrm{e}\sim 0.2$--0.3. 
In addition, we found the development of the magnetosphere near the spin axis of the remnant black hole as a result of the infall of the amplified magnetic flux from the disk into the black hole. The intensity of the outgoing Poynting flux in the magnetosphere is high and is consistent with that of short-hard gamma-ray bursts.

Although the results of our previous work showed a self-consistent evolution picture of black hole-neutron star binaries, some questions remain to be answered.
First, our previous work assumed initially strong magnetic fields with its maximum strength $\geq3\times10^{16}$\,G. 
The high magnetic-field strength was given in order to get a high field strength at the formation of the accretion disk and to enable numerical computation to resolve the magnetorotational instability (MRI)~\cite{balbus1991,balbus1998} in the disk right after the disk formation.
We should clarify whether, at least qualitatively, the same evolution is obtained even if we assume a much lower strength of the initial magnetic field.
Second, our previous work assumed a poloidal magnetic field confined in the neutron star as an initial condition. 
It is quite natural to ask whether the result is qualitatively changed or not if we employ different magnetic-field configuration in the neutron star as the initial condition. 
Third, in our previous work, we imposed the equatorial-plane symmetry on all the simulations. 
It is necessary to understand what happens in the absence of such a symmetry. 
Finally, our previous work used only the DD2 equation of state (EOS)~\cite{banik2014sep} to model the neutron star. 
We should investigate how the evolution process \addms{and properties of the ejecta} depend quantitatively on the EOSs employed. 

The paper is organized as follows.
In Sec.~\ref{sec:methods}, we briefly describe the method and initial setup for the numerical simulation.
In Sec.~\ref{sec:results}, we present the numerical results focusing on the entire evolution process, mass ejection mechanisms, and collimated electromagnetic outflow developed near the spin axis of the black hole. 
We pay particular attention to the quantitative difference among the numerical results with different initial \addms{magnetic-field profile, different EOSs,} and different computational setups. Finally, we conclude this work in Sec.~\ref{sec:conclusion}. 
Throughout this paper, we use the geometrical units in which $G = c = 1$, where $G$ and $c$ are the gravitational constant and the speed of light, respectively.

\section{Numerical methods} \label{sec:methods}

\begin{figure}[!th]
      \begin{center}
        \includegraphics[scale=0.2]{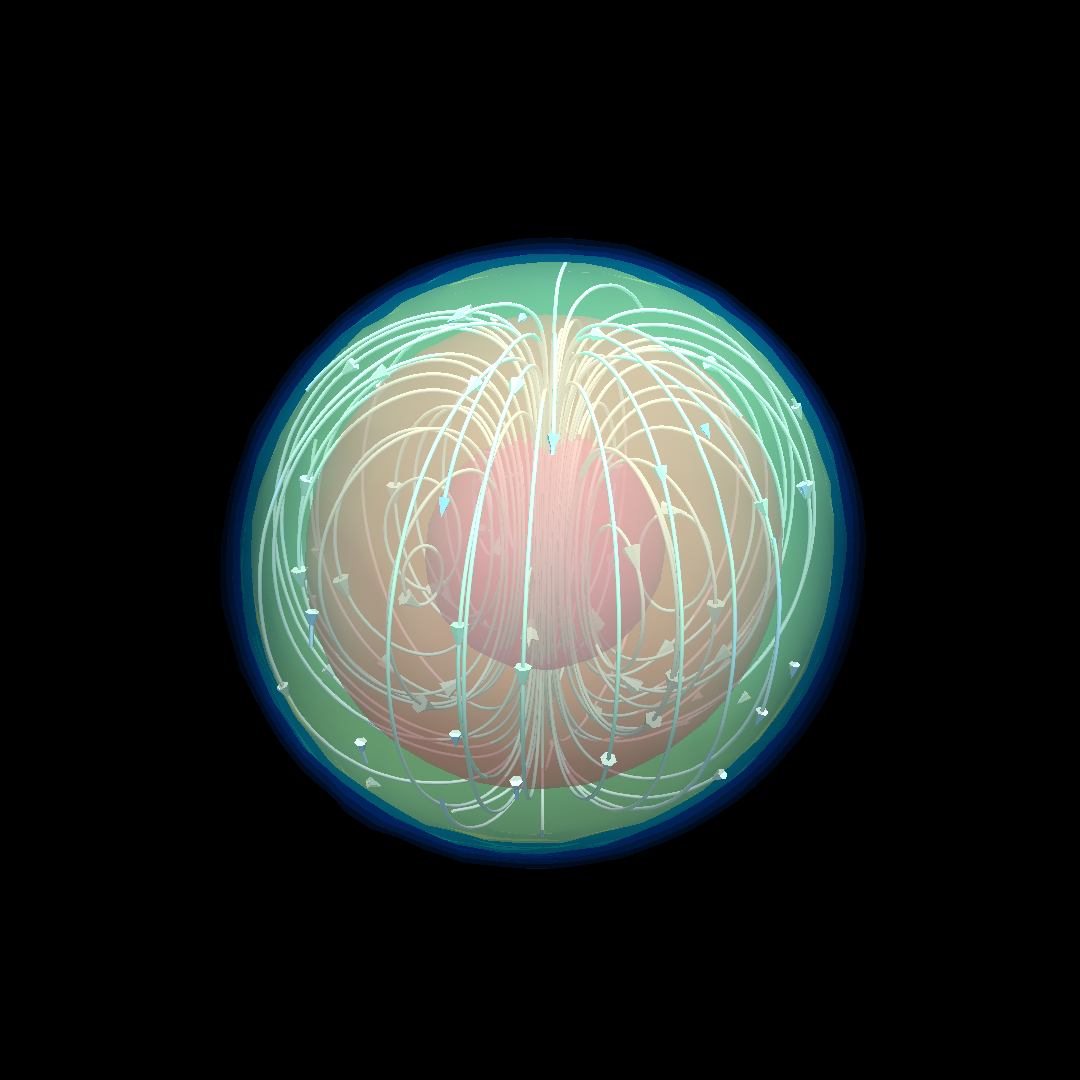} \\ 
        \vspace{1mm}
        \includegraphics[scale=0.2]{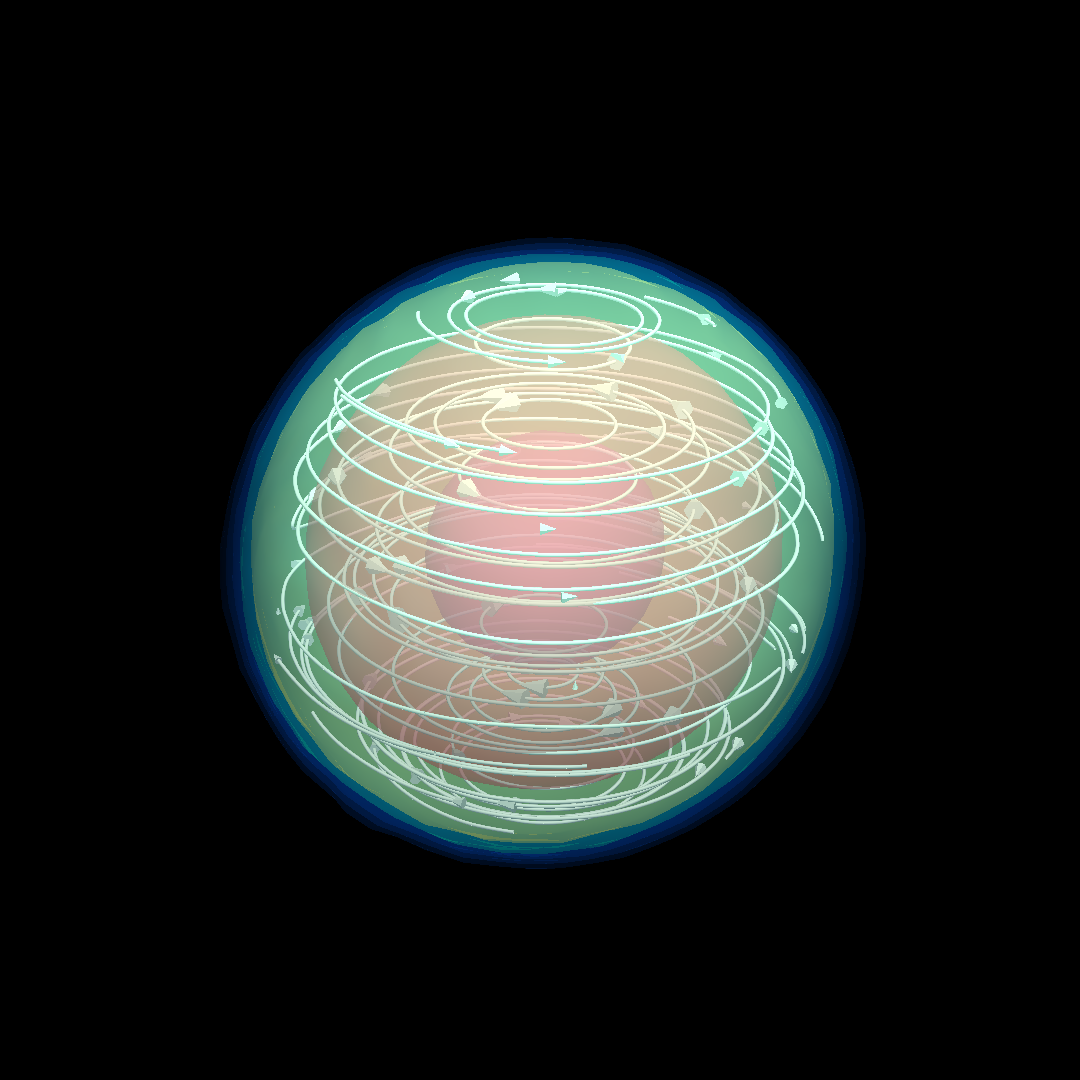} \\
        \caption{\addkh{The white lines show the magnetic-field line for the two types of the initial magnetic field configurations adopted in this work.
        The magnetic field is confined in the neutron star initially.
        The top and bottom panels show the cases for the poloidal and toroidal fields, respectively.}
        }
        \label{fig:init_b}
      \end{center}
\end{figure}

The numerical scheme for the present simulation is the same as that summarized in Ref.~\cite{hayashi2022jul} for which the readers may refer to.

For modeling the neutron-star matter, we employ two nuclear-theory-based finite-temperature EOSs referred to as DD2~\cite{banik2014sep} and SFHo~\cite{steiner2013sep} for a high-density range and Helmholtz EOS~\cite{timmes2000} for a low-density range. 
As initial data, we prepare black hole-neutron star binaries in a quasi-equilibrium state assuming the neutrinoless beta-equilibrium cold state~\cite{kyutoku2018jan}. 
The initial gravitational mass of the neutron star is set to be $M_{\rm NS}=1.35M_{\odot}$ following Ref.~\cite{hayashi2022jul}.
For the DD2 and SFHo EOSs, the circumferential radius of the isolated spherical neutron star of mass $1.3$--$1.4M_{\odot}$ is $\approx 13.2\,\mathrm{km}$ and $\approx 11.9\,\mathrm{km}$, respectively. 
These EOSs satisfy constraints imposed by the observation of gravitational waves for GW170817~\cite{abbott2017oct1} and by the X-ray observation by NICER~\cite{miller2019dec}.

For the initial black-hole mass, we choose $M_{\rm BH,0}=5.4M_{\odot}$; 
the mass ratio of the black hole to the neutron star is $Q:=M_{\rm BH,0}/M_{\rm NS}=4$. 
The initial dimensionless spin parameter of the black hole is set to be $\chi=0.75$. 
This setting is the same as one of our previous settings~\cite{hayashi2022jul}. 
With such a spin and a mass ratio, tidal disruption of the neutron star with $M_{\rm NS}=1.35M_\odot$ takes place for both EOSs. The initial orbital angular velocity $\Omega_0$ is set to be $m_0 \Omega_0 = 0.056$ for $Q=4$ where $m_0=M_{\rm BH,0}+M_{\rm NS}=6.75M_\odot$.
With this initial setup, the binary spends about three orbits before the merger. 

We initially superimpose a poloidal or toroidal magnetic field confined in the neutron star.
\addkh{Figure~\ref{fig:init_b} shows the magnetic-field lines for two types of initial field configurations adopted in this work.}
For the poloidal field case, following our previous work~\cite{kiuchi2015sep}, 
the magnetic field is given in terms of the vector potential as
\begin{eqnarray} \label{init_b_field_p}
  A_j &= & \{ -y_{\mathrm{NS}} \delta_j^{~x} + x_{\mathrm{NS}} \delta_j^{~y} \} \nonumber \\
  && \times  A_b \max(P/P_{\mathrm{max}} - 10^{-3} , 0)^2.
\end{eqnarray}
For the toroidal field case, the vector potential is given as
\begin{eqnarray} \label{init_b_field_t}
  A_j &= & \{ (x_{\mathrm{NS}}(z_{\mathrm{NS}}^2-R_{\mathrm{NS}}^2)) \delta_j^{~x} + (y_{\mathrm{NS}}(z_{\mathrm{NS}}^2-R_{\mathrm{NS}}^2)) \delta_j^{~y} \nonumber \\
  && -(z_{\mathrm{NS}}(x_{\mathrm{NS}}^2+y_{\mathrm{NS}}^2-R_{\mathrm{NS}}^2)) \delta_j^{~z}\} \nonumber \\
  && \times  A_b (1+\cos(r_{\mathrm{NS}}/0.95R_{\mathrm{NS}})) \nonumber \\
  && \qquad \qquad (r_{\mathrm{NS}}<0.95R_{\mathrm{NS}}).
\end{eqnarray}
Here, $(x_{\mathrm{NS}},y_{\mathrm{NS}},z_{\mathrm{NS}})$ denote the coordinates with respect to the neutron-star center (location of the maximum rest-mass density), $r_{\mathrm{NS}}$ is the \addkh{radial coordinate} with respect to the neutron-star center, and $R_{\mathrm{NS}}$ is the coordinate radius of the neutron-star.
$P$ is the pressure, $P_{\mathrm{max}}$ is the maximum pressure, and $j = x, y$, and $z$.
$A_b$ is a constant and is chosen so that the initial maximum magnetic-field strength $b_{\mathrm{0,max}}$ is $3 \times 10^{15}~\mathrm{G}$ or $5 \times 10^{16}~\mathrm{G}$.
These values are chosen to obtain a strong magnetic field in the remnant disk formed shortly after the tidal disruption of the neutron star. 
The strong magnetic field in the remnant disk is required to resolve the fastest growing mode of the MRI~\cite{balbus1991,balbus1998} in the limited grid resolution, because its wavelength is proportional to the magnetic-field strength. 
Although strong fields we prepare are not realistic for orbiting neutron stars, the resulting turbulent state in the accretion disk established by the MRI is not likely to depend strongly on the initial magnetic-field strength.
Thus, it would be reasonable to suppose that the turbulent state with a strong magnetic field will be established even for the case with much weaker initial magnetic-field strength if the grid resolution is sufficient. 
\addkh{As is often done in this research field, we implicitly assume that the magnetic-field strength would be increased by the MRI and magnetic winding, and a turbulent state would be eventually established even if we started a simulation from low magnetic-field strengths. This is just an assumption, but the result of a simulation, which is started with a low magnetic-field strength, indicates certain evidence for this (see Sec.~\ref{sec:results}).}
We also note that even with $b_{\mathrm{0,max}}=5\times 10^{16}$\,G, the electromagnetic energy (of order $10^{49}$\,erg) is much smaller than the internal energy and gravitational potential energy (of order $10^{53}$\,erg) of the neutron star, and thus, the inspiral and tidal-disruption stages are not affected by the strong field significantly. 

We only consider the magnetic field confined in the neutron star initially, and do not consider a pulsar-like dipole magnetic field extending to the outside of the neutron star. 
This is because only the magnetic field confined in the neutron star has a significant effect on the subsequent evolution of the system \addkh{for the realistic magnetic-field strength}.
In terms of the accretion disk evolution including the post-merger mass ejection, only the magnetic field in the disk, which originates from the magnetic field inside the neutron star, plays an important role.
In terms of the magnetosphere formation, a dipolar magnetic field initially located outside the neutron star may be amplified linearly due to winding. 
However in the disk, the magnetic field is amplified exponentially by the MRI, and the amplified magnetic field flux is ejected from the disk to the polar region by the MRI dynamo and subsequently forms the magnetosphere of a high field strength. 
For the realistic initial magnetic-field strength lower than $10^{12}$\,G, the magnetic field amplified by the MRI should dominantly come into play.

\begin{table*}[]
  \centering

  \caption{
   Key parameters and quantities for the initial conditions together with the parameters of grid setup for our numerical simulations. 
   $b_{\rm 0,max}$: the initial maximum magnetic-field strength, 
   $\Delta x_{i_{\rm max}}$: the grid spacing for the finest refinement level, $L_1$: the location of the outer boundaries along each axis, and the values of $N$ and $i_{\rm max}$. 
   For all the models, the neutron-star mass is $1.35M_\odot$, the initial black-hole mass is $5.4M_\odot$, the initial dimensionless spin of the black hole is 0.75, and the initial ADM mass $M_{\rm ADM,0}$ is $6.679M_\odot \approx 0.9894m_0$. 
   The models from our previous paper (Q4B5L and Q4B5H) are also shown for comparison. 
  }
  \label{tab:init_cond}
  
  \begingroup
  \setlength{\tabcolsep}{4pt} 
  \renewcommand{\arraystretch}{1.2} 
  
  \begin{tabular}{ccccccccc}
    \hline
    \hline
    model name & EOS & $b_{\rm 0,max} ~[{\rm G}]$ & $b_{\rm 0}$ configuration & plane sym. & 
    $\Delta x_{i_{\rm max}} ~[{\rm m}]$  & $L_1 ~[{\rm km}]$ & ~~$N$~~   & ~$i_{\rm max}$~ \\
    \hline
Q4B3e15   & DD2   & $3\times 10^{15}$  & poloidal & yes  & 400 & $6.98\times 10^4$ & 170 & 11 \\
Q4B5tn    & DD2   & $5\times 10^{16}$  & toroidal & no   & 400 & $6.98\times 10^4$ & 170 & 11 \\
Q4B5n     & DD2   & $5\times 10^{16}$  & poloidal & no   & 400 & $6.98\times 10^4$ & 170 & 11 \\
SFHoQ4B5  & SFHo  & $5\times 10^{16}$  & poloidal & yes  & 250 & $3.10\times 10^4$ & 243 & 10 \\
Q4B5L~\cite{hayashi2022jul}  & DD2   & $5\times 10^{16}$  & poloidal & yes  & 400 & $1.74\times 10^4$ & 170 & 9 \\
Q4B5H~\cite{hayashi2022jul}  & DD2   & $5\times 10^{16}$  & poloidal & yes  & 270 & $1.62\times 10^4$ & 234 & 9 \\
    \hline
  \end{tabular}

  \endgroup

\end{table*}

We do not consider the effect of the neutrino viscosity on the MRI supposing that the magnetic-field strength could be enhanced to be $\agt 10^{14}$\,G due to the rapid winding in the main region of the accretion disk (see Sec.~\ref{sec:results-disk}) even if the early growth of the MRI could be suppressed~\cite{masada2008,guilet2016}. 

The simulation is performed using a fixed-mesh refinement (FMR) algorithm. The $i$-th refinement level covers a half or full cubic box of 
$[-L_i:L_i] \times [-L_i:L_i] \times [0:L_i]$ or $[-L_i:L_i] \times [-L_i:L_i] \times [-L_i:L_i]$, where $L_i=N\Delta x_i$ and $\Delta x_i$ is the grid spacing for the $i$-th level. For the half-cubic box case, the plane-symmetric boundary condition on the $z=0$ plane (equatorial plane) is imposed. 
The grid spacing for each level is determined by $\Delta x_i=2\Delta x_{i+1}$ ($i=1,2, \cdots , i_{\mathrm{max}}-1$) with $\Delta x_{i_{\mathrm{max}}}=400\,\mathrm{m}$ for the DD2 models and $\Delta x_{i_{\mathrm{max}}}=250\,\mathrm{m}$ for the SFHo model. 
That is, we perform lower-resolution simulations for the DD2 EOS models, because high-resolution simulations require an extremely high computational cost for simulating  the seconds-long merger processes. 
In addition, as we showed in our previous paper~\cite{hayashi2022jul}, the results for the lower-resolution runs are quantitatively similar to those for the corresponding high-resolution runs (with $\Delta x_{i_{\mathrm{max}}}=270\,\mathrm{m}$), and hence, we consider that a fair convergence would be achieved even with the present choice. 
$i_{\rm max}$ is chosen to be 11 for the DD2 models and 10 for the  SFHo model. 
The values of $N$ are $170$ for the DD2 models and 243 for the SFHo model, respectively (cf.~Table~\ref{tab:init_cond}). 

In this paper, we perform four new simulations varying the EOS, the value of $b_{\mathrm{0,max}}$, the magnetic-field configuration, and the equatorial-plane symmetry. The parameters and quantities for the four models are summarized in Table~\ref{tab:init_cond}. We compare the new results with those obtained in our previous paper~\cite{hayashi2022jul} for the models with the same values of $M_{\rm BH,0}$, $M_\mathrm{NS}$, and $\chi$.

As we already mentioned in our previous paper~\cite{hayashi2022jul}, during the merger stage, the black hole is kicked mainly by the back reaction of the dynamical mass ejection and the resulting velocity is $v_{\rm kick}=200$--400\,km/s in our present setting. 
For the SFHo model for which the ejecta mass is smaller, the kick velocity is lower.
In order to avoid the black hole from running into the FMR boundary, we control the shift vector with the prescription proposed in our previous paper~\cite{hayashi2022jul}.

Following our previous work~\cite{hayashi2022jul}, we stop the time evolution of the gravitational field at a certain moment after the ratio of the rest mass of the remnant disk to the black-hole mass drops below $10^{-2}$. This prescription is reasonable because the self-gravity of the matter located outside the black hole can be safely neglected and the gravitational field is approximately stationary in such a low-mass disk stage.
\addms{For stabilizing magnetohydrodynamics computation, we introduce floor density of $\rho=10^3\,{\rm g/cm}^3$ for $r \leq r_0 \approx 10^2$\,km and max$(10^3(r_0/r)^3, 0.17)\,{\rm g/cm}^3$ for $r > r_0$ where $0.17\,{\rm g/cm^3}$ is the lowest value of the rest-mass density in our EOS table.
}

\section{Results} \label{sec:results}

\subsection{Overview} \label{sec:overview}

\begin{table*}[]
  \centering

  \caption{ 
   \addkh{The first two columns list $M_{\mathrm{eje,dyn}}$: the dynamical ejecta mass evaluated at $t=20$\,ms, and $M_{\mathrm{eje,pm}}$: the lower bound of the post-merger ejecta mass in units of $M_\odot$. 
   Since the mass of the post-merger ejecta is still increasing at the termination of all the runs, we here list the lower bound for it \addms{(see footnote 1)}. 
   The next two columns list $Y_{e \ \mathrm{eje,dyn}}$: the averaged electron fraction of the dynamical ejecta evaluated at $t=20$\,ms, and $Y_{e \ \mathrm{eje,pm}}$: that of the post-merger ejecta.
   The last two columns list $v_{\mathrm{eje,dyn}}$: the averaged velocity of the dynamical ejecta evaluated at $t=20$\,ms, and $v_{\mathrm{eje,pm}}$: that of the post-merger ejecta.}
  }
  \label{tab:result_ave}
  
  \begingroup
  \setlength{\tabcolsep}{4pt} 
  \renewcommand{\arraystretch}{1.2} 
  
  \begin{tabular}{ccccccc}
    \hline
    \hline
    model name & $M_{\mathrm{eje,dyn}}$ & $M_{\mathrm{eje,pm}}$ & $Y_{e \ \mathrm{eje,dyn}}$ & $Y_{e \ \mathrm{eje,pm}}$ & $v_{\mathrm{eje,dyn}}$ & $v_{\mathrm{eje,pm}}$\\
    \hline
Q4B3e15   & 0.045 & 0.030 & 0.062 & 0.24 & 0.17 & 0.044 \\
Q4B5tn    & 0.045 & 0.030 & 0.062 & 0.21 & 0.17 & 0.051 \\
Q4B5n     & 0.046 & 0.033 & 0.062 & 0.22 & 0.17 & 0.053 \\
SFHoQ4B5  & 0.013 & 0.019 & 0.051 & 0.27 & 0.16 & 0.048 \\
Q4B5L~\cite{hayashi2022jul}  & 0.046 & 0.035 & 0.061 & 0.23 & 0.17 & 0.051 \\
Q4B5H~\cite{hayashi2022jul}  & 0.046 & 0.028 & 0.056 & 0.25 & 0.17 & 0.047\\
    \hline
  \end{tabular}

  \endgroup

\end{table*}

\begin{figure*}[h]
        \includegraphics[scale=0.26]{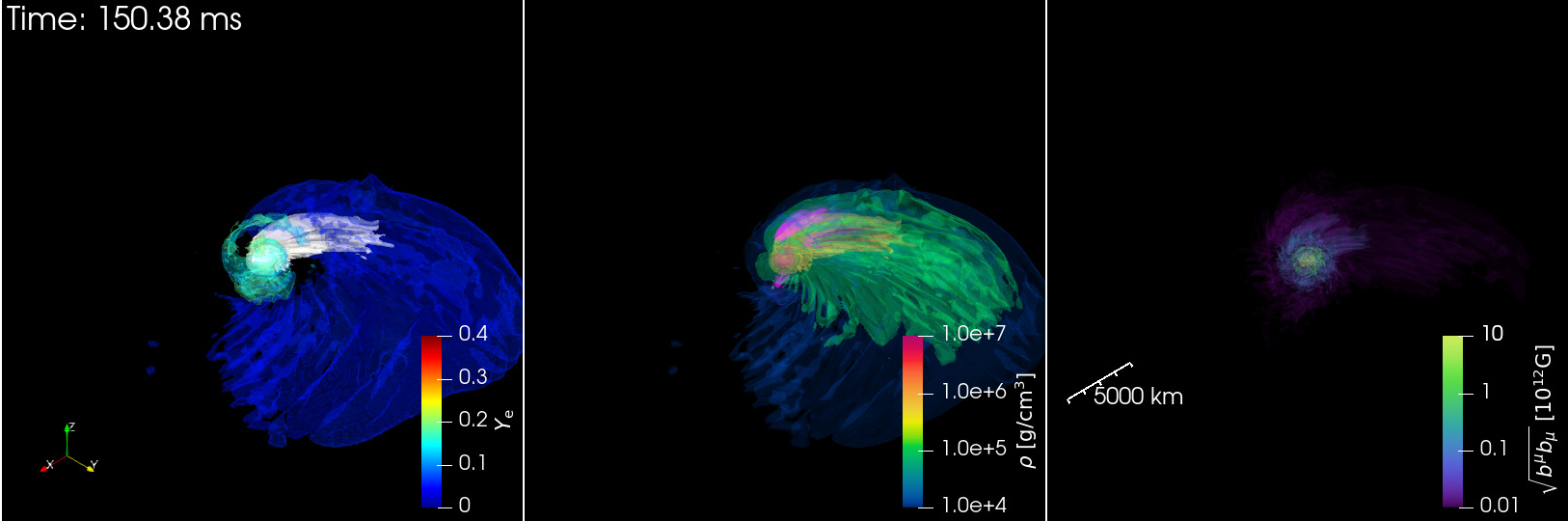} \\ 
        \vspace{1mm}
        \includegraphics[scale=0.26]{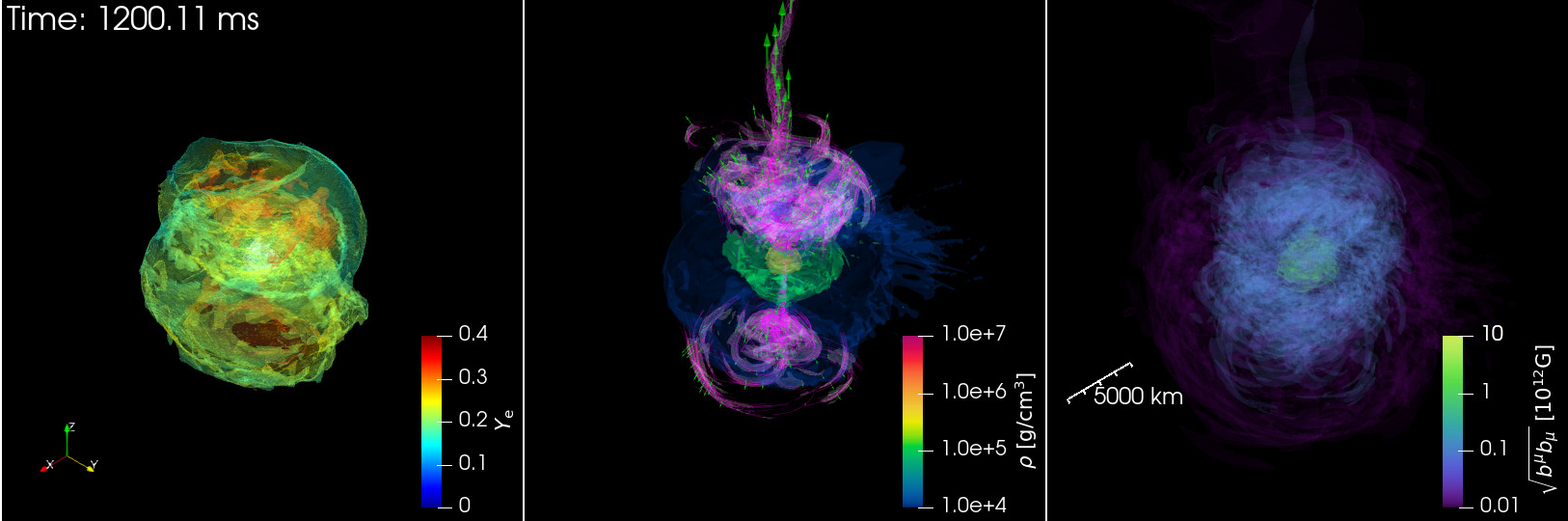} \\
        \vspace{1mm}
        \includegraphics[scale=0.26]{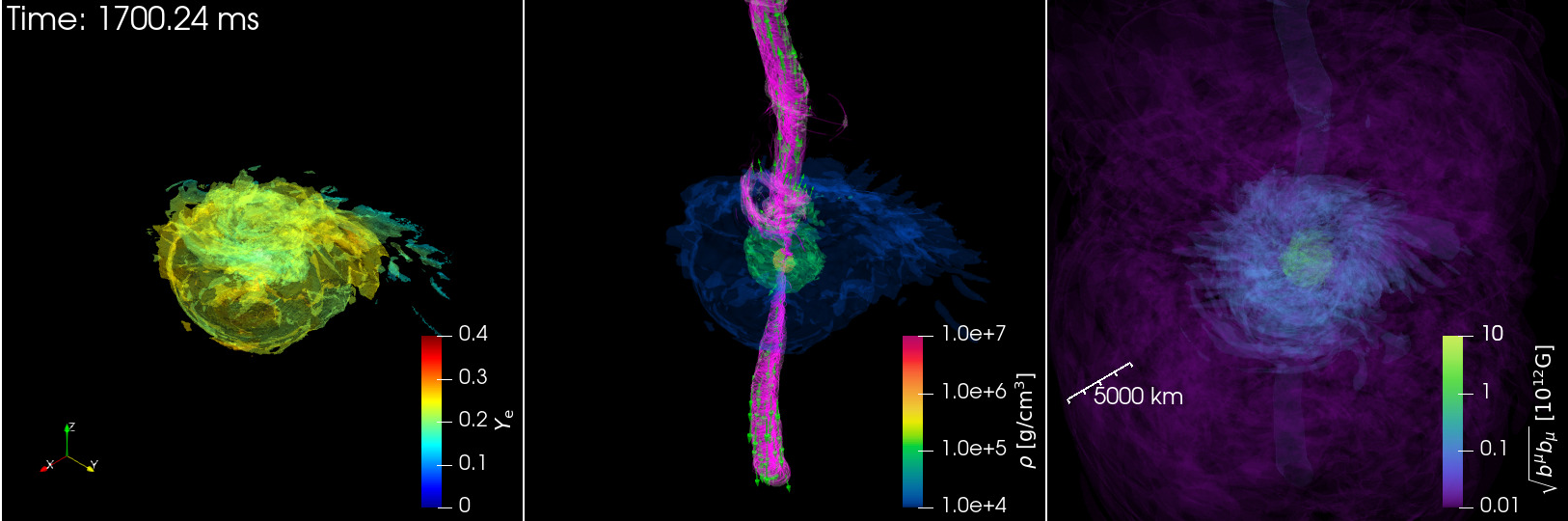} \\
        \vspace{1mm}
        \includegraphics[scale=0.26]{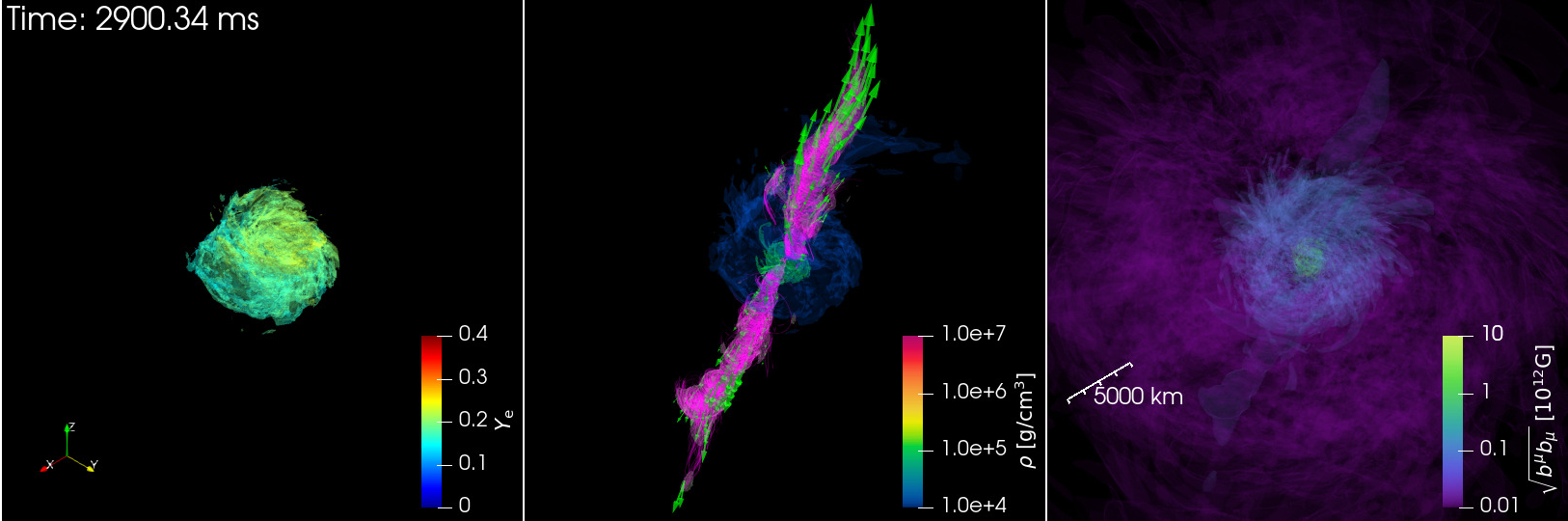} \\
        \caption{The three-dimensional snapshots for model Q4B5tn in a domain with the length scale of $\sim 10^{4}$\,km at $t\approx150$, $1200$, $1700$, and $2900$\,ms. For each time slice, 
        the left panel shows the ejecta, which is colored for the electron fraction $Y_{\mathrm{e}}$; the middle panel shows the rest-mass density $\rho$\,$\mathrm{(g/cm^3)}$ (contours) with magnetic-field lines (pink lines), unbound outflow (white color) and its velocity (green arrows); the right panel shows the magnetic-field strength $b=\sqrt{b^{\mu}b_{\mu}}$\,$\mathrm{(G)}$.  
        See also the following link for the animation: \url{https://www2.yukawa.kyoto-u.ac.jp/~kota.hayashi/Q4B5tn-3D.mp4}.
        }
        \label{fig:3D4_Q4B5tn}
\end{figure*}

\begin{figure*}[h]
      \begin{center}
        \includegraphics[scale=0.18]{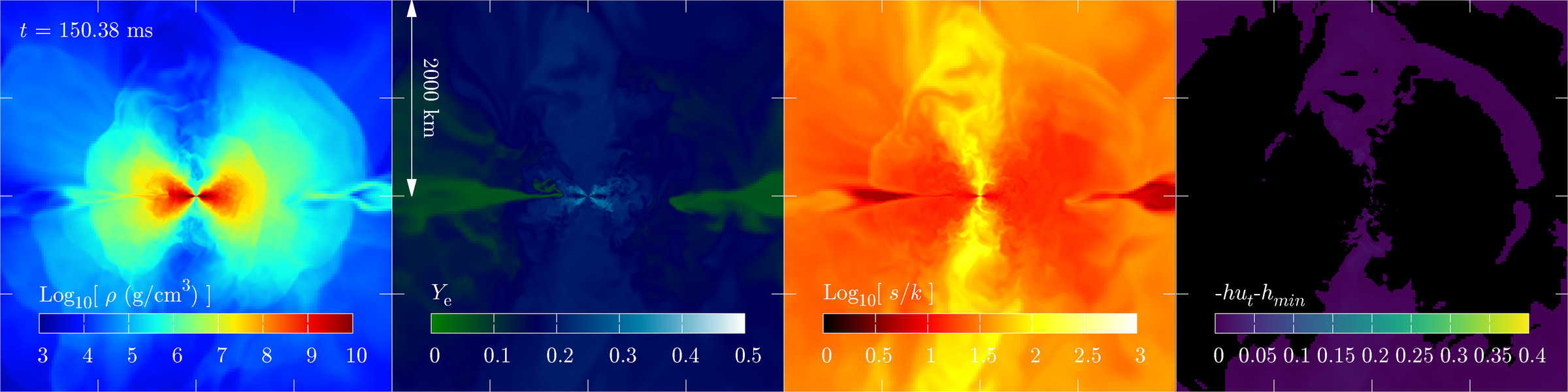} \\ 
        \vspace{2mm}
        \includegraphics[scale=0.18]{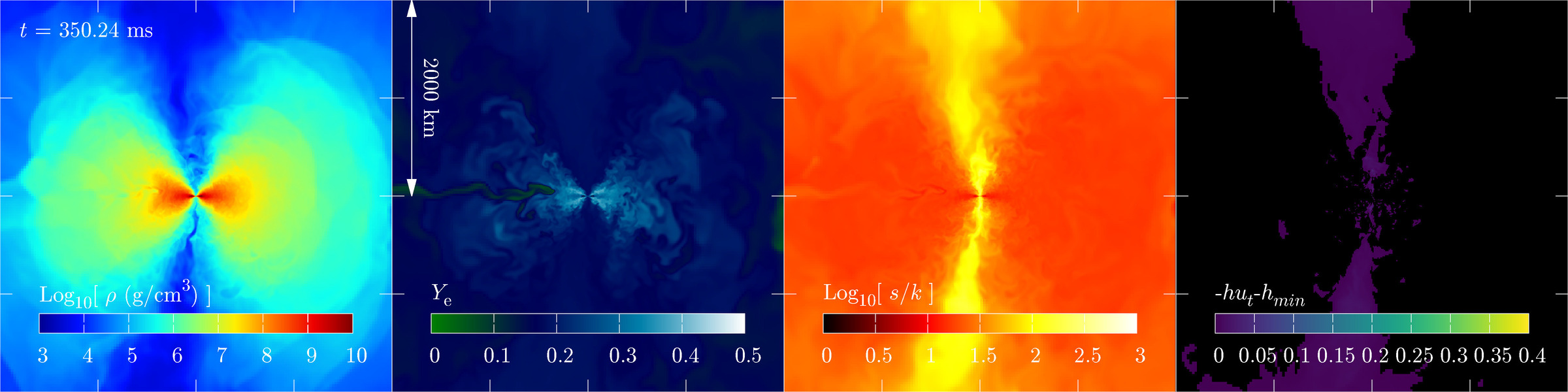} \\
        \vspace{2mm}
        \includegraphics[scale=0.18]{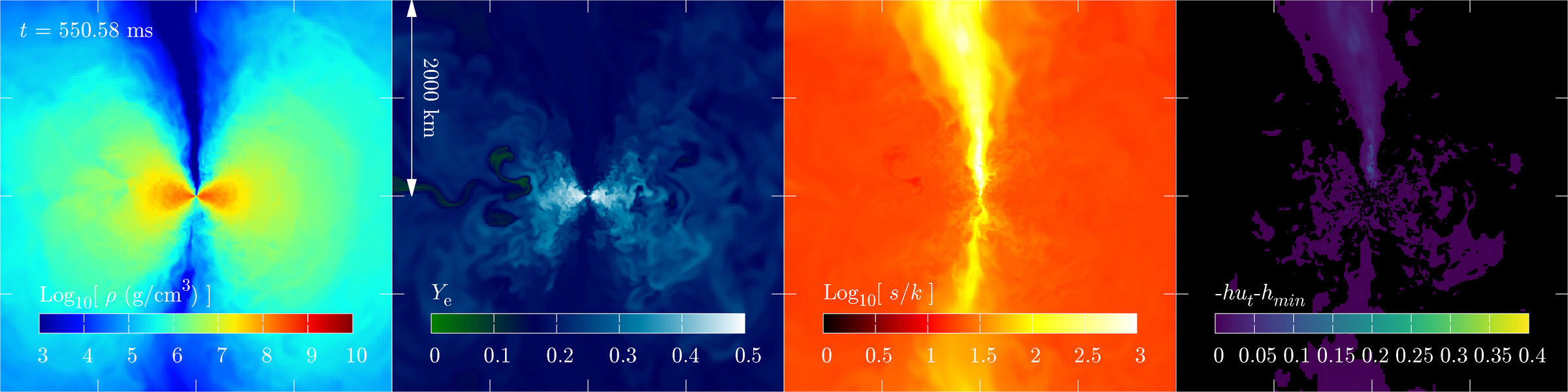} \\
        \vspace{2mm}
        \includegraphics[scale=0.18]{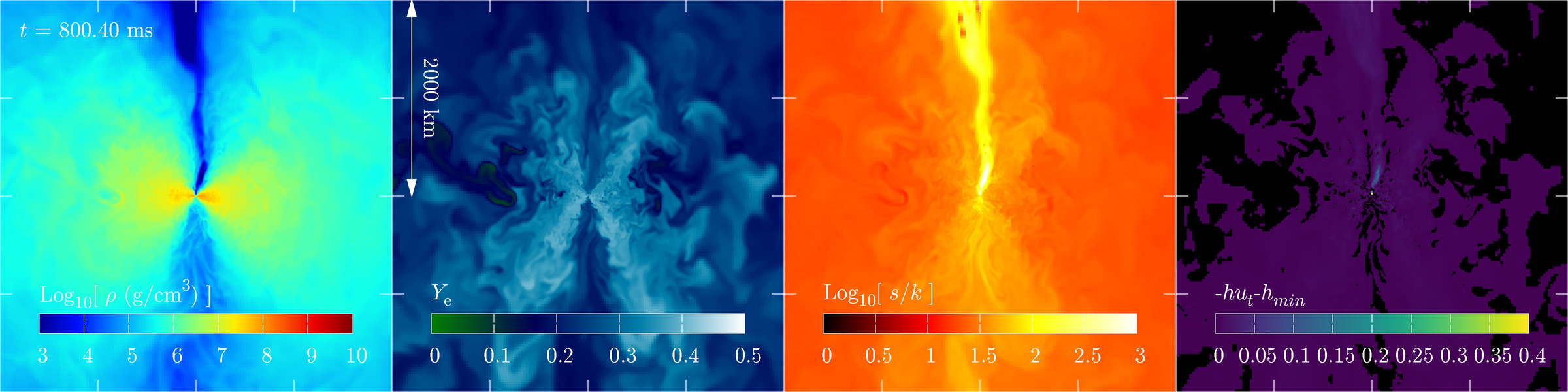} \\
        \vspace{2mm}
        \includegraphics[scale=0.18]{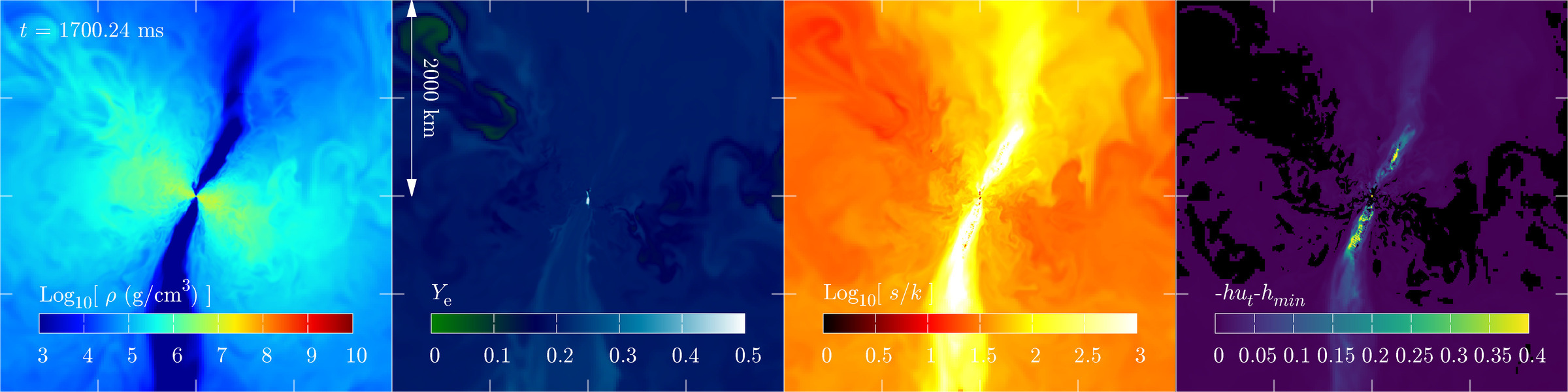} \\
        \caption{The 2D snapshots for model Q4B5tn on the $y$-$z$ plane with a region of  $[-2000~{\rm km}:2000~{\rm km}]$ for both $y$ and $z$ at $t\approx150$, $350$, $550$, $800$, and $1700$\,ms. For each time slice, the first, second, third, and fourth panels show the rest-mass density $\rho$\,$\mathrm{(g/cm^3)}$, the electron fraction $Y_{\mathrm{e}}$, the entropy per baryon $s$ in units of $k$, and $-hu_{t}-h_{\mathrm{min}}$, respectively.
        In the fourth panel, unbound matter is (non-black) colored and bound matter is colored by black. 
        See also the following link for the animation: \url{https://www2.yukawa.kyoto-u.ac.jp/~kota.hayashi/Q4B5tn-2Dyz.mp4}.
        }
        \label{fig:2D4_Q4B5tn}
      \end{center}
\end{figure*}

\begin{figure*}[!th]
      \begin{center}
        \includegraphics[scale=0.2]{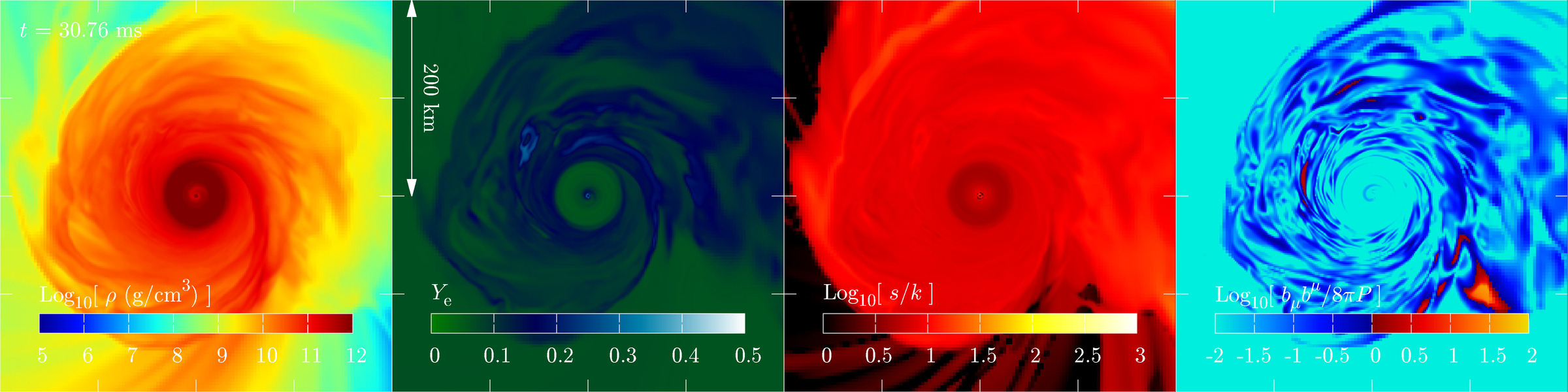} \\ 
        \vspace{2mm}
        \includegraphics[scale=0.2]{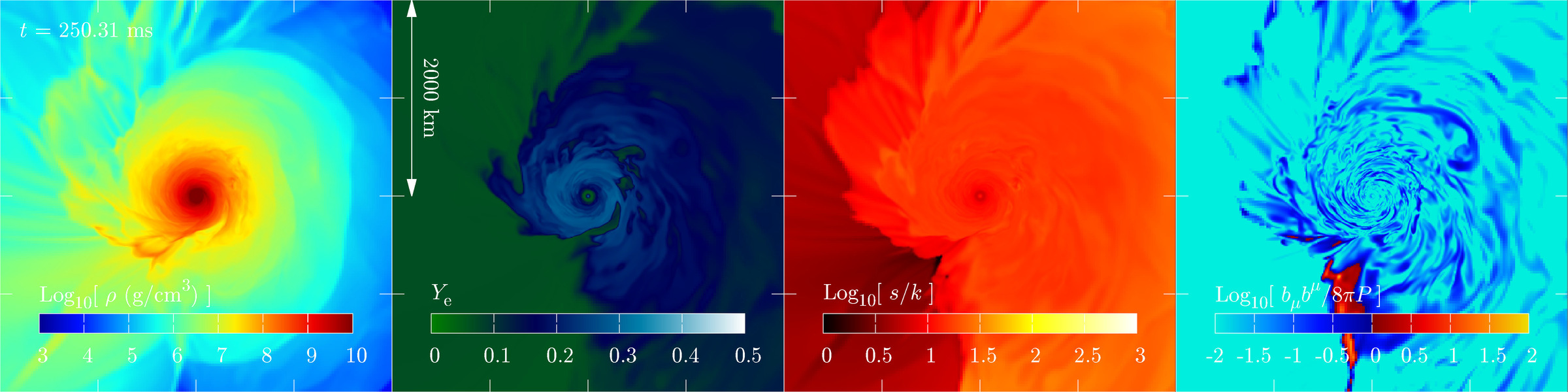} \\
        \vspace{2mm}
        \includegraphics[scale=0.2]{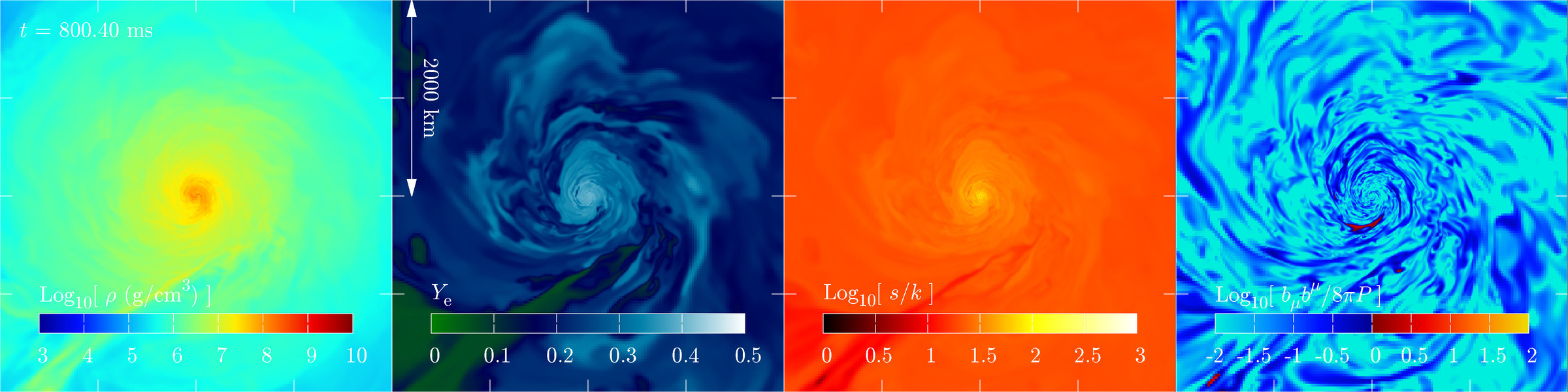} \\
        \caption{The 2D snapshots for model Q4B5tn on the $x$-$y$ plane with $[-200~{\rm km}:200~{\rm km}]$ for both $x$ and $y$ at $t\approx30$\,ms, and with $[-2000~{\rm km}:2000~{\rm km}]$ at $t\approx250$ and $800$\,ms. For each time slice, 
        the first, second, third, and fourth panels show the rest-mass density $\rho$\,$\mathrm{(g/cm^3)}$, the electron fraction $Y_{\mathrm{e}}$, the entropy per baryon $s$ in units of $k$, and the inverse of the plasma beta $b^2/8 \pi P$, respectively. 
        See also the following link for the animation: \url{https://www2.yukawa.kyoto-u.ac.jp/~kota.hayashi/Q4B5tn-2Dxy.mp4}. 
        }
        \label{fig:2D4xy_Q4B5tn}
      \end{center}
\end{figure*}

\addkh{
First, we show that the overall evolution process in the merger and post-merger stages are qualitatively the same as our previous findings in Ref.~\cite{hayashi2022jul} irrespective of initial magnetic field strengths, configurations, neutron-star EOSs, and equatorial-plane symmetry.
Table~\ref{tab:result_ave} shows a summary of the outcome quantities such as the mass, the averaged electron fraction and the velocity of the dynamical and post-merger ejecta.

Figure~\ref{fig:3D4_Q4B5tn} displays the three-dimensional snapshots for model Q4B5tn in a domain with the length scale of $\sim 10^{4}$\,km at $t\approx150$, $1200$, $1700$, and $2900$\,ms. For this model, we initially provide a toroidal magnetic field with its maximum strength of $5\times10^{16}$\,G, and do not impose the equatorial-plane symmetry. For each time slice, the left panel shows the ejecta, which is colored for the electron fraction $Y_{\mathrm{e}}$; the middle panel shows the rest-mass density $\rho$\,$\mathrm{(g/cm^3)}$ (contours) with magnetic-field lines (pink lines), unbound outflow (white color) and its velocity (green arrows); the right panel shows the magnetic-field strength defined by $b=\sqrt{b^{\mu}b_{\mu}}$\,$\mathrm{(G)}$ where $b^{\mu}$ is the magnetic field in the frame comoving with fluid (see Eq.~(\ref{tmunu}) for the definition of the energy-momentum tensor in our work). 
Figure~\ref{fig:2D4_Q4B5tn} displays the two-dimensional (2D) snapshots for the same model (Q4B5tn) on the $y$-$z$ plane with a region of $[-2000~{\rm km}:2000~{\rm km}]$ for both $y$ and $z$ at $t\approx150$, $350$, $550$, $800$, and $1700$\,ms. For each time slice, 
the first, second, third, and fourth panels show the rest-mass density, the electron fraction, the entropy per baryon $s$ in units of the Boltzmann constant $k$, and $-hu_{t}-h_{\mathrm{min}}$, respectively. 
Here, $u_t$ is the lower time component of the four velocity, $h$ is the specific enthalpy, and $h_{\mathrm{min}}$ is the minimum specific enthalpy for a given electron fraction.
In the fourth panel, unbound matter, which is identified by $-hu_{t}-h_{\mathrm{min}}>0$, is (non-black) colored and bound matter is colored by black. Figure~\ref{fig:2D4xy_Q4B5tn} displays the 2D snapshots for the same model (Q4B5tn) on the $x$-$y$ plane at $t\approx30$, $250$, and $800$\,ms. For each time slice, the first, second, third, and fourth panels show the rest-mass density, the electron fraction, the entropy per baryon, and the inverse of the plasma beta, $b^2/8 \pi P$, respectively. 
}

\addms{These figures show that the merger and post-merger processes are qualitatively quite similar to those found in our previous paper~\cite{hayashi2022jul}. The neutron star is tidally disrupted but about 80\% of the neutron-star matter falls immediately into the black hole. On the other hand, a part of the neutron-star matter becomes unbound from the system to be the dynamical ejecta and the rest of the matter forms an accretion disk around the remnant black hole. In the accretion disk which has shear layer due to the presence of the differential rotation, the magnetic field is amplified by the winding, MRI, and Kelvin-Helmholtz instability. Here the Kelvin-Helmholtz instability plays an important role for the region in which spiral arms collide with each other (see, e.g., the first snapshots of Fig.~\ref{fig:2D4xy_Q4B5tn}). After a sufficient amplification of the magnetic field, a turbulent state is developed, and as a result, the effective viscosity associated with the turbulent state is established. Then, the angular momentum is efficiently transported outward, and the density and temperature of the disk decrease with time. In an early stage of the disk evolution, the temperature is so high that the neutrino cooling is efficient,  and hence, the thermal energy generated by the effective viscous processes is liberated by the neutrino emission. However, after the disk temperature drops, the neutrino cooling is inefficient, and thus, the viscous heating can be used to eject a part of the matter from the disk, inducing the post-merger mass ejection. After the amplification of the magnetic field in the disk, the magnetic field along the spin axis of the black hole is also enhanced (see Fig.~\ref{fig:3D4_Q4B5tn}) forming a magnetosphere along the spin axis of the black hole, in which the density is low (see, e.g., Fig.~\ref{fig:2D4_Q4B5tn}) and \addkh{the magnetization parameter $b^2/4\pi\rho$ is above unity.}   
}

\addkh{
Although the overall qualitative evolution process is mostly independent of the initial and numerical setups, we find some differences in the post-merger mass ejection and the evolution of the magnetosphere.
In the absence of the equatorial-plane symmetry, the distribution of the post-merger ejecta is not equatorially symmetric (see, e.g., the third and fourth snapshots of Fig.~\ref{fig:2D4_Q4B5tn}).
We also observe that the accretion disk and the magnetosphere tilt with time in the late evolution stage of $t > 2$\,s (see, e.g., the fifth snapshots of Fig.~\ref{fig:2D4_Q4B5tn}). 
In addition, another model with no equatorial-plane symmetry (model Q4B5n) shows that magnetic-field polarity in the magnetosphere reverses (see Sec.~\ref{sec:pflux_and_mag}).
These differences are likely due to the fact that the post-merger evolution is determined by the stochastic turbulent processes in the disk.
The properties of the magnetosphere mentioned here play a role in the decreasing stage of the isotropic-equivalent Poynting luminosity (see Sec.~\ref{sec:time_duration_pflux}).

In the following, thus, we present the results for a more detailed analysis of the system paying attention to the quantitative dependence on the numerical setting. 
In addition, the anisotropic part of the Maxwell and the Reynolds stresses are carefully evaluated to discuss the qualitative aspect of the effective viscosity induced by the magnetohydrodynamical turbulence (see Sec.~\ref{sec:results-viscosity}).
We also propose a new method to assess whether the black hole has the ability to form a magnetosphere and launch a jet by evaluating magnetohydrodynamics properties near the horizon (see Sec.~\ref{sec:madness}).
}

\subsection{The evolution of the accretion disk and post-merger mass ejection} \label{sec:results-disk}

\begin{figure}[!th]
      \begin{center}
        \includegraphics[scale=0.4]{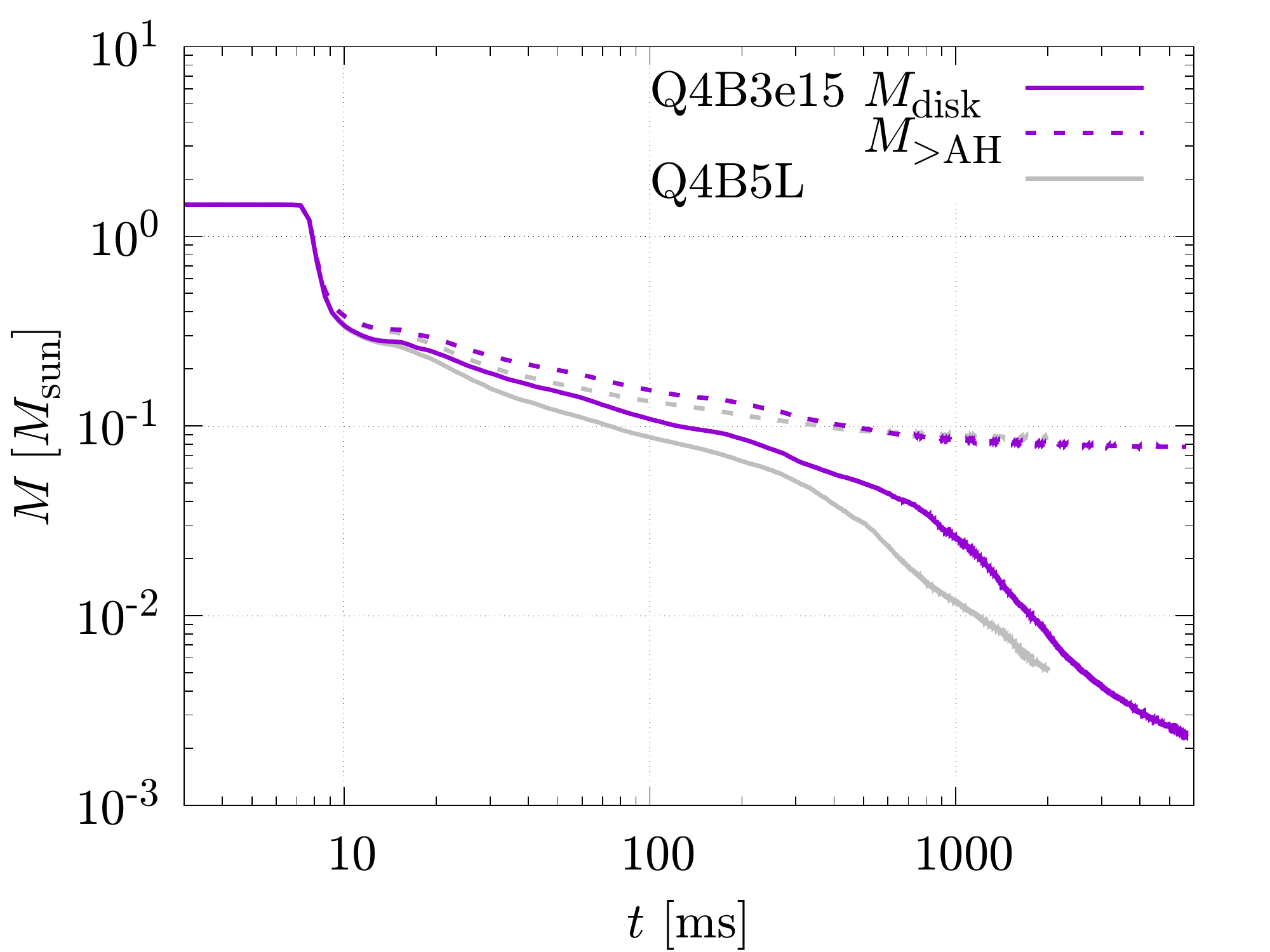} \\
        \includegraphics[scale=0.4]{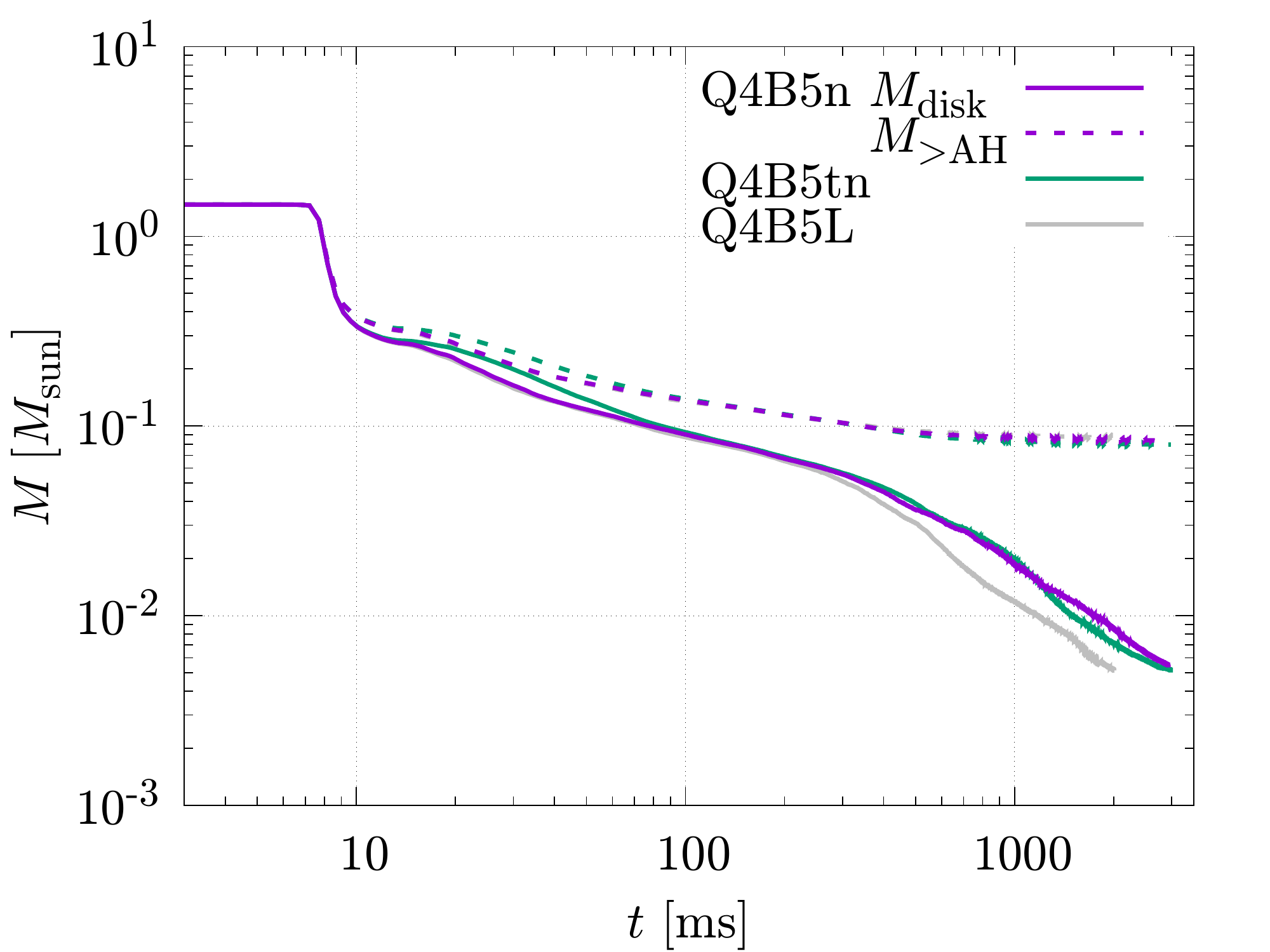} \\
        \includegraphics[scale=0.4]{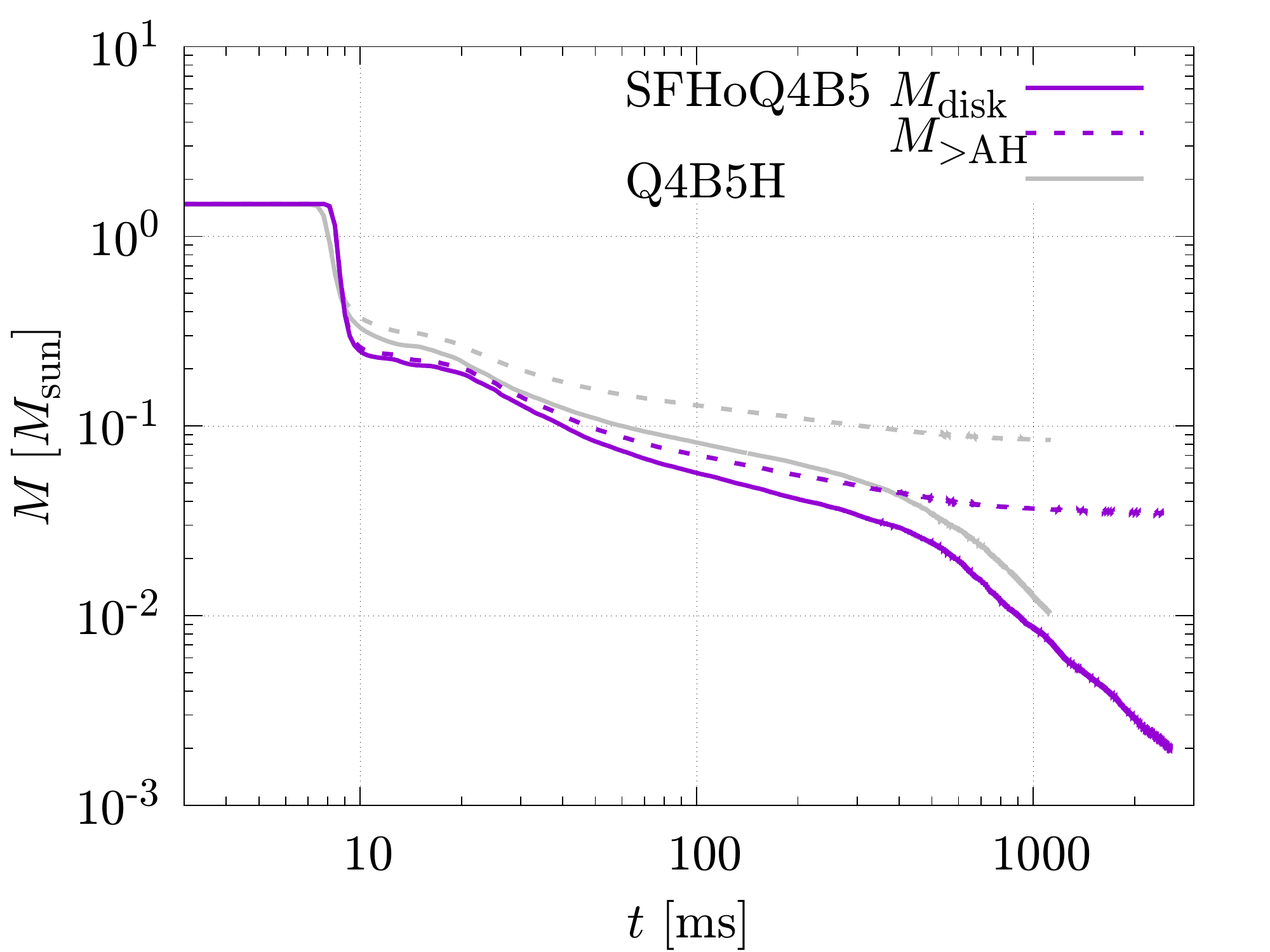}
        \caption{The time evolution of the rest mass of the matter located outside the apparent horizon (dotted curves) and the accretion-disk mass (solid curves) 
        for models Q4B3e15 (top panel), Q4B5n and Q4B5tn (middle panel), and SFHoQ4B5 (bottom panel).
        The results for models Q4B5L and Q4B5H of our previous paper~\cite{hayashi2022jul} are also shown for comparison (in grey color).
        }
        \label{fig:mrem}
      \end{center}
\end{figure}

\begin{figure}[!th]
      \begin{center}
        \includegraphics[scale=0.4]{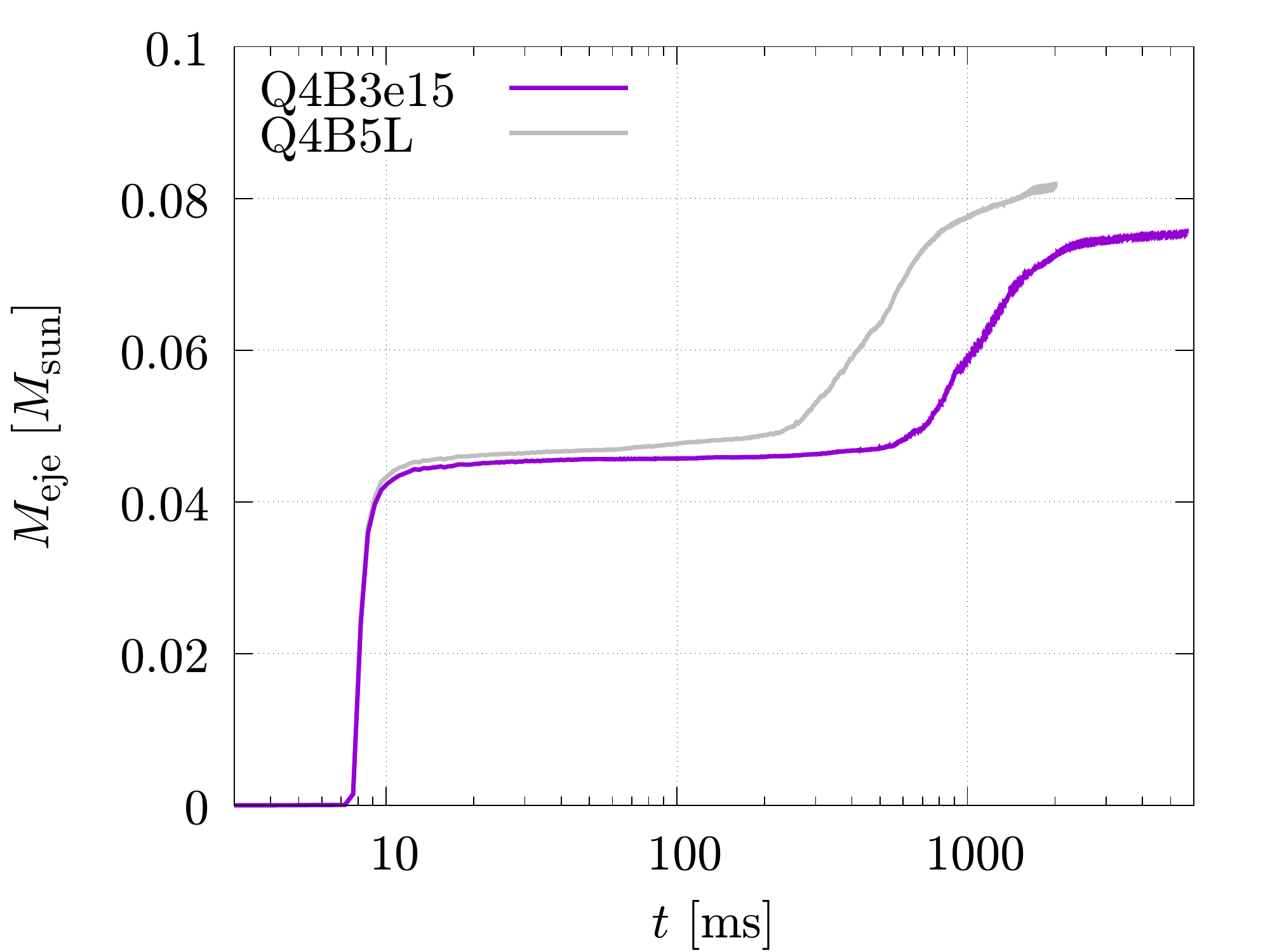} \\
        \includegraphics[scale=0.4]{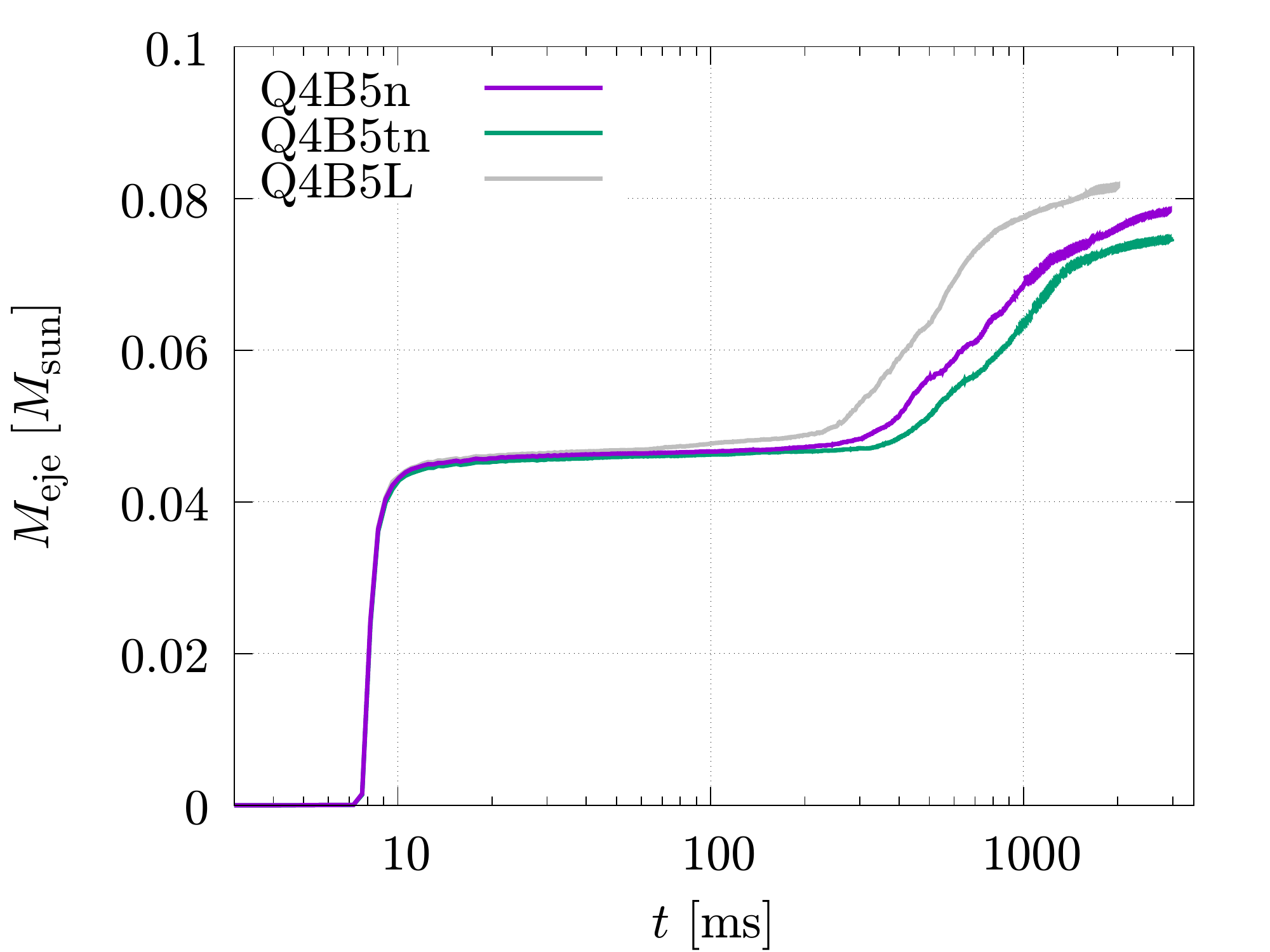} \\
        \includegraphics[scale=0.4]{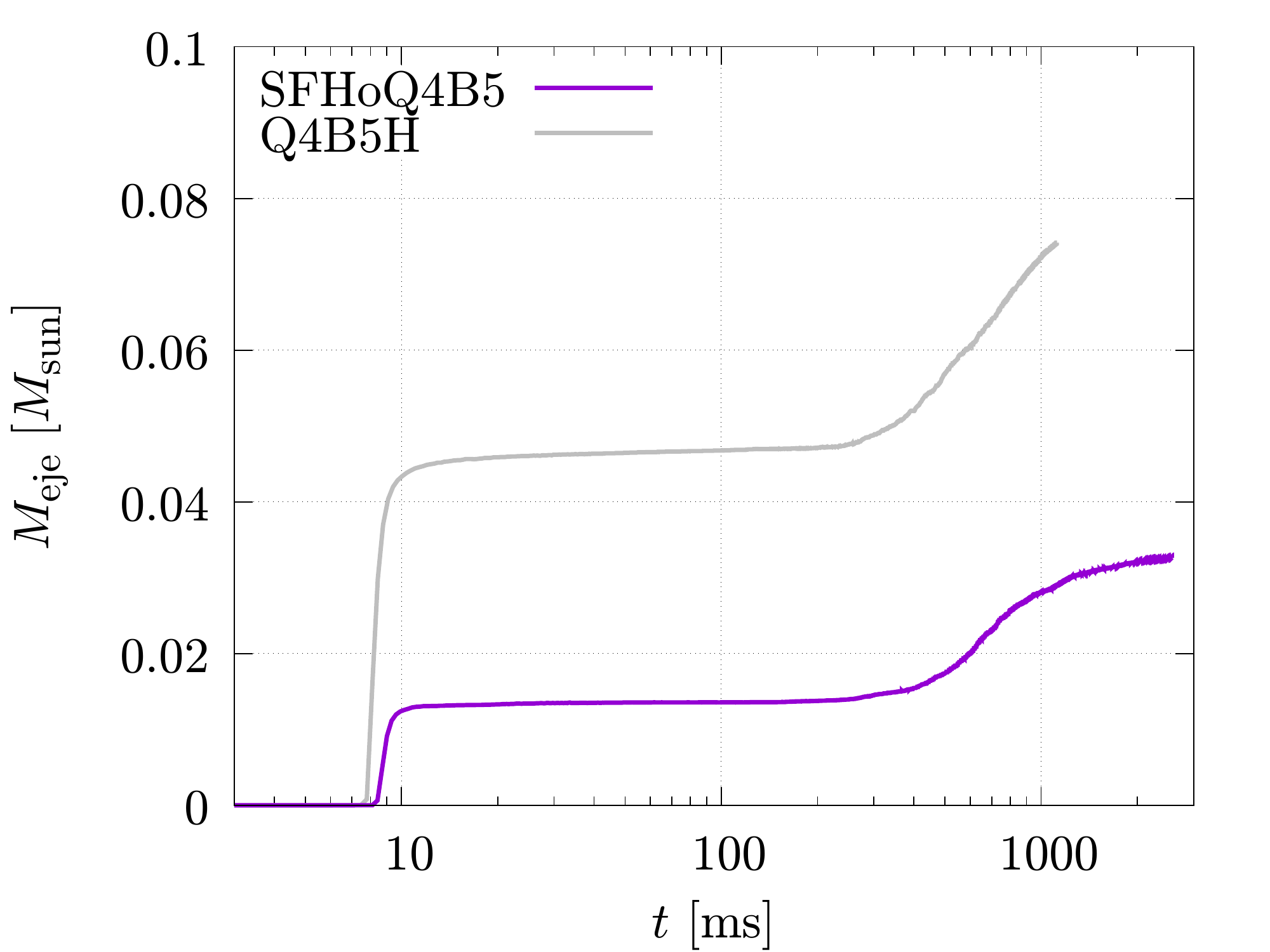}
        \caption{The same as Fig.~\ref{fig:mrem} but for the time evolution of the rest mass of the unbound matter. 
        }
        \label{fig:meje}
      \end{center}
\end{figure}

\begin{figure}[!th]
      \begin{center}
        \includegraphics[scale=0.4]{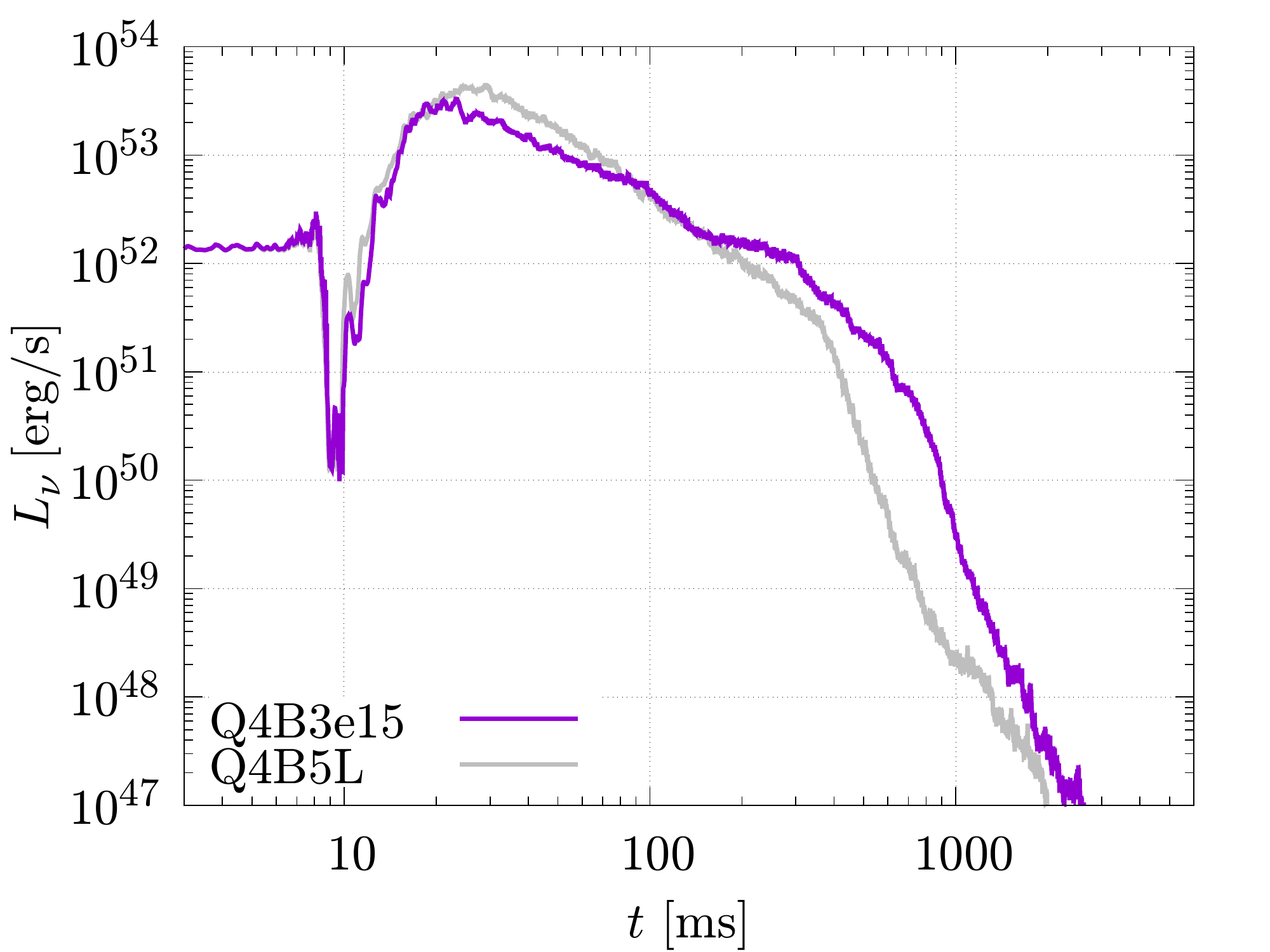} \\
        \includegraphics[scale=0.4]{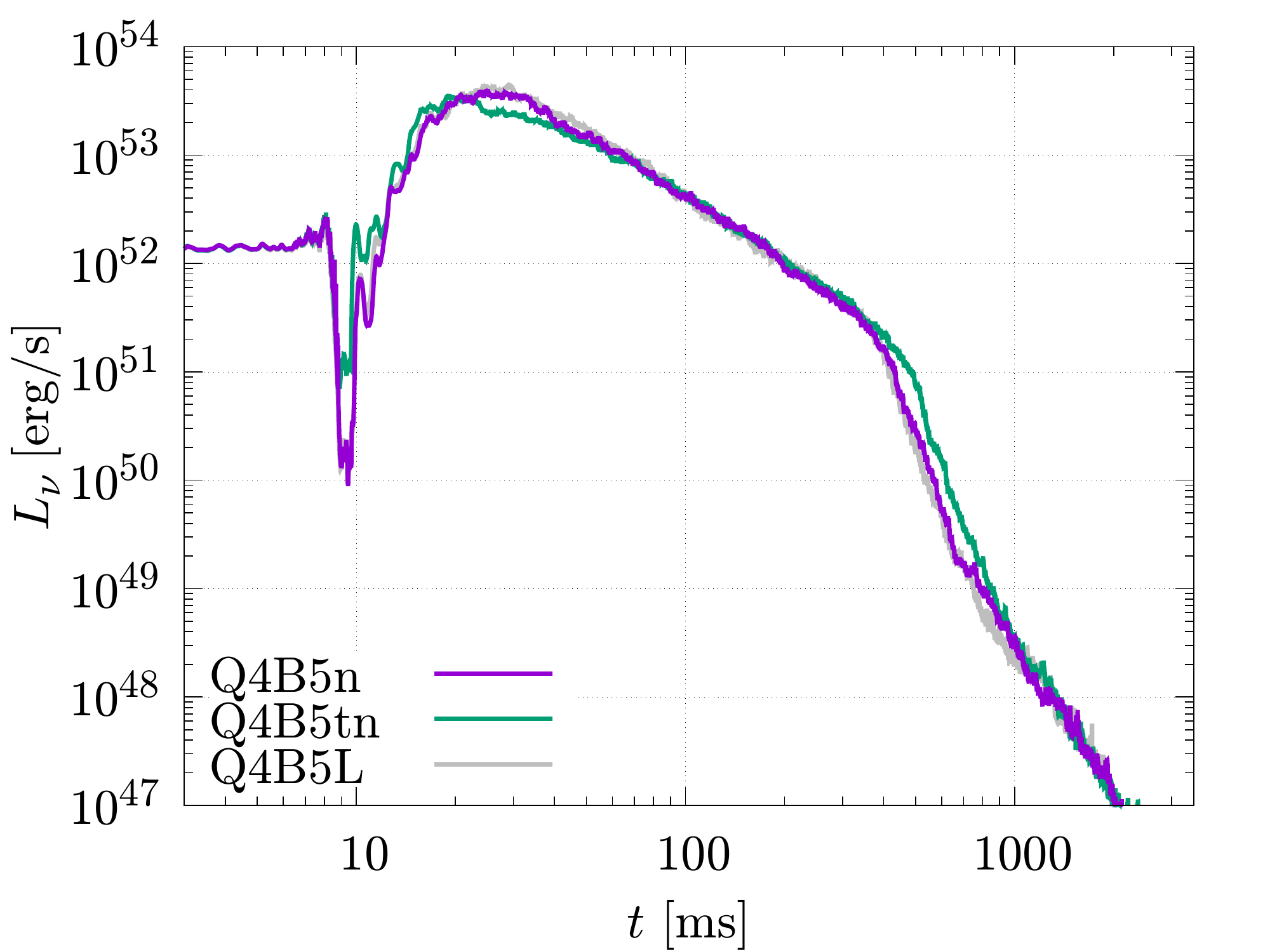} \\
        \includegraphics[scale=0.4]{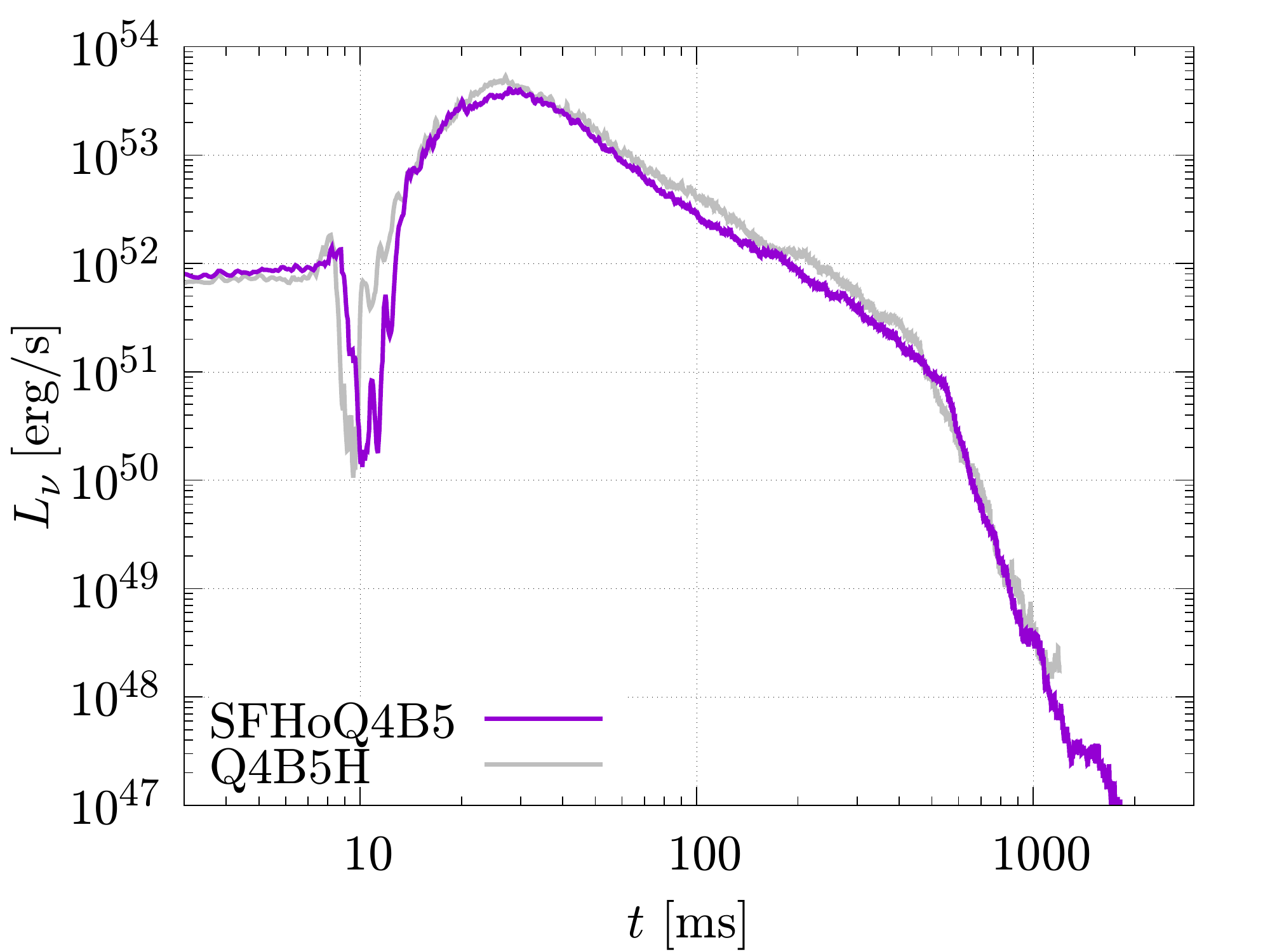} 
        \caption{The same as Fig.~\ref{fig:mrem} but for the time evolution of the total neutrino luminosity. The post-merger mass ejection sets in at $t \sim 300$--600\,ms at which $L_\nu \sim 10^{51}$--$10^{51.5}~{\rm erg/s}$. 
        }
        \label{fig:nlum}
      \end{center}
\end{figure}

\subsubsection{Disk evolution and ejecta}

In this subsection, we present details on the evolution of the accretion disk and on the post-merger mass ejection \addms{focusing particularly on the quantitative dependence on the models.} Figure~\ref{fig:mrem} shows the time evolution of the rest mass of the matter located outside the apparent horizon $M_{>\mathrm{AH}}$ 
(dashed curves) and the accretion disk mass $M_{\rm disk}$ (solid curves). 
Figure~\ref{fig:meje} shows the time evolution of the rest mass of the unbound matter (ejecta) $M_{\mathrm{eje}}$. 
For both figures, the new results obtained in this paper are compared with the previous results of Ref.~\cite{hayashi2022jul}. 

The quantities shown in Figs.~\ref{fig:mrem} and \ref{fig:meje} are defined by
\begin{eqnarray}
  M_{> {\rm AH}}&:=&\int_{r>r_{\rm AH}}\rho_*d^3x + M_{\rm esc}, \\
  M_{\rm eje}&:=&\int_{-hu_{t}>h_{\rm min},r>r_{\rm AH}}\rho_*d^3x + M_{\rm esc}, \\
  M_{\rm disk}&:=&M_{> {\rm AH}}-M_{\rm eje},
\end{eqnarray}
where $\rho_*:=\rho \sqrt{-g} u^t$ with $g$ the determinant of the spacetime metric, $g_{\mu\nu}$, $u^t$ the time component of the four velocity, $u^\mu$, and $r_{\rm AH}$ denotes the coordinate radius of the apparent horizon with respect to the position of the black-hole puncture.  
$M_{\rm esc}$ denotes the rest mass of the matter escaping from the computational domain calculated by
\begin{eqnarray}
  \dot{M}_{\rm esc}&:=&\oint \rho \sqrt{-g} u^{i} dS_{i}, \\
  M_{\rm esc}&:=&\int^{t} \dot{M}_{\rm esc} dt.
\end{eqnarray}
The surface integral is performed at the outer boundaries of the computational domain.
The ejecta component is identified by considering the Bernoulli criterion; we identify the matter located outside the apparent horizon that satisfies $hu_t<-h_{\rm min}$ as the unbound component.  

Our new results for $M_{>\mathrm{AH}}$, $M_\mathrm{disk}$, and $M_\mathrm{eje}$ show that the dependence of these quantities on the initial magnetic field setup and equatorial symmetry imposed is weak. 
Also, the results for the SFHo EOS model are qualitatively similar to those for the DD2 model, although for the SFHo model, $M_{>\mathrm{AH}}$, $M_\mathrm{disk}$, and $M_\mathrm{eje}$ are smaller than those for the DD2 model reflecting the smaller neutron-star radius for the SFHo model. 

$M_{>\mathrm{AH}}$ decreases steeply at $\sim 10$\,ms at which the merger occurs and $\sim 80$--$85\%$ of the neutron-star matter plunges into the black hole.
After that, $M_{>\mathrm{AH}}$ continues to decrease gradually due to the matter accretion into the black hole induced by the angular-momentum transport from the magnetohydrodynamics effect. $M_\mathrm{disk}$ right after the merger is $M_\mathrm{disk,0}\approx 0.28 M_{\odot}$ and $\approx 0.22 M_{\odot}$ for the DD2 and SFHo models, respectively.

The evolution of $M_\mathrm{eje}$ clearly shows that two distinct components of the ejecta exist. One is the dynamical ejecta, for which $M_\mathrm{eje}$ steeply increases right after the merger spending only for a few ms.
The rest mass for this component is $\approx 0.046 M_{\odot}$ and $\approx 0.013 M_{\odot}$ for DD2 and SFHo models, respectively.
After this increase by the dynamical mass ejection, $M_\mathrm{eje}$ remains approximately constant for several hundred ms.
Then, $M_\mathrm{eje}$ starts increasing again at $t \sim 300$--$600$\,ms. 
This component is the post-merger ejecta driven by the heating associated with MRI turbulence after the neutrino luminosity decreases below the heating rate (see Fig.~\ref{fig:nlum}), i.e., $L_\nu$ decreases below $10^{51}$--$10^{51.5}$\,erg/s. The rest mass for this component is $\approx 0.030 M_{\odot}$ and $\approx 0.019 M_{\odot}$ for the DD2 and SFHo models at the termination of the simulations, respectively.~\footnote{Strictly speaking we can provide only the lower bound of the post-merger ejecta mass because at the termination of the simulations, the mass is still increasing slightly. Note, however, that the possible additional increase is less than $0.01M_{\odot}$ because the disk mass at the termination of the simulation is less than $0.01M_{\odot}$.}
These values are about 10\% of $M_\mathrm{disk,0}$. The result for the DD2 model shows good agreement with our previous results~\cite{hayashi2022jul}. 

\addkh{
Compared with the simulations starting from a black hole-disk system such as Refs.~\cite{daniel2018may,fernandez2018oct,christie2019sep}, the ratio of the post-merger ejecta mass to the remnant disk mass for the present simulations is smaller by a factor of several.
One reason for this is that the early-post-merger mass ejection (which could occur within 100--200 ms after the onset of the merger) is not found in the present simulations. 
This is because the strong poloidal magnetic field, which is the key for this, is not developed in the disk remaining after the merger. 
Another reason is that the mass accretion onto the black hole is enhanced due to the highly non-axisymmetric structure of the remnant disk from black hole-neutron star mergers.}

The only significant quantitative dependence on the computational setting is found on the onset time of the post-merger mass ejection. For example, for model Q4B3e15 which has the low initial-magnetic-field strength, the onset time of the post-merger mass ejection is $t \sim 600$\,ms, i.e., $\sim 200$\,ms behind a high initial-magnetic-field strength model Q4B5L. 
The reason for this is that for the model with the low initial field strength it takes a longer time until the magnetic-field strength is enhanced enough for the disk to be in the equipartition state and for numerical computation to resolve the fastest growing mode of the MRI.
It results in the delay of the development of the MRI-induced turbulence in the accretion disk. Note, however, that this delay \addms{is likely due to the insufficient grid resolution and} may not be present in the realistic case, in which the MRI is resolved even for the weak fields.
Besides this difference, \addms{the quantities of the post-merger mass ejection such as the ejecta mass (see Table~\ref{tab:result_ave}) are similar to those} found in our previous paper~\cite{hayashi2022jul}.

\subsubsection{Magnetic-field evolution}

\begin{figure}[!th]
      \begin{center}
        \includegraphics[scale=0.4]{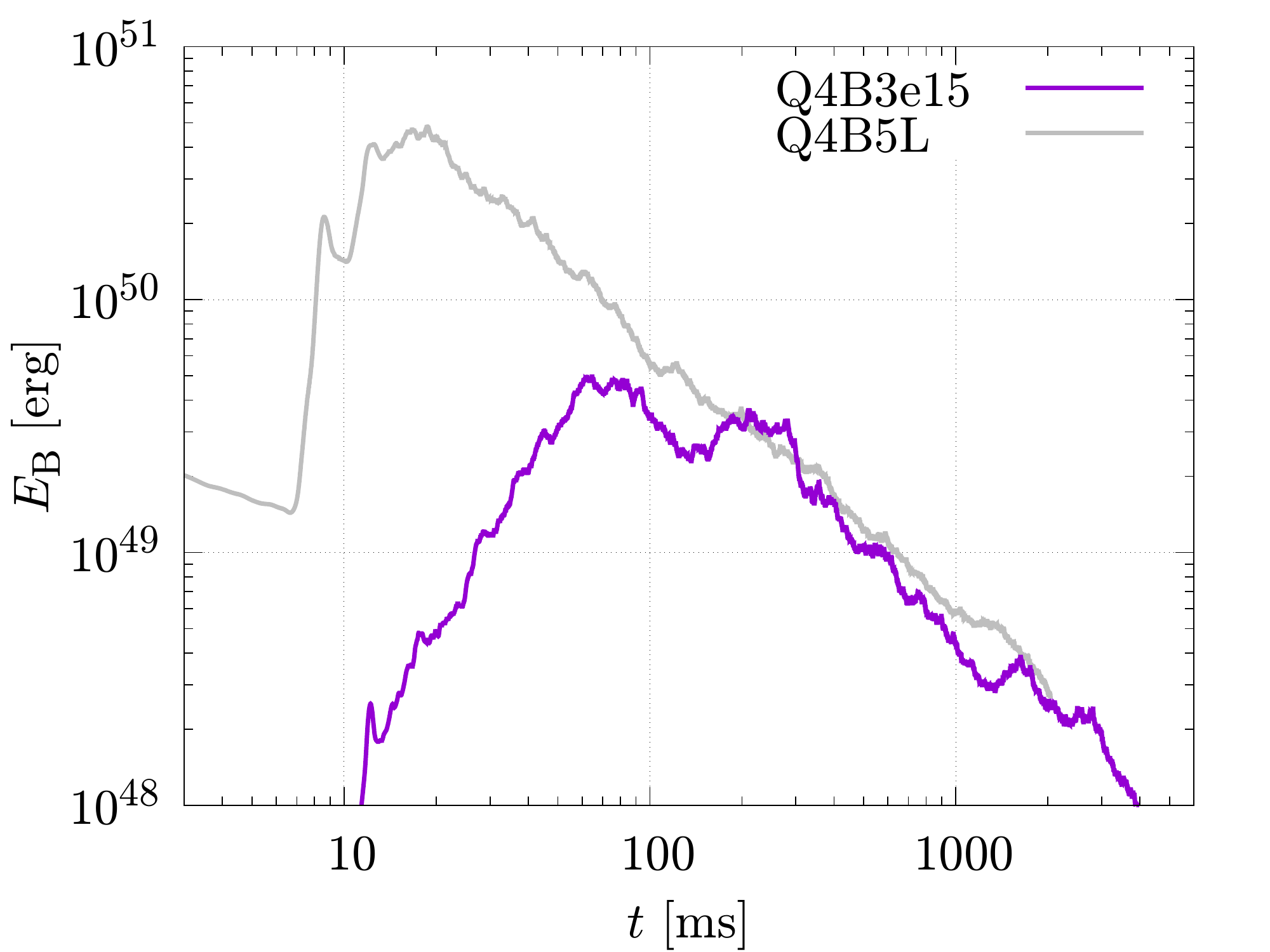} \\
        \includegraphics[scale=0.4]{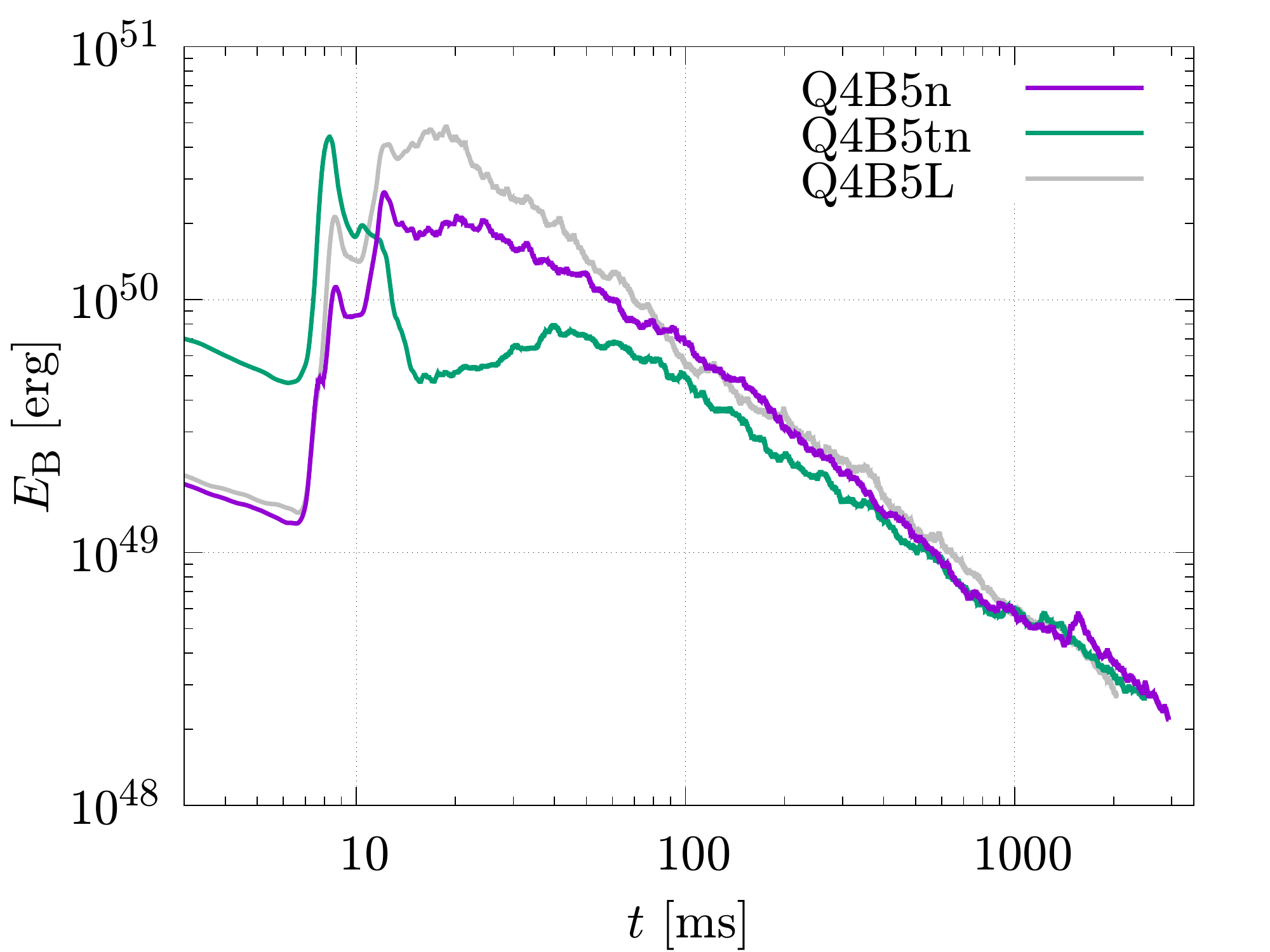} \\
        \includegraphics[scale=0.4]{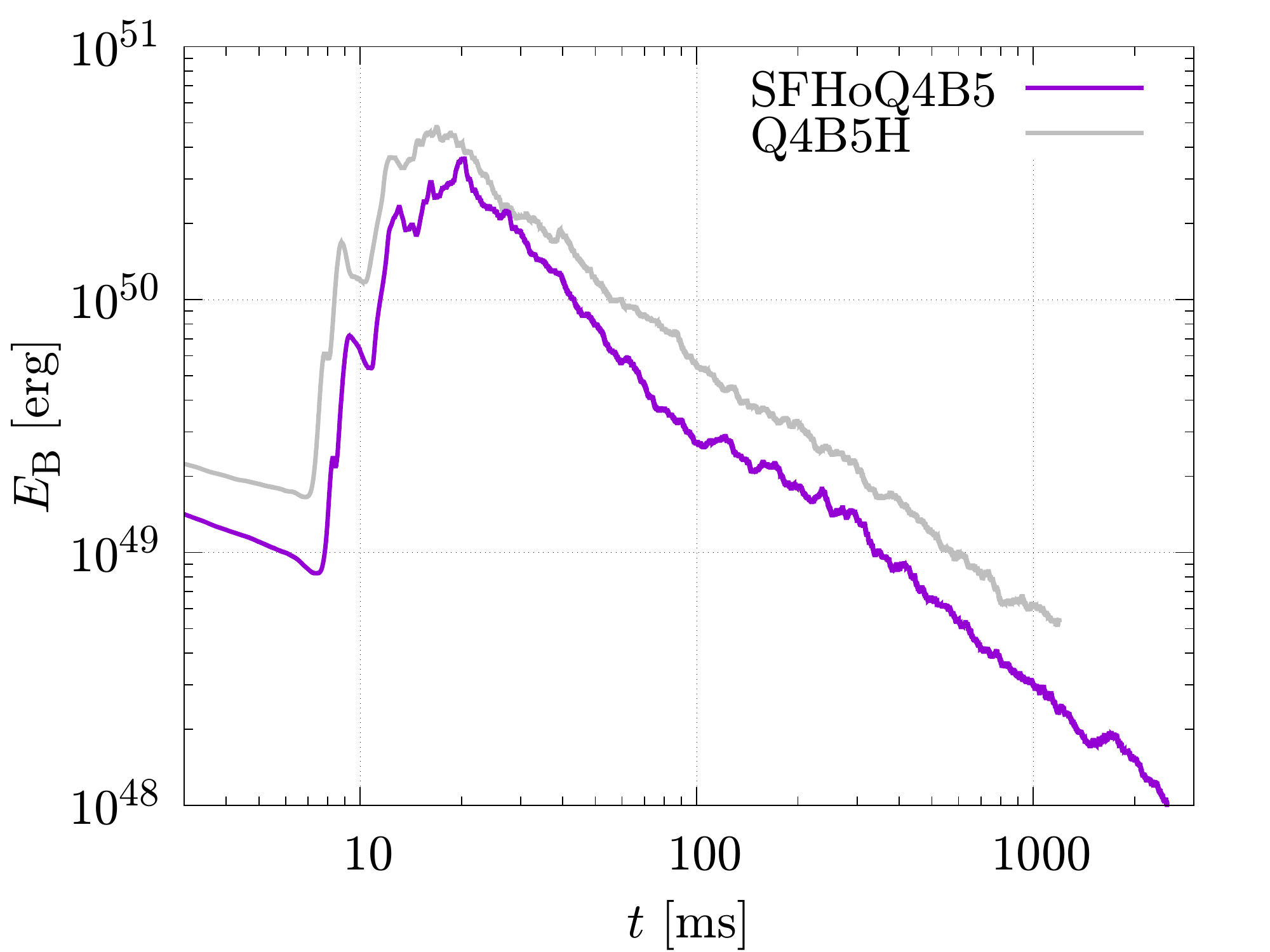} 
        \caption{The time evolution of the electromagnetic energy evaluated 
        for models Q4B3e15 (top panel), Q4B5n and Q4B5tn (middle panel), and SFHoQ4B5 (bottom panel).
        The results for models Q4B5L and Q4B5H of our previous paper~\cite{hayashi2022jul} are also shown for comparison (in grey color).
	}
        \label{fig:eb}
      \end{center}
\end{figure}

\begin{figure}[!th]
      \begin{center}
        \includegraphics[scale=0.4]{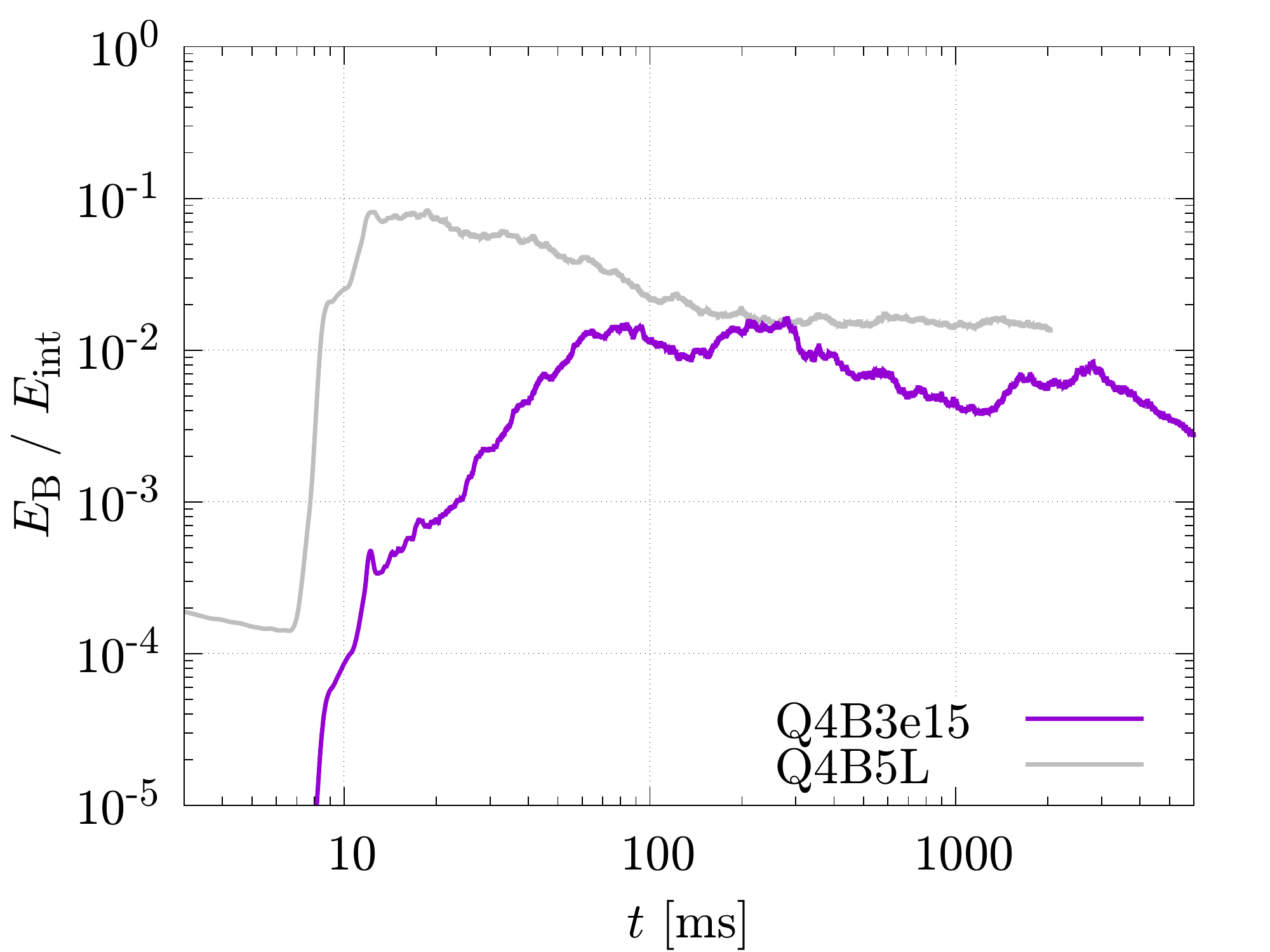} \\
        \includegraphics[scale=0.4]{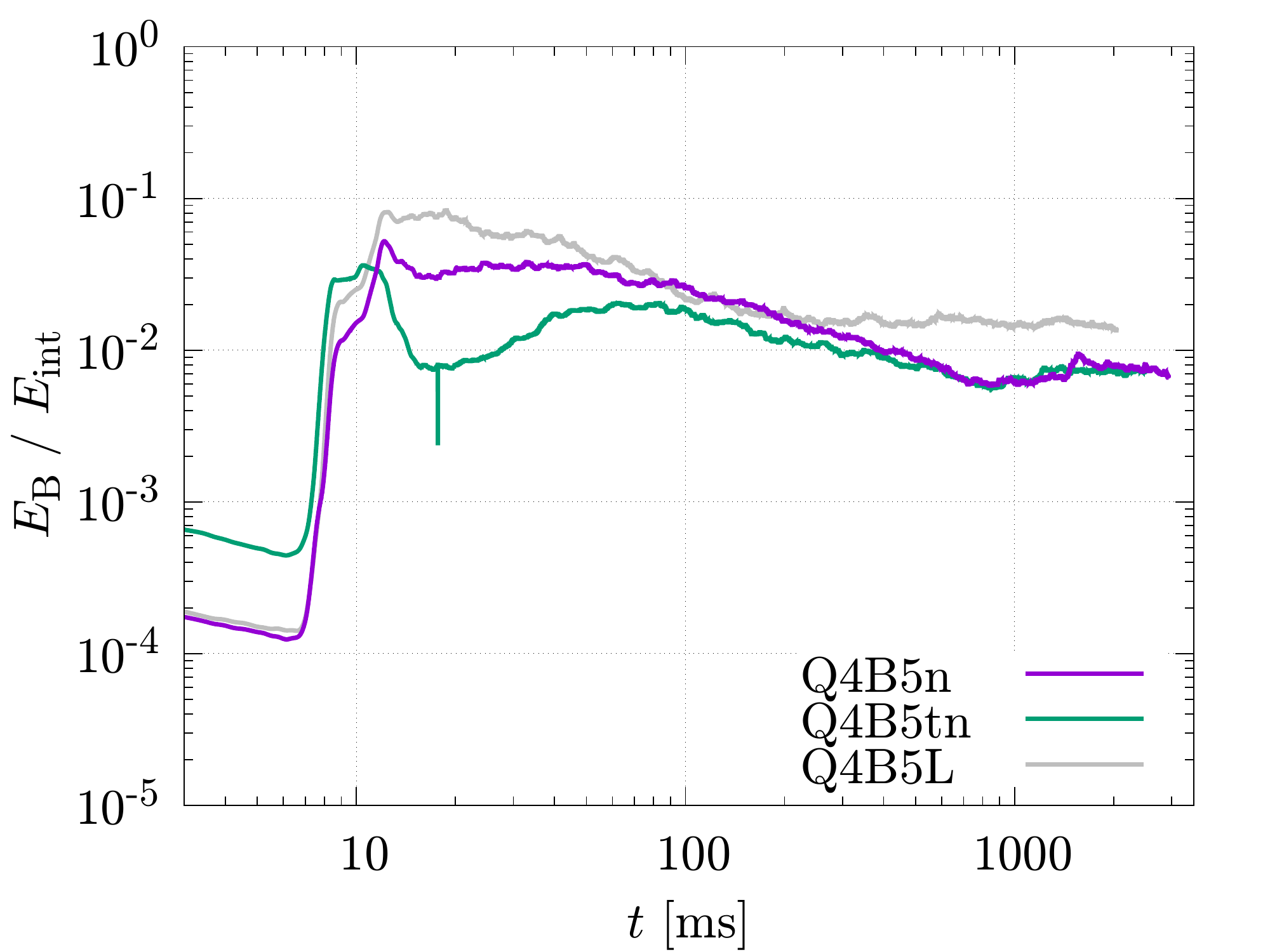} \\
        \includegraphics[scale=0.4]{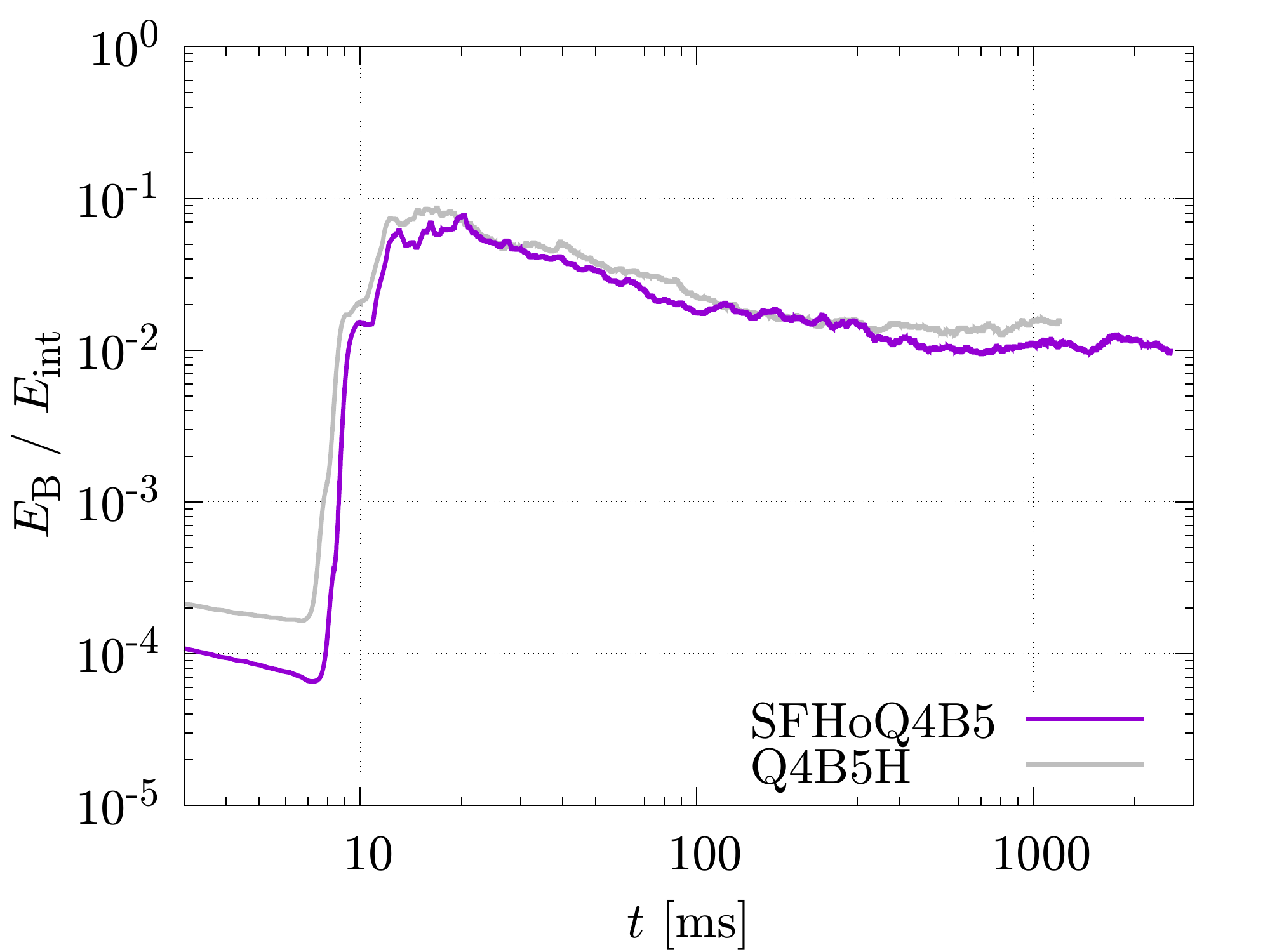} 
        \caption{The same as Fig.~\ref{fig:eb} but for the time evolution of the ratio of the electromagnetic energy 
        to the internal energy evaluated for the outside of the apparent horizon. 
	}
        \label{fig:eb_eint}
      \end{center}
\end{figure}

Figures~\ref{fig:eb} and \ref{fig:eb_eint} show the time evolution of the electromagnetic energy, $E_{\rm B}$, and the ratio of the electromagnetic energy to the internal energy, $E_{\rm int}$, respectively.
Here, $E_{\rm B}$ and $E_{\rm int}$ are defined, respectively, by
\begin{eqnarray}
  E_{\rm B}&:=&\frac{1}{8\pi}\int_{r>r_{\rm AH}}u^t \sqrt{-g}\,b_{\mu}b^{\mu}d^3x, \\
  E_{\rm int}&:=&\int_{r>r_{\rm AH}}\rho_* \varepsilon d^3x,
\end{eqnarray}
where $\varepsilon$ denotes the specific internal energy. 
In this paper the energy-momentum tensor for the ideal magnetohydrodynamics is written as
\begin{eqnarray}
T_{\mu\nu}&=&\rho h u_\mu u_\nu + P g_{\mu\nu} \nonumber \\
&+& {1 \over 4\pi} \left(b^\alpha b_\alpha u_\mu u_\nu + {1 \over 2}b^\alpha b_\alpha g_{\mu\nu}-b_\mu b_\nu\right). 
    \label{tmunu}
\end{eqnarray}

A quantitative difference is found for models Q4B3e15 and Q4B5nt at $t \alt 100$\,ms, which corresponds to the stage where the disk is not yet in the equipartition state.
For model Q4B3e15, the initial magnetic field is weak, and hence, \addms{the MRI is not well resolved and moreover the magnetic-field amplification by winding is insufficient to reach the saturation in the early stage after the merger}. 
For this case, the magnetic-field energy keeps increasing until $t \sim 60$\, ms at which the disk settles eventually into an equipartition state, and thus, the magnetic-field amplification saturates.
The main contributor to this amplification is clearly understood to be the magnetic winding because $E_{B}$ increases in proportion to $t^{2}$ approximately for $t \approx 20$--$60$\,ms. 
In this model, the MRI is not well resolved until $t \sim 200$\,ms because of the insufficient grid resolution and/or the insufficient magnetic-field strength (see below), so that we can only find the effect of winding in the magnetic-field amplification.
However, for a real system (or in an ideal computation with an infinite grid resolution), the MRI should take place significantly from an earlier stage after the merger.
We may expect that in such realistic cases, the disk would achieve the equipartition state earlier. \addms{It is worthy to note that the onset time of the post-merger mass ejection numerically found correlates with the time at which the equipartition is established (cf.~Fig.~\ref{fig:meje}). Thus, the present results illustrate that it is not easy to strictly identify the onset time of the post-merger mass ejection by numerical simulation (see also a discusion in Sec.~\ref{sec:results-viscosity}).}

For model Q4B5tn, we find another remarkable behavior right after the merger: 
During the merger stage $E_\mathrm{B}$ is amplified up to $\sim 4\times10^{51}$\,$\mathrm{erg}$, but it rapidly drops by an order of magnitude to $\sim 4\times10^{50}$\,$\mathrm{erg}$. We do not see this drop for other models.
Our interpretation for this drop is that the magnetic-field dissipation by reconnection near the equatorial plane occurs. 
For this model, we initially embed a strong magnetic field with opposite polarities across the equatorial plane. This magnetic-field configuration is the source of the efficient magnetic-field reconnection. After this drop, $E_\mathrm{B}$ starts increasing again by winding, although we do not find the clear power law proportional to $t^2$ because the magnetic-field energy is close to saturation. 
At $t \approx 40$\,ms $E_\mathrm{B}$ reaches saturation and starts decreasing again. After the saturation, $E_\mathrm{B}/E_\mathrm{int}$ approaches asymptotically $\sim 10^{-2}$ as in other models.

The evolution process after the magnetic field saturates is qualitatively identical  irrespective of the initial magnetic-field strength, configuration, neutron-star EOS, and equatorial-plane symmetry. Thus, we conclude that the evolution process shown here is the universal one for black hole-neutron star mergers that experience tidal disruption of neutron stars.

\begin{figure*}[!th]
      \begin{center}
        \includegraphics[scale=0.4]{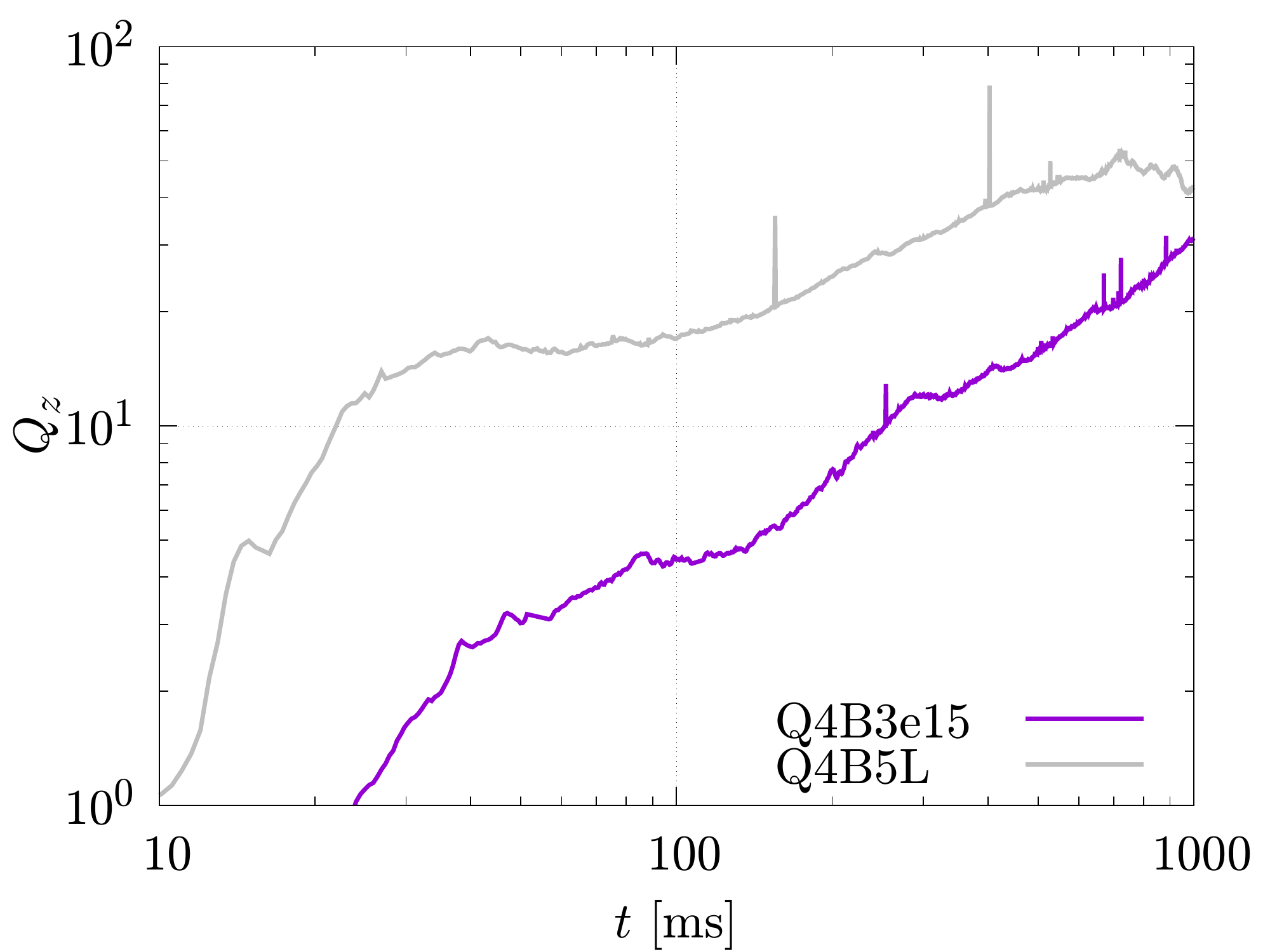} 
        \includegraphics[scale=0.4]{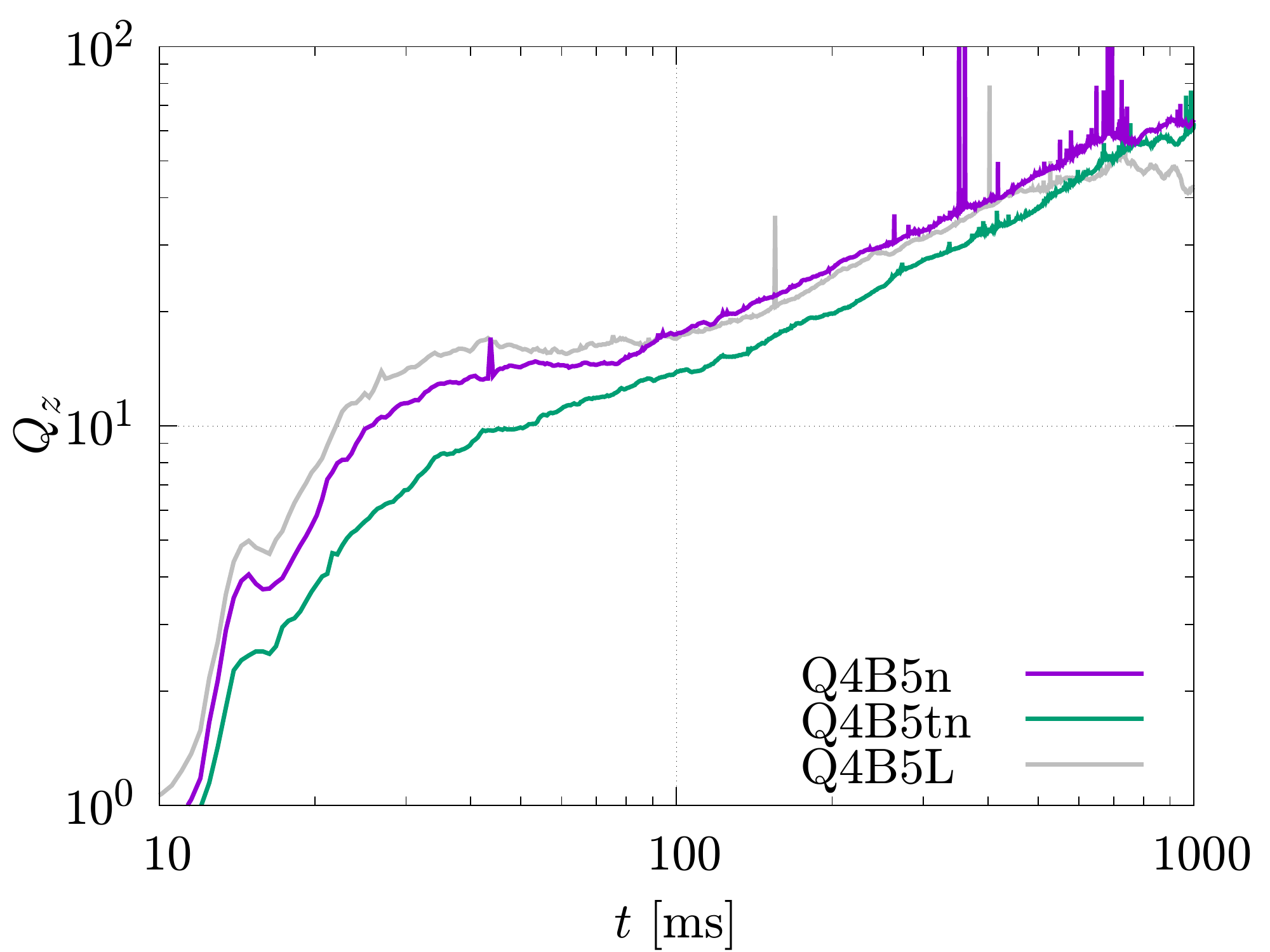} \\
        \caption{The time evolution of the MRI quality factor 
        for models Q4B3e16 (left panel) and Q4B5n and Q4B5tn (right panel).
        The results for model Q4B5L of our previous paper~\cite{hayashi2022jul} are also shown for comparison (in grey color).
	}
        \label{fig:mri_qual}
      \end{center}
\end{figure*}

Figure~\ref{fig:mri_qual} shows the evolution of an MRI quality factor, defined by
\begin{eqnarray}
    Q_z:=\left\langle \left|  \lambda_z/\Delta x \right|\right\rangle_{\rm ave}, \\
    {\lambda_z}:= \frac{b_{z}}{\sqrt{4{\pi}{\rho}h+b^{\mu}b_{\mu}}}\frac{2{\pi}}{\Omega},
\end{eqnarray}
where $\Omega$ is the local angular velocity, $z$-direction is the direction of the rotation axis, and 
$\langle \cdots \rangle_{\rm ave}$ denotes the spatial average with the weight of the rest-mass density for the region with $\rho \geq 10^6\,{\rm g/cm^3}$.
For $Q_z > 10$, we interpret that the MRI is numerically well-resolved. 
It is found that for most of the models, $Q_z>10$ is achieved for $t \agt 20$--$50$\,ms, while for model Q4B3e15, $Q_z>10$ is achieved only for $t \agt 250$\,ms.\footnote{A steep increase of $Q_z$ takes place at $t \sim 200$\,ms, and thus, the MRI activity is also  partly visible already for $t \sim 200$\,ms in this model.}
This illustrates that for the model with lower initial magnetic-field strengths, it takes a longer time until the fastest growing mode of the MRI can be well resolved. As we already remarked, this is an artifact due to the insufficient grid resolution in numerical computation, and hence, in real systems, the MRI turbulence would be developed from an earlier stage. 

\subsubsection{Effective viscosity} \label{sec:results-viscosity}

\begin{figure*}[!th]
      \begin{center}
        \includegraphics[scale=0.4]{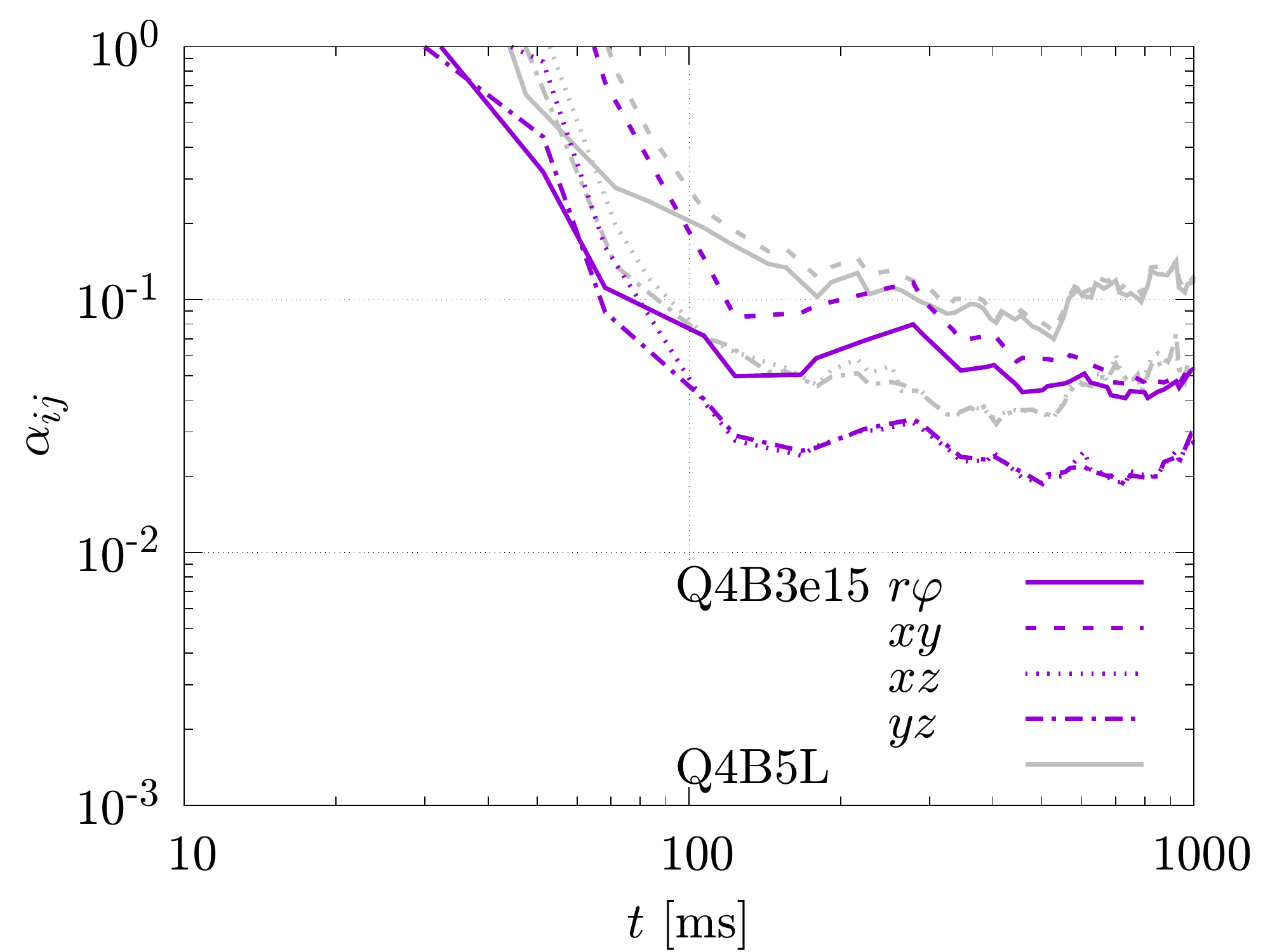} 
        \includegraphics[scale=0.4]{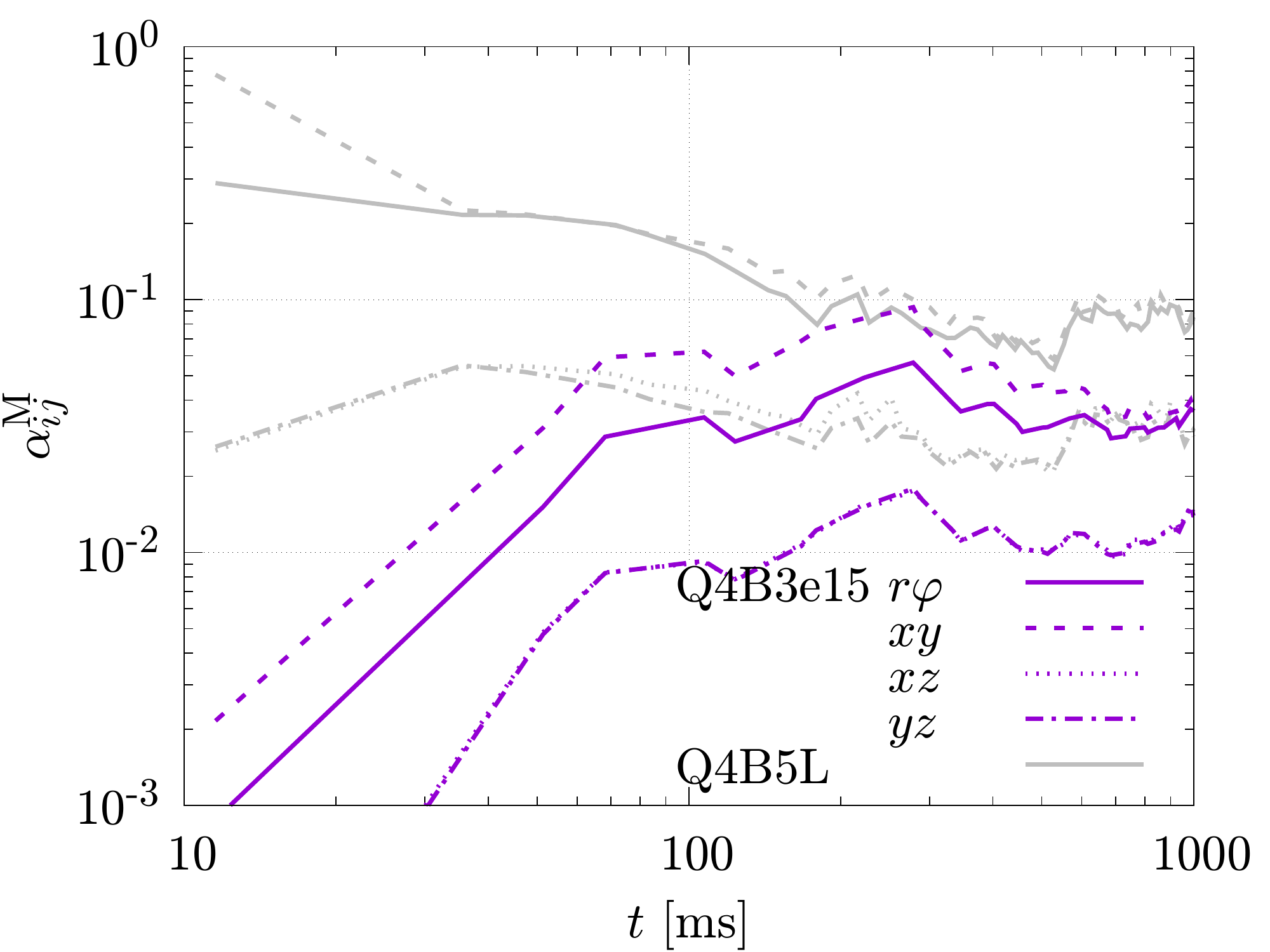} \\
        \includegraphics[scale=0.4]{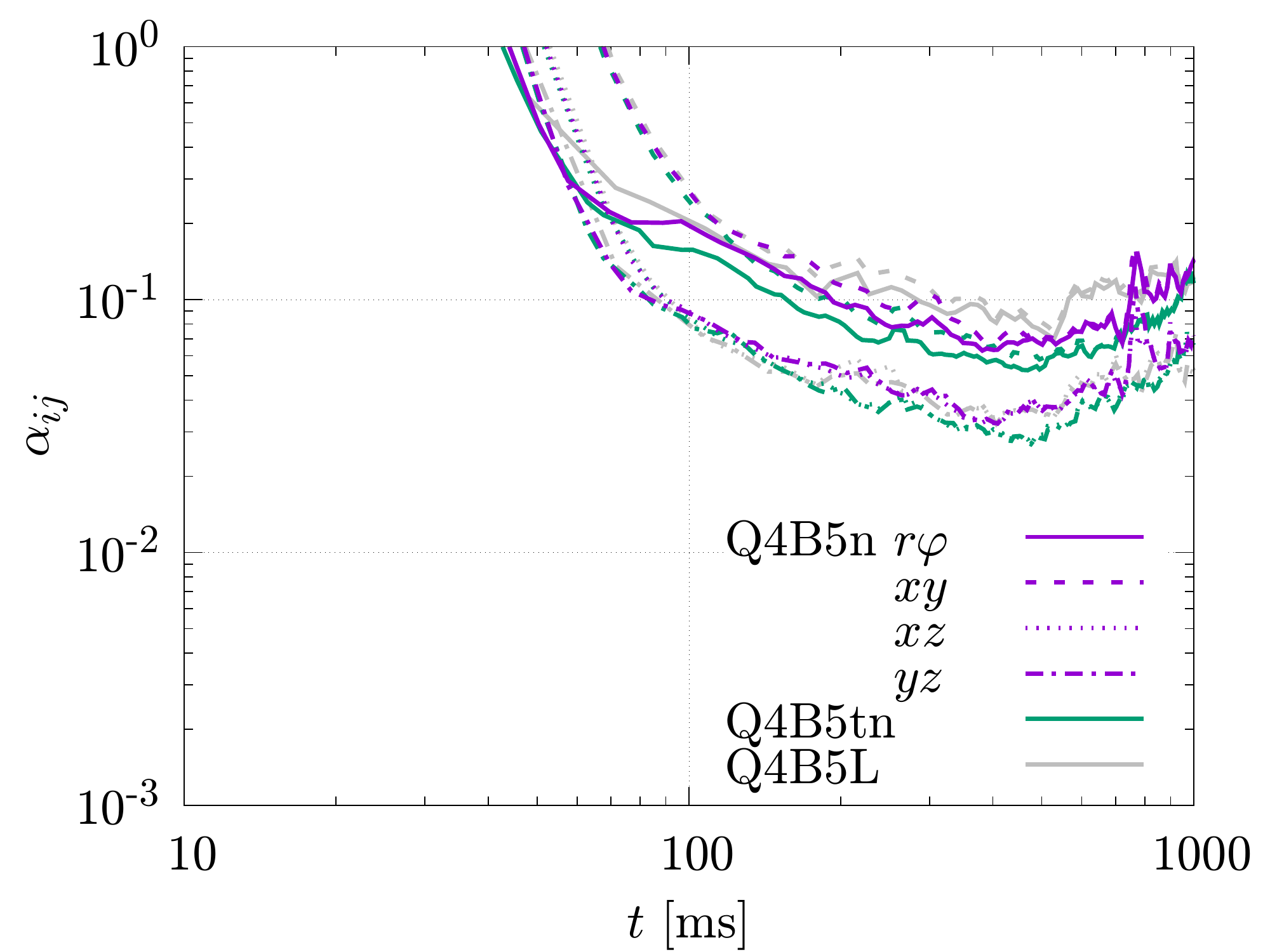}
        \includegraphics[scale=0.4]{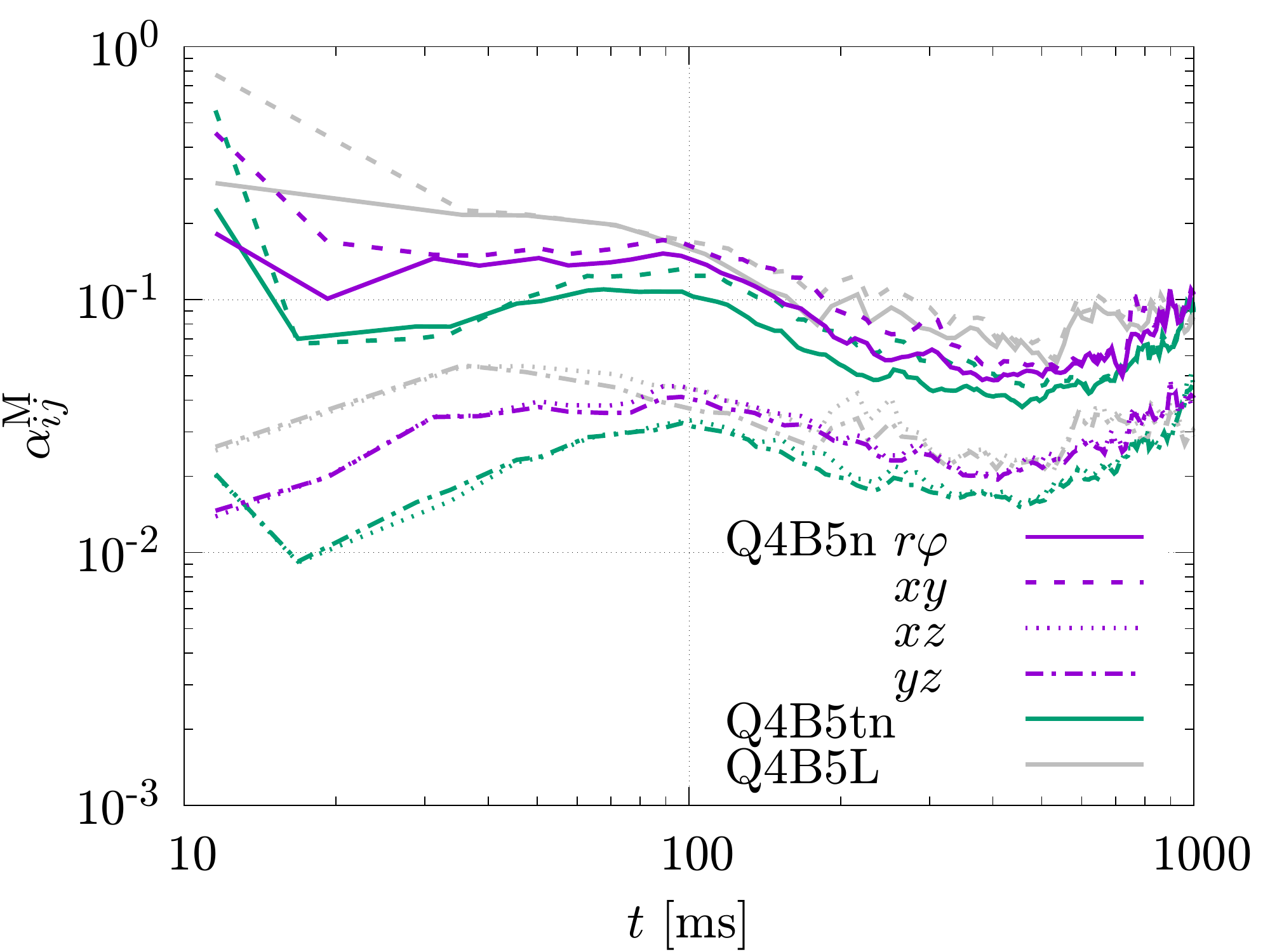} \\
        \caption{The time evolution of the ratio of the magnetohydrodynamical anisotropic stress (left) and Maxwell stress (right) to the pressure, $\alpha_{ij}$ and $\alpha^\mathrm{M}_{ij}$, respectively. 
        The results for $r\varphi$, $xy$, $xz$, and $yz$ components are shown.
        The top two panels show the results for model Q4B3e15, and the bottom two panels show the results for models Q4B5n and Q4B5tn.
	}
        \label{fig:alpha_ij}
      \end{center}
\end{figure*}

Associated with the development of the MRI turbulence and resulting dynamo action, effective viscosity is enhanced in the accretion disk. 
We here analyze an effective viscosity tensor by evaluating the ratio of the magnetohydrodynamical anisotropic stress to the pressure, which is defined by
\begin{eqnarray}
    \alpha_{ij}:=\left\langle \left|  {1 \over P}
    \left(\rho h \hat u_i \hat u_j -{1 \over 4\pi}b_i b_j \right)  \right|\right\rangle_{\rm ave}. 
\end{eqnarray}
We also evaluate the ratio of the Maxwell stress to the pressure defined by
\begin{eqnarray}
    \alpha^\mathrm{M}_{ij}:=\left\langle \left|  {1 \over P}
    \left( -{1 \over 4\pi}b_i b_j \right)  \right|\right\rangle_{\rm ave}. 
\end{eqnarray}
Here, $i\not=j$ ($i$, $j=x, y, z$) and $\langle \cdots \rangle_{\rm ave}$ denotes the spatial average with the weight of the rest-mass density for the region with $\rho \geq 10^7\,{\rm g/cm^3}$.
$\hat u_i$ is defined by $u_i-\langle u_i \rangle_{t,{\rm ave}}$ where $\langle u_i \rangle_{t,{\rm ave}}$ denotes the local time average of $u_i$. 
The time average needs to be subtracted from $u_i$ to eliminate the contribution of coherent motion (not random motion; \addms{e.g., the orbital motion around the black hole}) for evaluating the anisotropic stress associated with the turbulent motion.

Figure~\ref{fig:alpha_ij} plots the time evolution of the off-diagonal components of $\alpha_{ij}$ and $\alpha^\mathrm{M}_{ij}$. 
This shows that for $t \alt 100$\,ms $\alpha_{ij} > O(0.1)$, but we interpret that this is not a physical value, nor associated with the magnetohydrodynamics effect: The remnant matter shows the non-axisymmetric structure for $\alt 100$\,ms after the merger, and in such a case, the value of $\alpha_{ij}$, specifically the contribution from the Reynolds stress part, cannot be evaluated properly. 
Thus we focus only on the stage for $t \agt 100$\,ms for which the non-axisymmetric structure is not very appreciable and the MRI turbulence is developed. 

When the disk is in a MRI turbulent stage, we find that the $r\varphi$ and $xy$ components are $\approx 0.05$--$0.1$, and $xz$ and $yz$ components are $\approx 0.02$--$0.05$. Hence, the order of the magnitude of $\alpha_{ij}$ agrees with the often-used value of the alpha viscous parameter for the accretion disk~\cite{SS1973}, although the magnitude for each component of $\alpha_{ij}$ has anisotropy. 
Our interpretation for this anisotropy is that not only the MRI turbulence but also the effects by the global magnetic fields such as magneto-centrifugal effects~\cite{blandford1982} contribute to the angular momentum transport because $\alpha_{ij}$ for $r\varphi$ and $xy$ components are larger than the others. 
We also note that the dominant part of $\alpha_{ij}$ stems from the Maxwell stress; the contribution of the Reynolds stress, which originates from the fluid turbulent motion, is $\sim 0.01$ irrespective of the model and component. 
This trend is universally found for all the models.

Model Q4B3e15, which has a low initial magnetic-field strength, shows a factor of $\sim 2$ smaller values of $\alpha_{ij}$ than for the other models, but this is relatively minor compared to the difference in the initial magnetic field strength (the initial magnetic field is smaller than the other models by a factor of $\sim 17$). This result suggests that for this model, the fastest growing mode of the MRI might be only partly resolved. 
Indeed, for this model, a delay with $\sim 200$\,ms in the magnetic-field amplification as well as in all the processes of the disk evolution is found. This indicates that the disk expansion due to the angular momentum transport is delayed due to the weaker turbulence viscosity. As a result of this delay, the drop of the disk temperature and neutrino luminosity delays, and thus, the post-merger mass ejection is delayed by $\sim 200$\,ms.
Note, however, that the delay could be the artifact due to the insufficient grid resolution.

\addkh{Similar analyses were done for the simulations starting from a black hole-disk system to evaluate the effective viscosity $\alpha_{r\varphi}$ in Ref.~\cite{de2021nov}, and the value of $\alpha_{r\varphi}$ found in their analysis were comparable to our present results. 
This is because the magnetic field amplification is saturated as the disk reaches to equipartition state. 
We note, however, that the result from the present simulation has a slightly higher value. 
The existence of the additional magnetic field amplification mechanism due to the non-axisymmetric structure of the disk and the minor difference in the definition of $\alpha_{ij}$ are interpreted as the reasons.}

\subsubsection{Property of ejecta} \label{sec:ejecta_property}

\begin{figure}[!th]
      \begin{center}
        \includegraphics[scale=0.4]{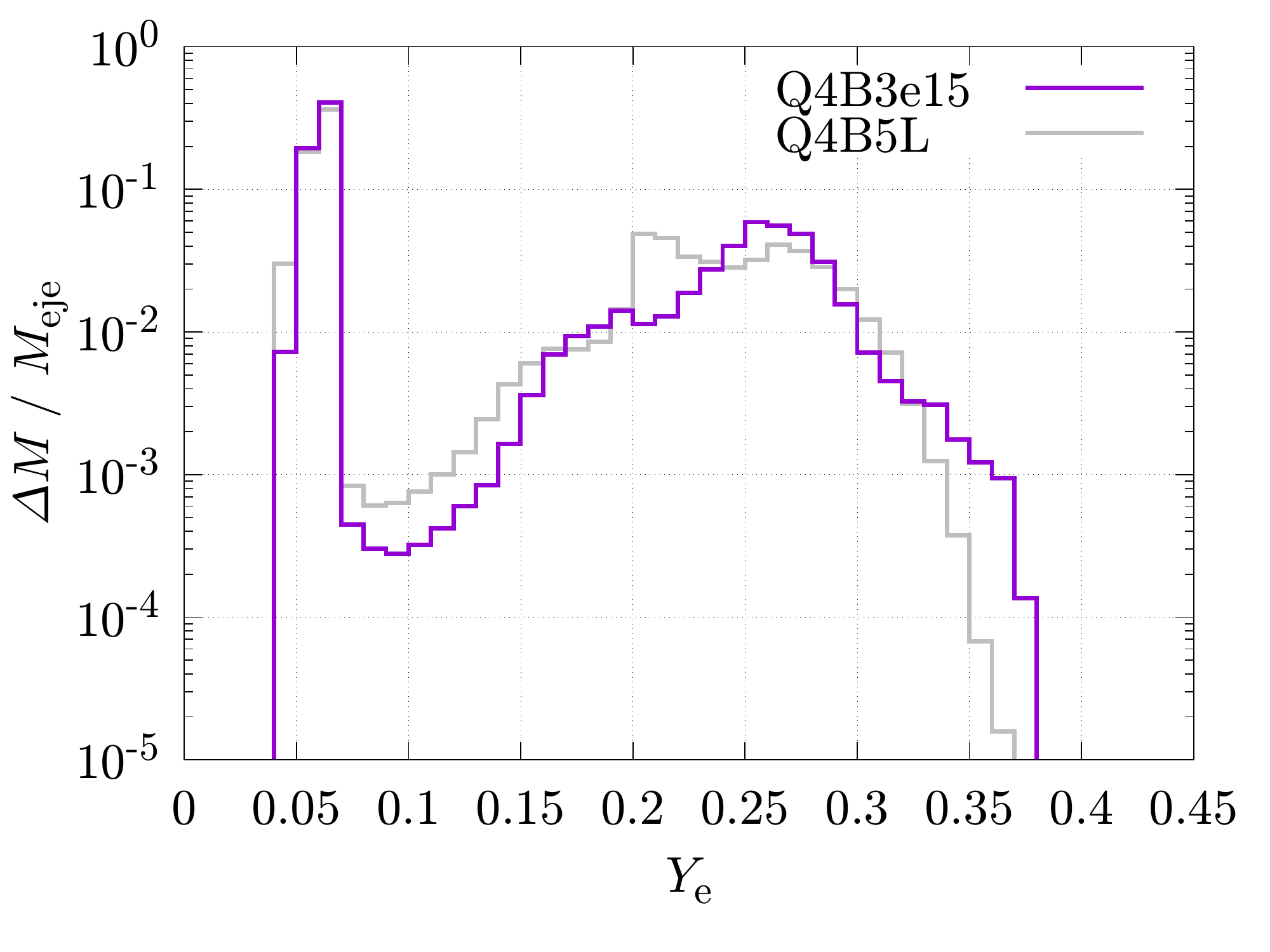} \\
        \includegraphics[scale=0.4]{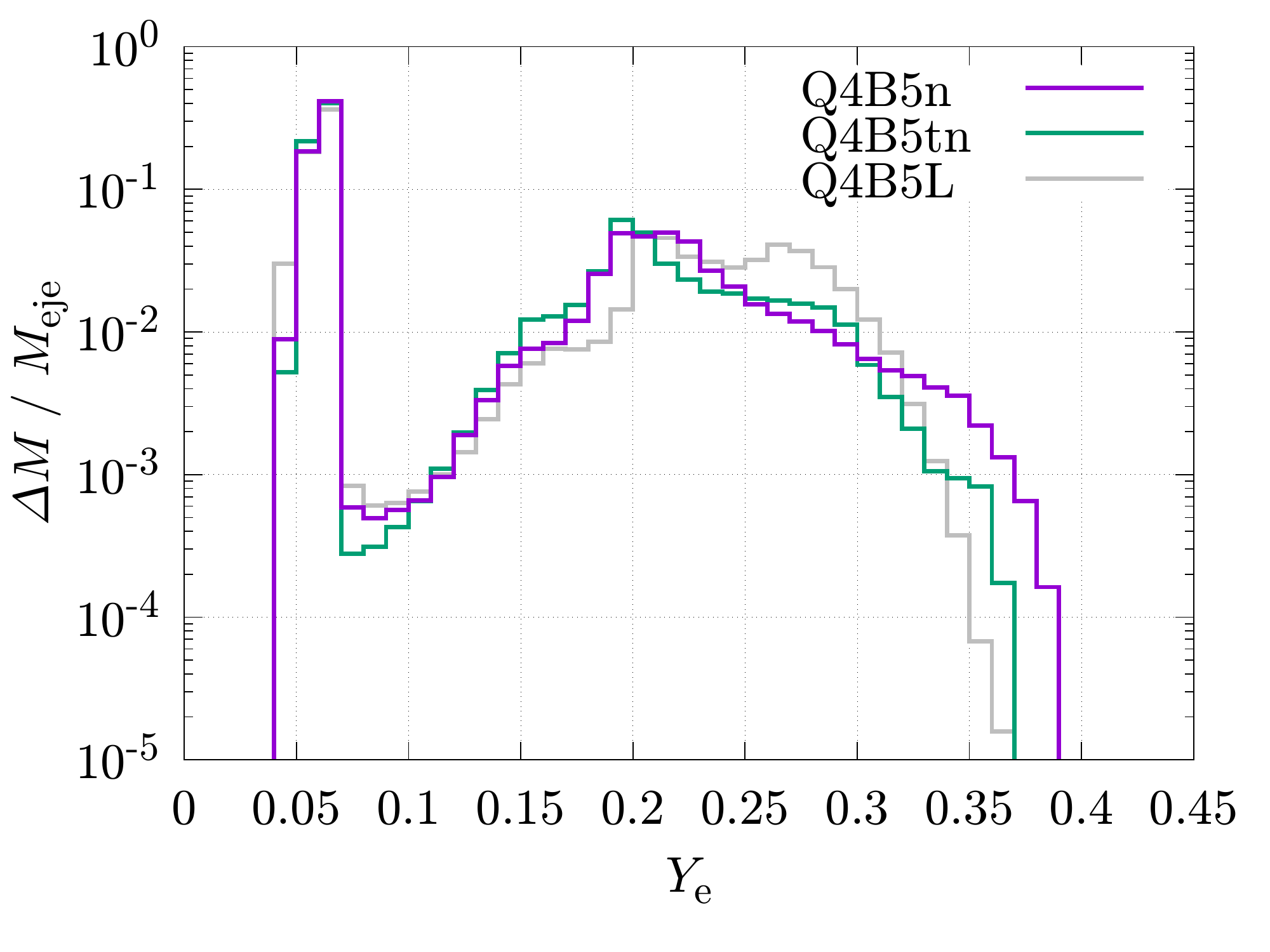} \\
        \includegraphics[scale=0.4]{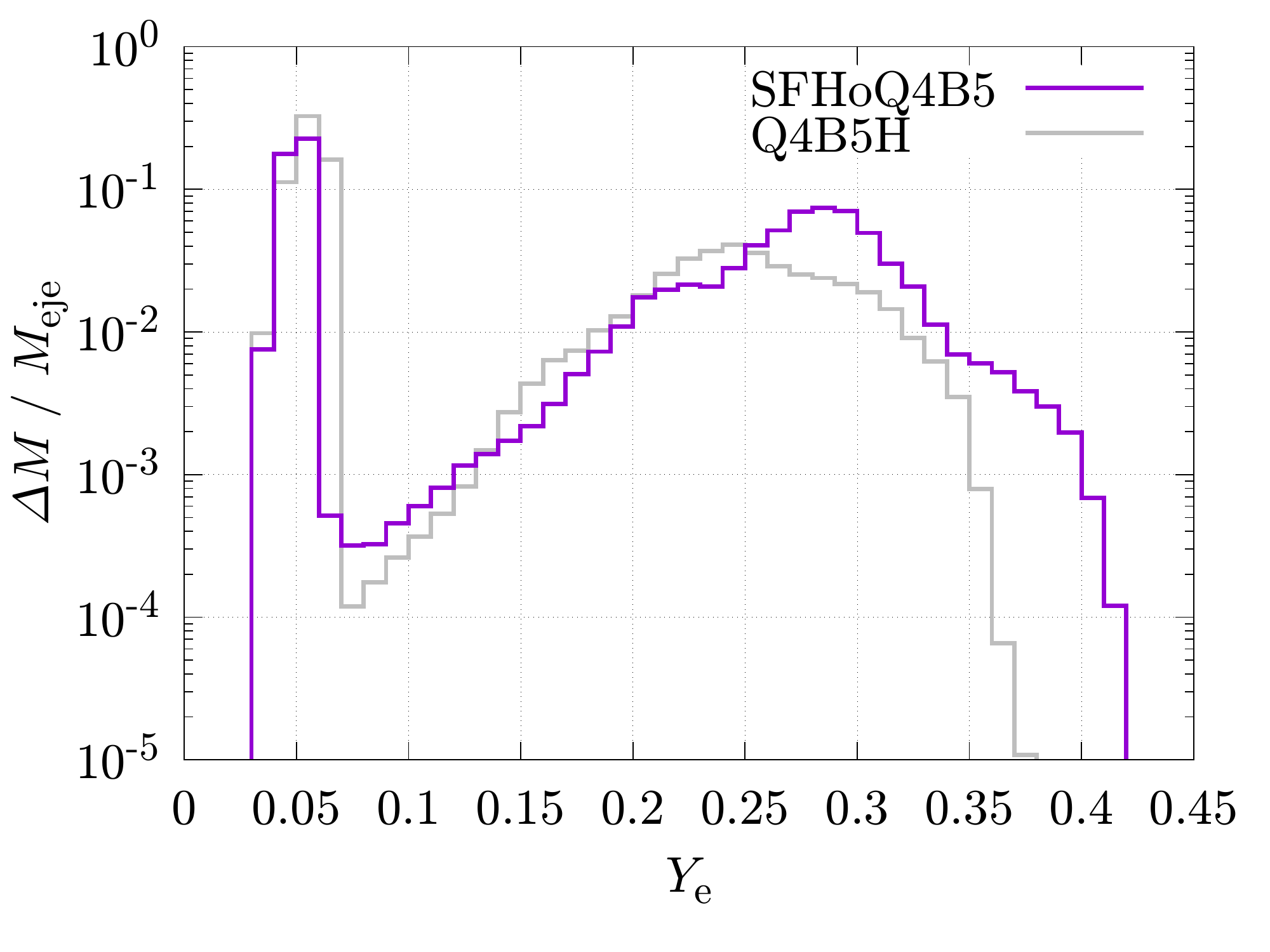} 
        \caption{Mass histogram as a function of the electron fraction of the ejecta
        for models Q4B3e15 (top panel), Q4B5n and Q4B5tn (middle panel), and SFHoQ4B5 (bottom panel).
        The results for models Q4B5L and Q4B5H of our previous paper are also shown for comparison (in grey color). Note that the vertical axis is normalized by the total ejecta mass. 
	}
        \label{fig:ye_hist}
      \end{center}
\end{figure}

\begin{figure}[!th]
      \begin{center}
        \includegraphics[scale=0.4]{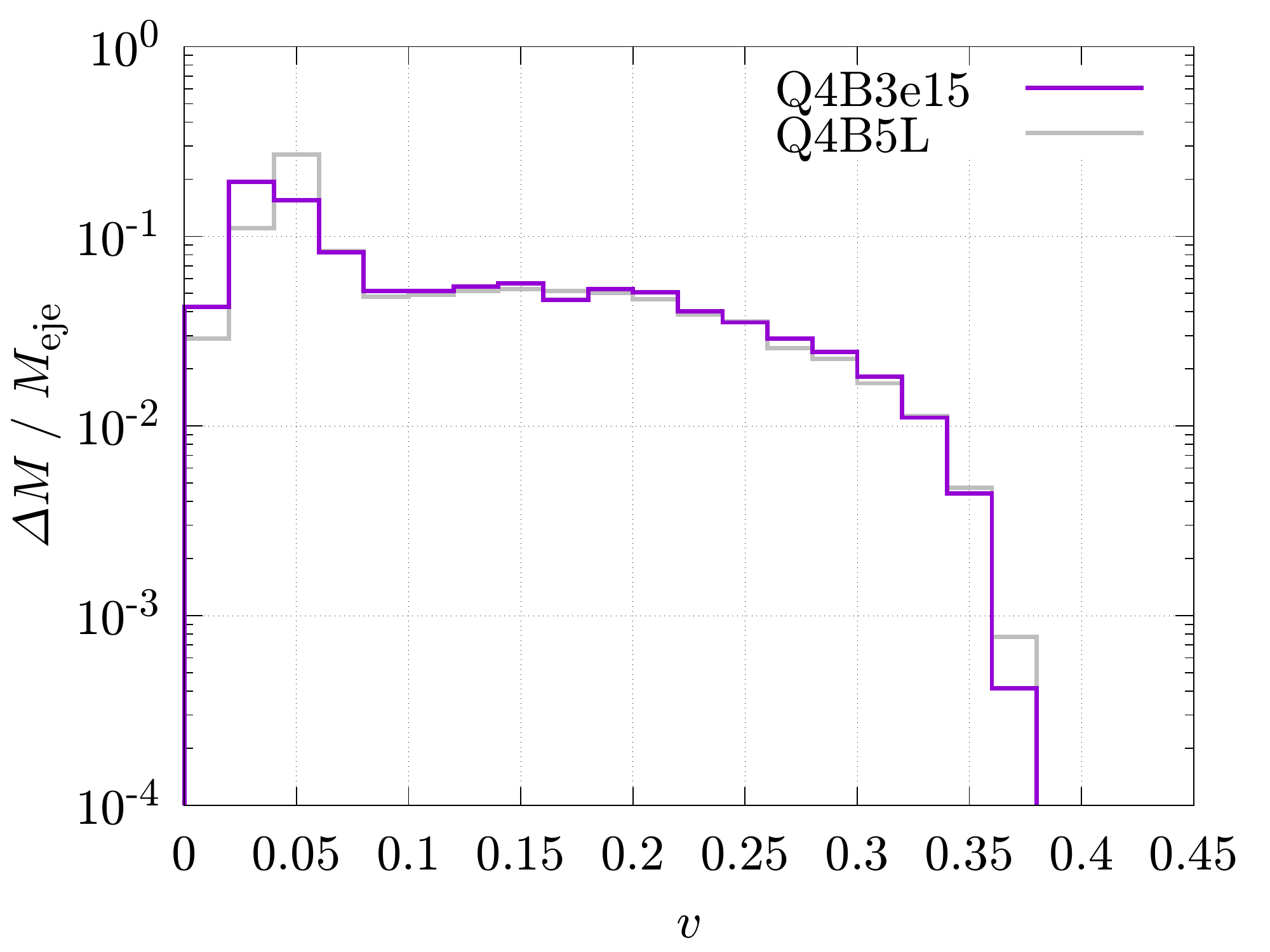} \\
        \includegraphics[scale=0.4]{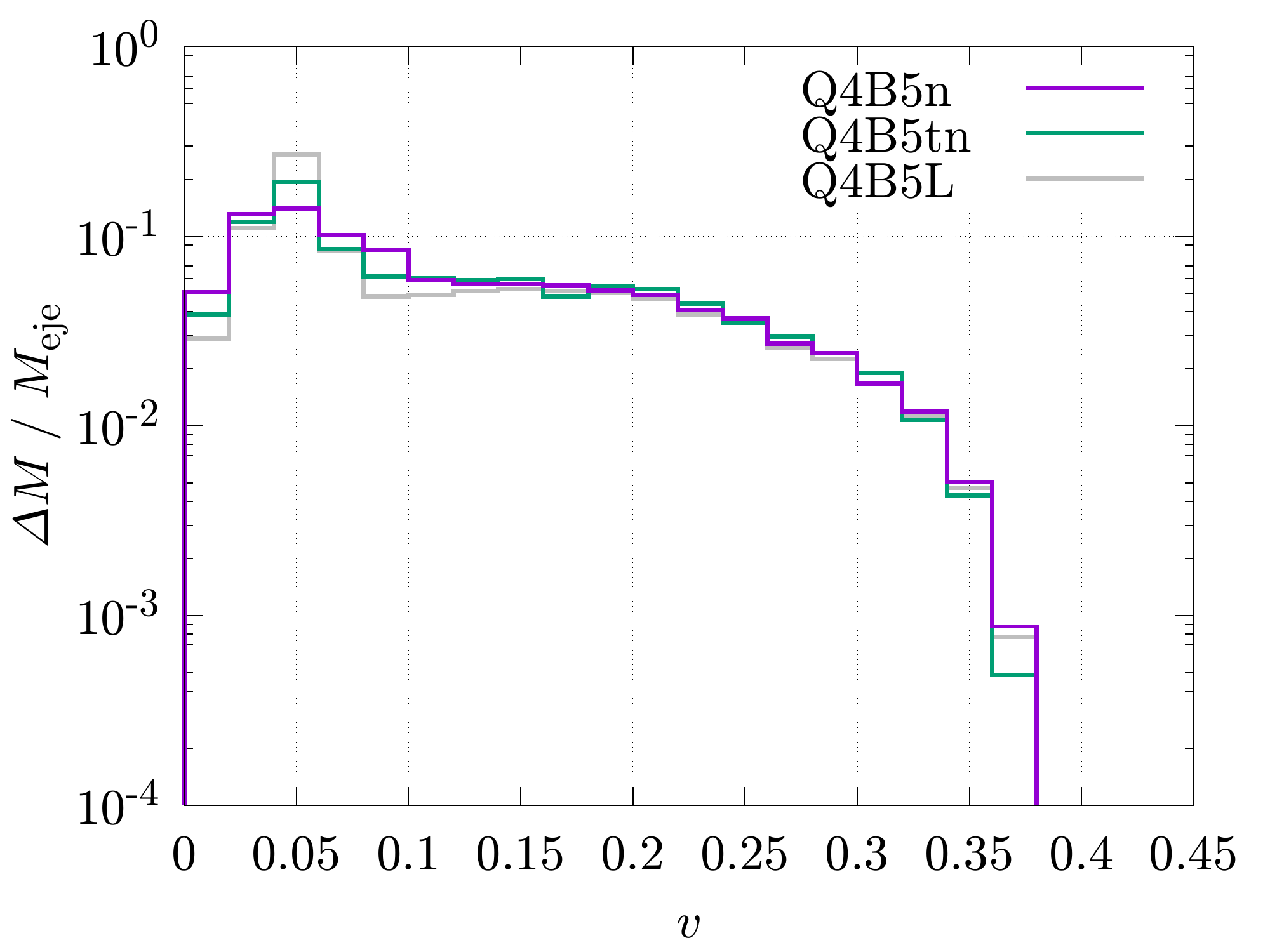} \\
        \includegraphics[scale=0.4]{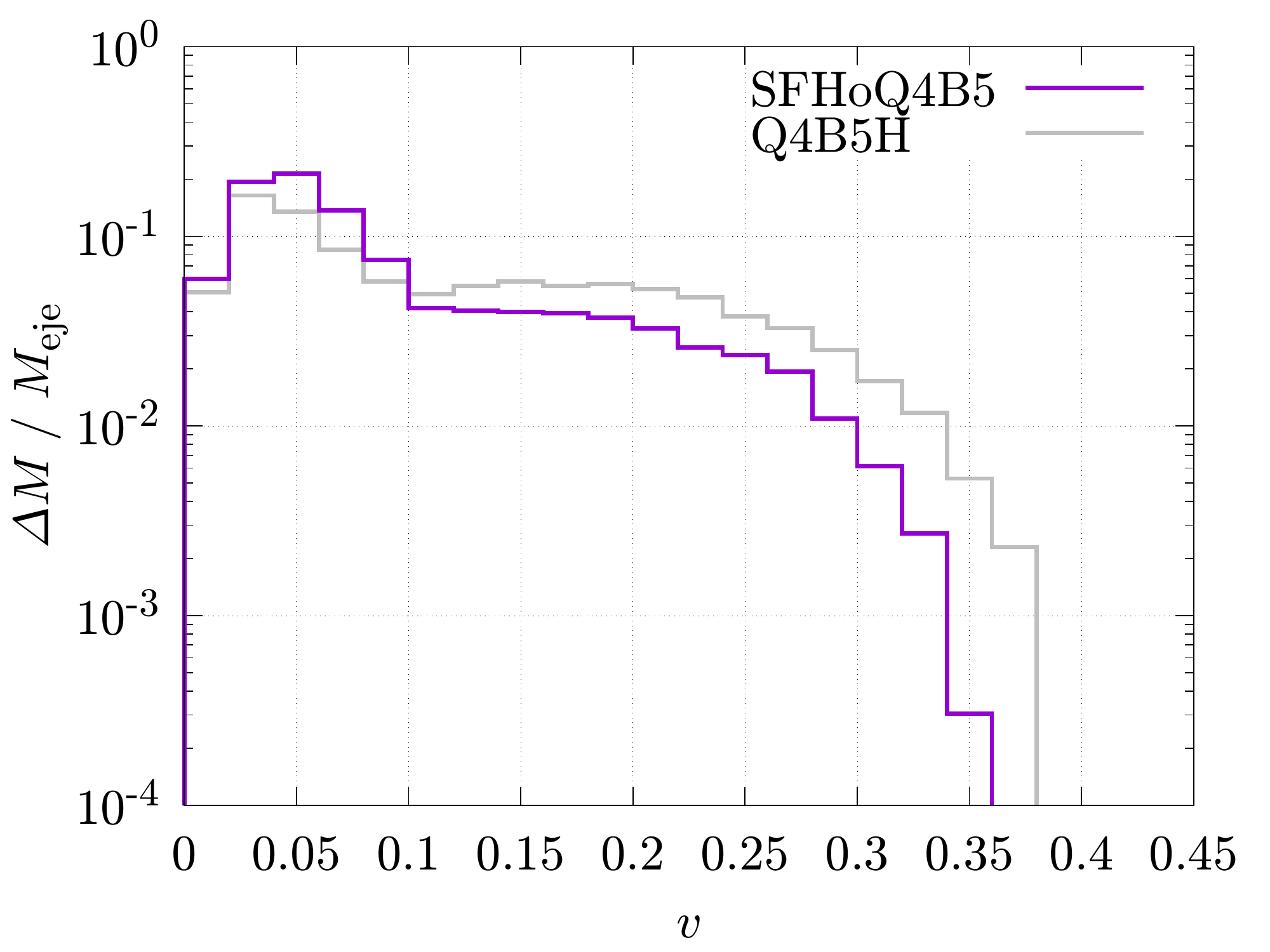} 
        \caption{The same as Fig.~\ref{fig:ye_hist} but for the mass histogram as a function of the velocity of the ejecta. 
	}
        \label{fig:v3_hist}
      \end{center}
\end{figure}

Now, we turn our attention to the properties of the ejecta.
Figures~\ref{fig:ye_hist} and \ref{fig:v3_hist} show the rest-mass histogram as a function of the electron fraction $Y_{\mathrm{e}}$ and velocity $v$ for the ejecta component, respectively. The ejecta velocity $v$ is defined by
\begin{eqnarray}
v &:=& \sqrt{1-\Gamma_{\infty}^{-2}}, \nonumber \\
\Gamma_{\infty} &:=& -hu_{t}/h_{\mathrm{min}},
\end{eqnarray}
where $\Gamma_{\infty}$ is interpreted as the terminal Lorentz factor under the assumption that the internal energy is converted completely to the kinetic energy of the ejecta in the far region. 

There are two distinctive $Y_{\mathrm{e}}$ components for the ejecta as found in our previous work~\cite{hayashi2022jul}\addkh{, where the comparison with the results in the literature is also made}. 
One is the dynamical ejecta for which  $Y_{\mathrm{e}}\approx0.03$--$0.07$ irrespective of the simulation setups. 
However, the range of $Y_\mathrm{e}$ for the dynamical ejecta depends slightly on the EOS: For model SFHoQ4B5, $Y_{\mathrm{e}}\approx0.03$--$0.06$, while for model Q4B5H with DD2 EOS,  $Y_{\mathrm{e}}\approx0.03$--$0.07$ ~\cite{kyutoku2018jan}.
The electron fraction of the dynamical ejecta directly reflects the neutron richness of the neutron star because the dynamical ejecta is affected only weakly by the thermal and weak-interaction process in the merger and post-merger stages.
Thus, the difference in the distribution of $Y_{\mathrm{e}}$ for the dynamical ejecta between models SFHoQ4B5 and Q4B5H comes directly from the difference in the neutron star EOSs. \addms{This difference is reflected in the product of the heavy elements in the $r$-process nucleosynthesis~\cite{Wanajo:2022jgw}.}

\addms{
For the post-merger ejecta, the electron fraction is as higher as $0.1 \alt Y_{\mathrm{e}}\alt 0.4$. Irrespective of the initial magnetic-field setups and equatorial-plane symmetry, similar distribution is found for the DD2 models, while the difference in the EOS makes a quantitative difference. Comparing the high-resolution models SFHoQ4B5 and Q4B5H, we find that model SFHoQ4B5 has a distribution with higher $Y_{\mathrm{e}}$ values: e.g, the 
maximum value of $Y_\mathrm{e}$ is smaller than 0.4 for all the DD2 models while it is $\approx 0.42$ for the SFHo model. 
Our speculation for the reason to this is as follows:} The neutron star modeled by the SFHo EOS is tidally disrupted at an orbit closer to the black hole because it has a smaller neutron-star radius. Then, the matter that forms the one-armed structure right after the merger and subsequently forms an accretion disk experiences stronger compression and shock heating between the inner and the outer spiral arms. As a result, the disk temperature is enhanced right after the tidal disruption and the value of $Y_{\mathrm{e}}$ is also increased (see also Ref.~\cite{kyutoku2018jan}).

There are also two components in the mass histogram as a function of the ejecta velocity  (see Fig.~\ref{fig:v3_hist}). 
The high-velocity component with $v/c\gtrsim0.1$ stems primarily from the dynamical ejecta, while the low-velocity component stems primarily from the post-merger ejecta.
For the DD2 models, the velocity histogram has approximately identical distributions  irrespective of the setups, and it also agrees with previous results~\cite{hayashi2022jul}.
For the SFHo model, the low-velocity component has a higher fraction than that for the DD2 models, because the dynamical ejecta mass is smaller while the post-merger ejecta mass is comparable with that of the DD2 models.

\subsection{Magnetic field in the funnel region and the relation to the short-hard gamma-ray burst} \label{sec:results-outflow}

\subsubsection{Poynting luminosity and magnetic field in the funnel}  \label{sec:pflux_and_mag}

\begin{figure}[!th]
      \begin{center}
        \includegraphics[scale=0.4]{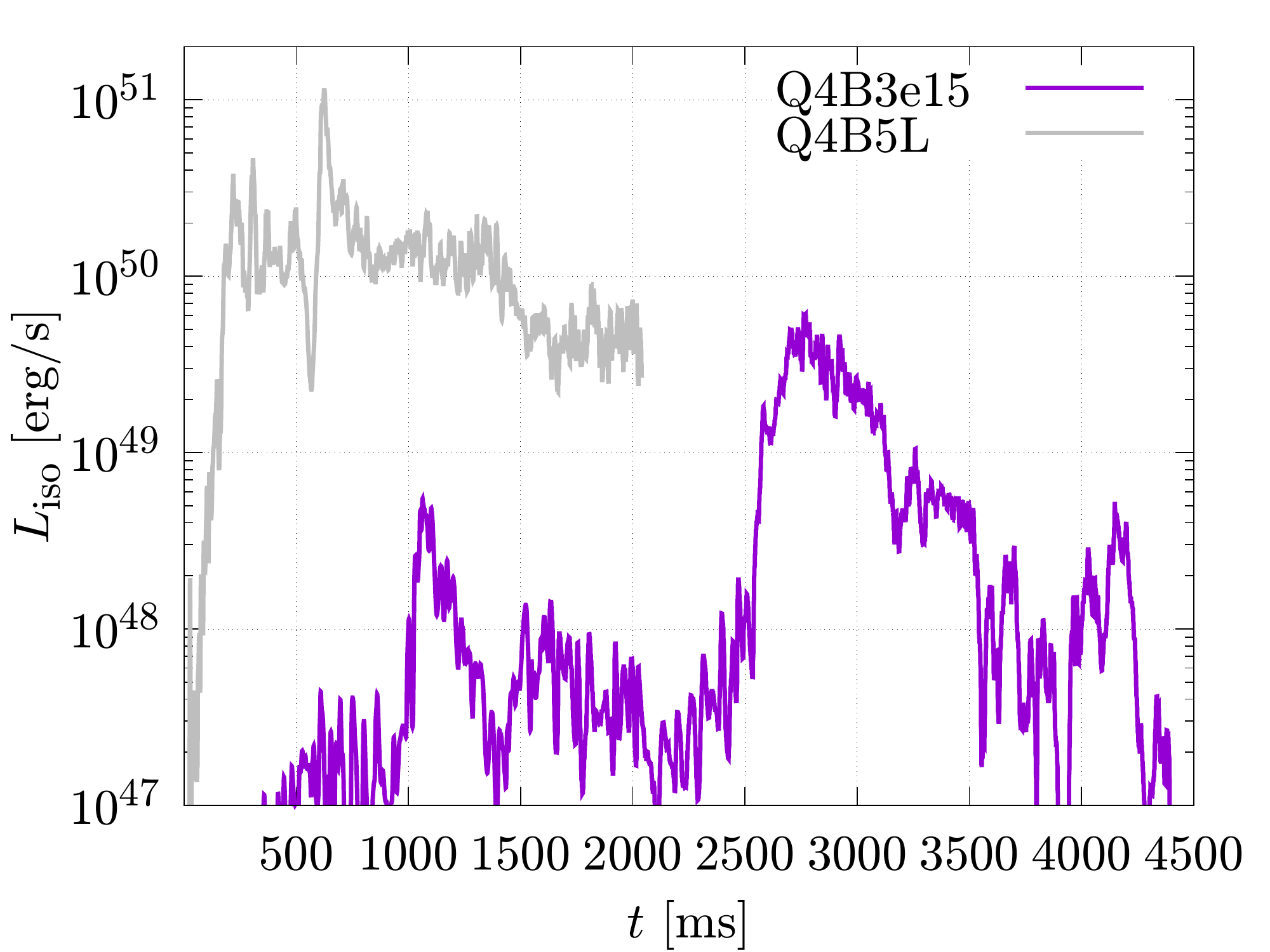} \\
        \includegraphics[scale=0.4]{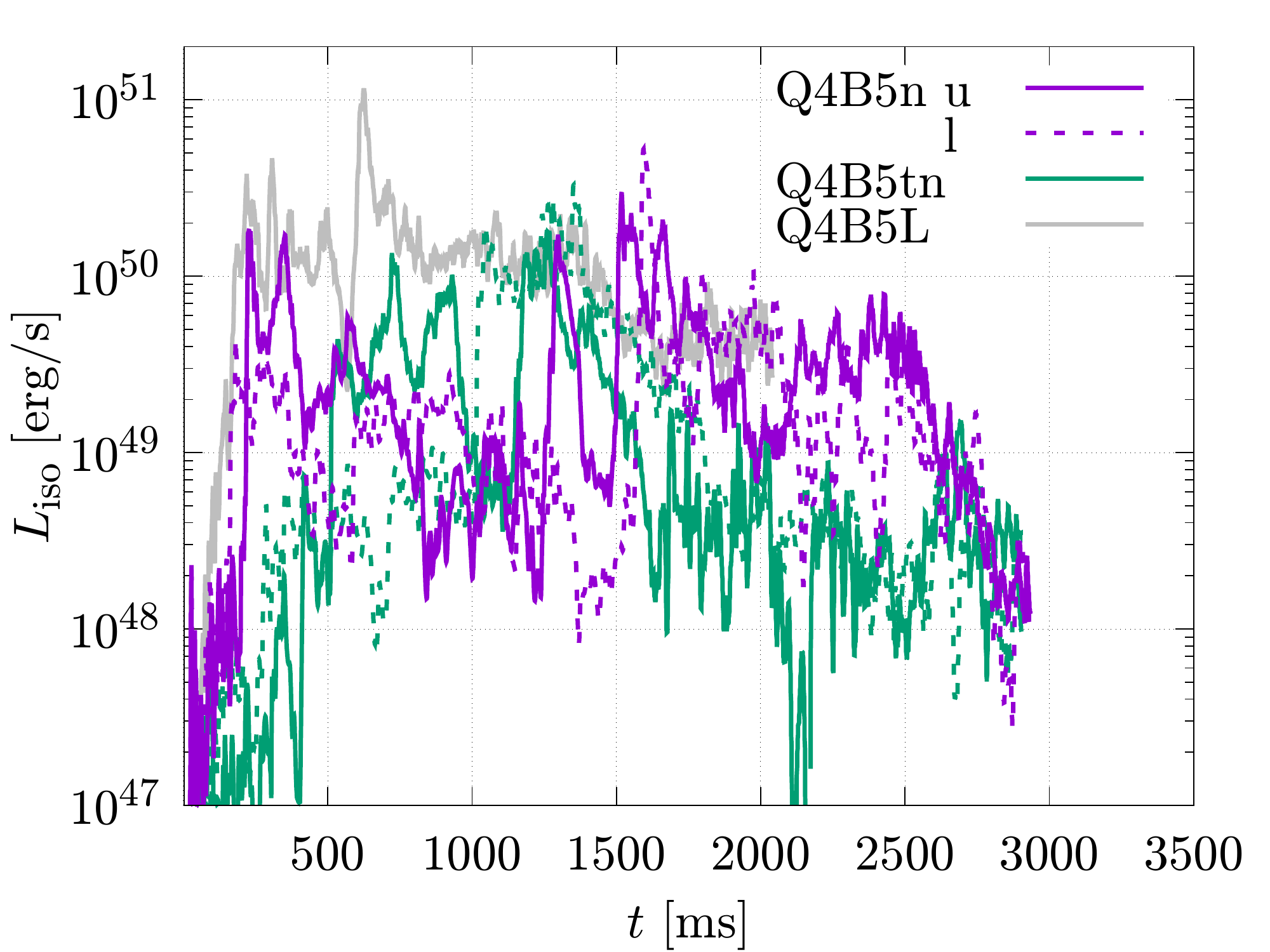} \\
        \includegraphics[scale=0.4]{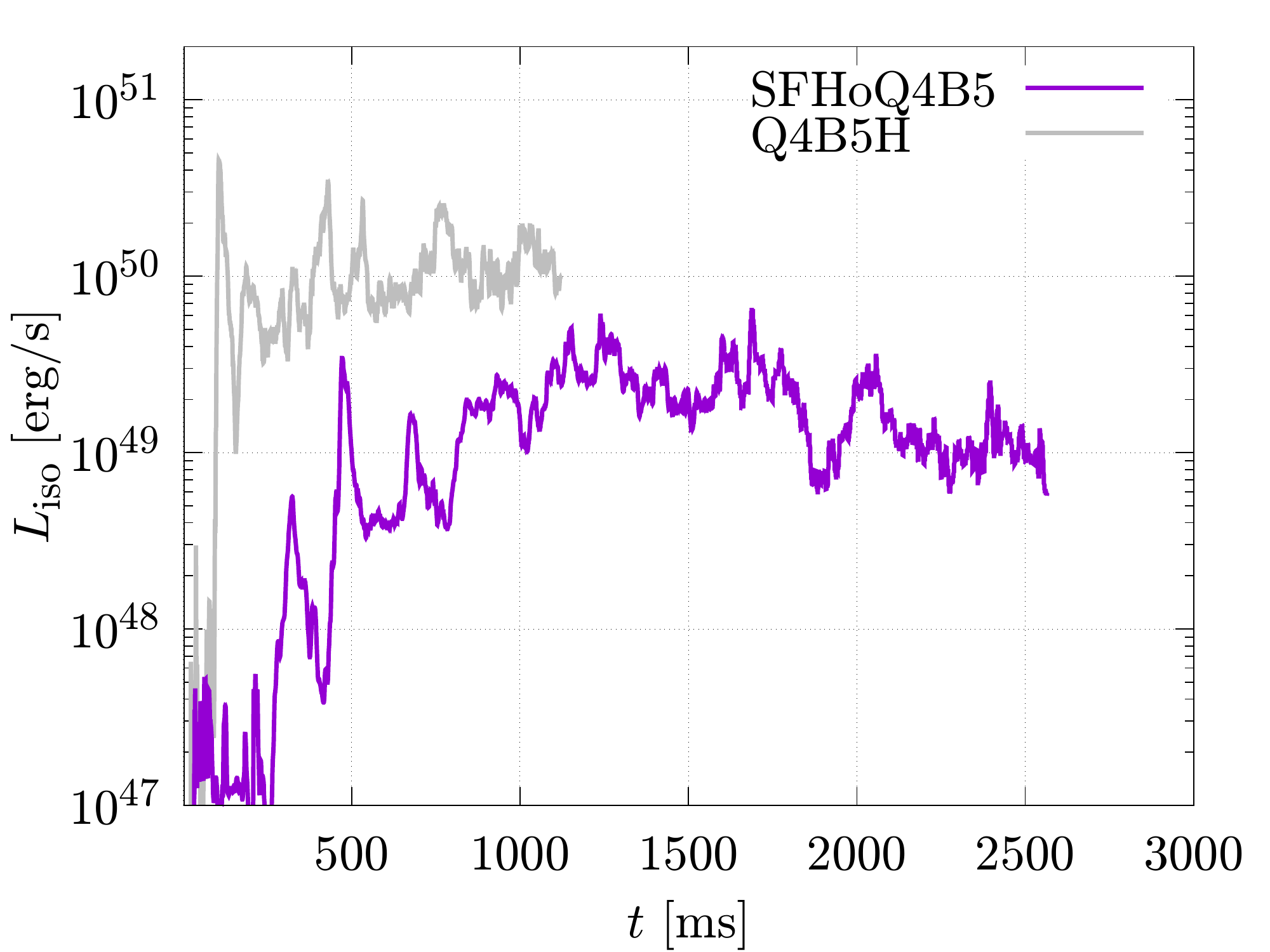} 
        \caption{The time evolution of isotropic-equivalent Poynting luminosity $L_\mathrm{iso}$ 
        for models Q4B3e15 (top panel), Q4B5n and Q4B5tn (middle panel), and SFHoQ4B5 (bottom panel).
        The results for models Q4B5L and Q4B5H of our previous paper \cite{hayashi2022jul} is also shown for comparison (in grey color).
        The characters ``u'' and ``l'' denote the upper and lower hemispheres, respectively.
	}
        \label{fig:plum}
      \end{center}
\end{figure}

\begin{figure*}[!th]
      \begin{center}
        \includegraphics[scale=0.6]{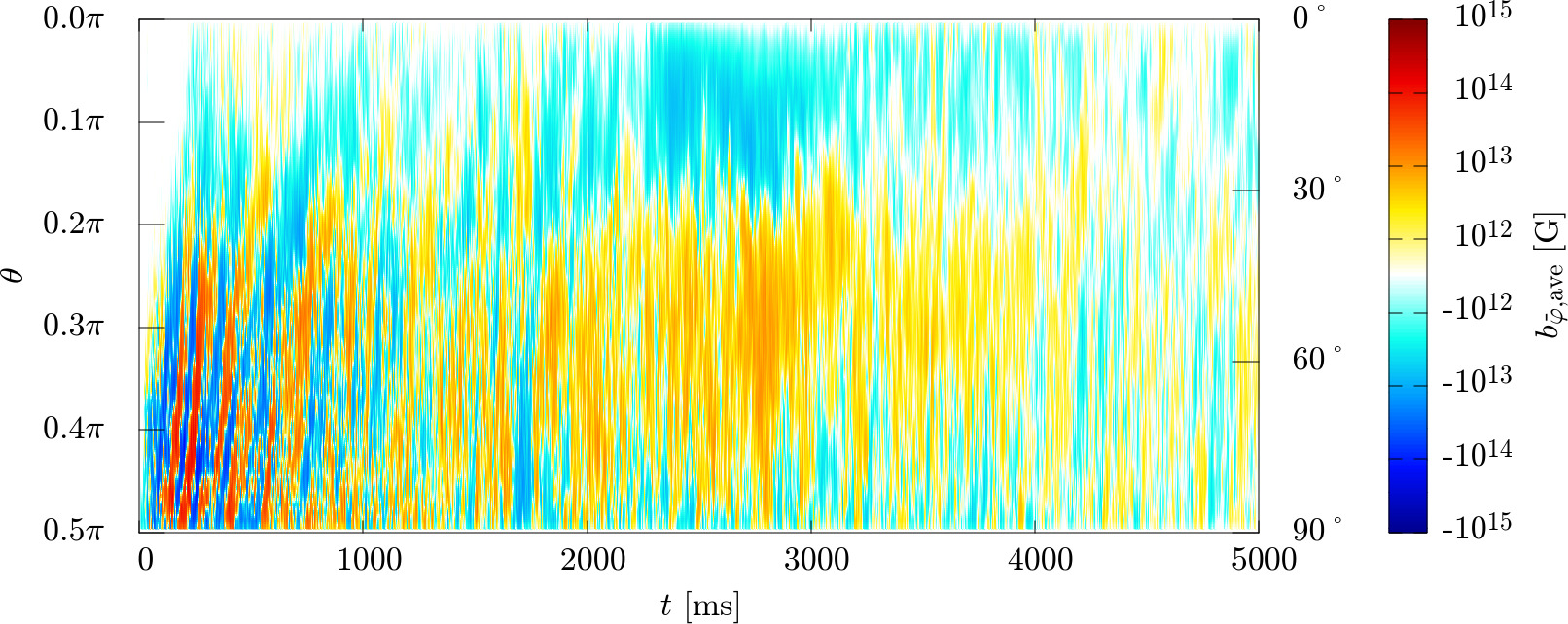} \\ 
        \vspace{2mm}
        \includegraphics[scale=0.6]{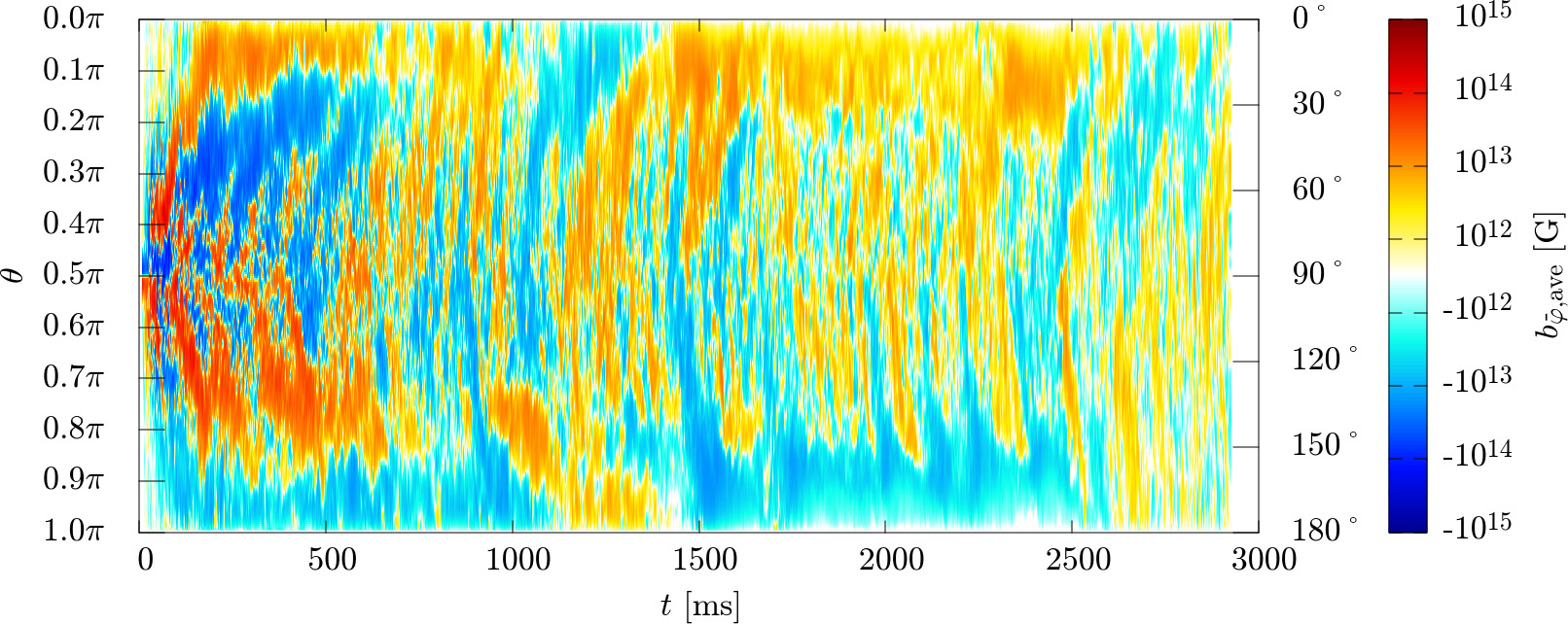} \\
        \vspace{2mm}
        \includegraphics[scale=0.6]{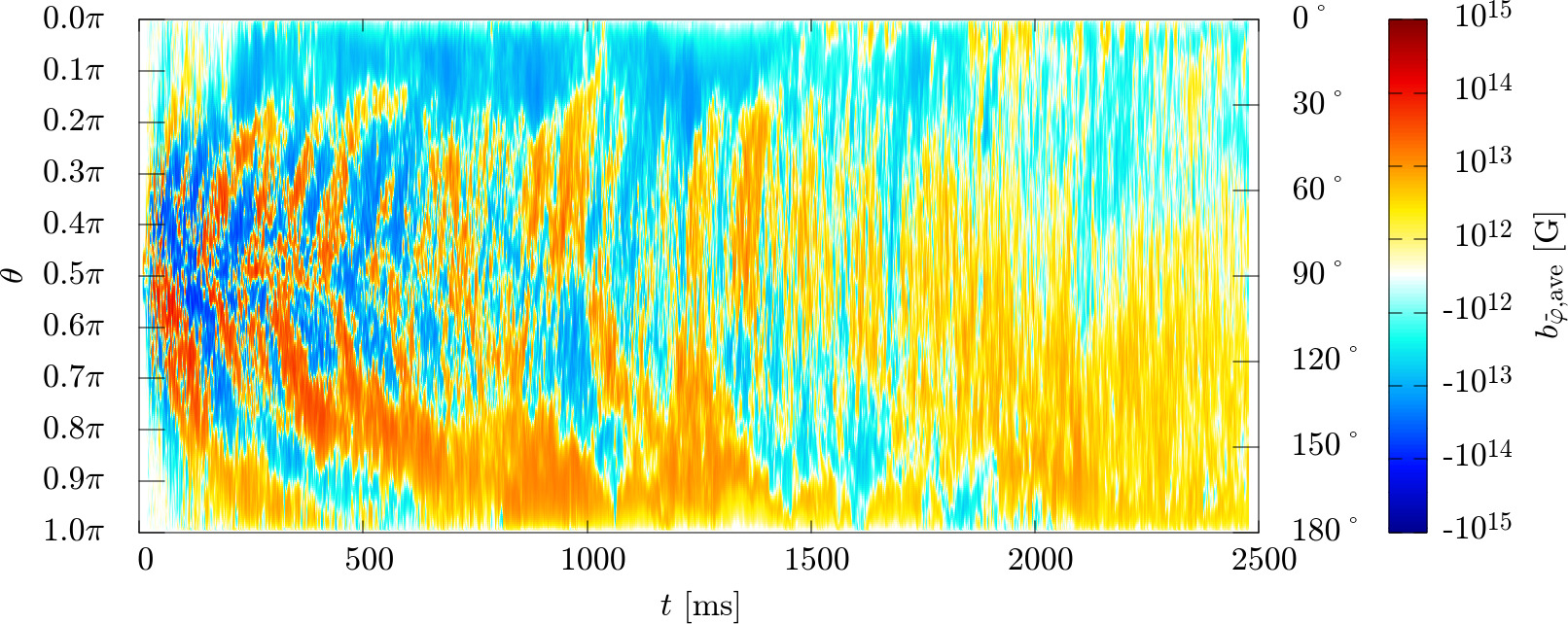} \\
        \caption{The profile of the average toroidal magnetic field along the polar direction ($\theta$) at $r \approx 150~{\rm km}$ as a function of time 
        for models Q4B3e15 (top panel), Q4B5n (middle), and Q4B5tn (bottom).
	}
        \label{fig:BUT}
      \end{center}
\end{figure*}

\begin{figure*}[!th]
      \begin{center}
        \includegraphics[scale=0.36]{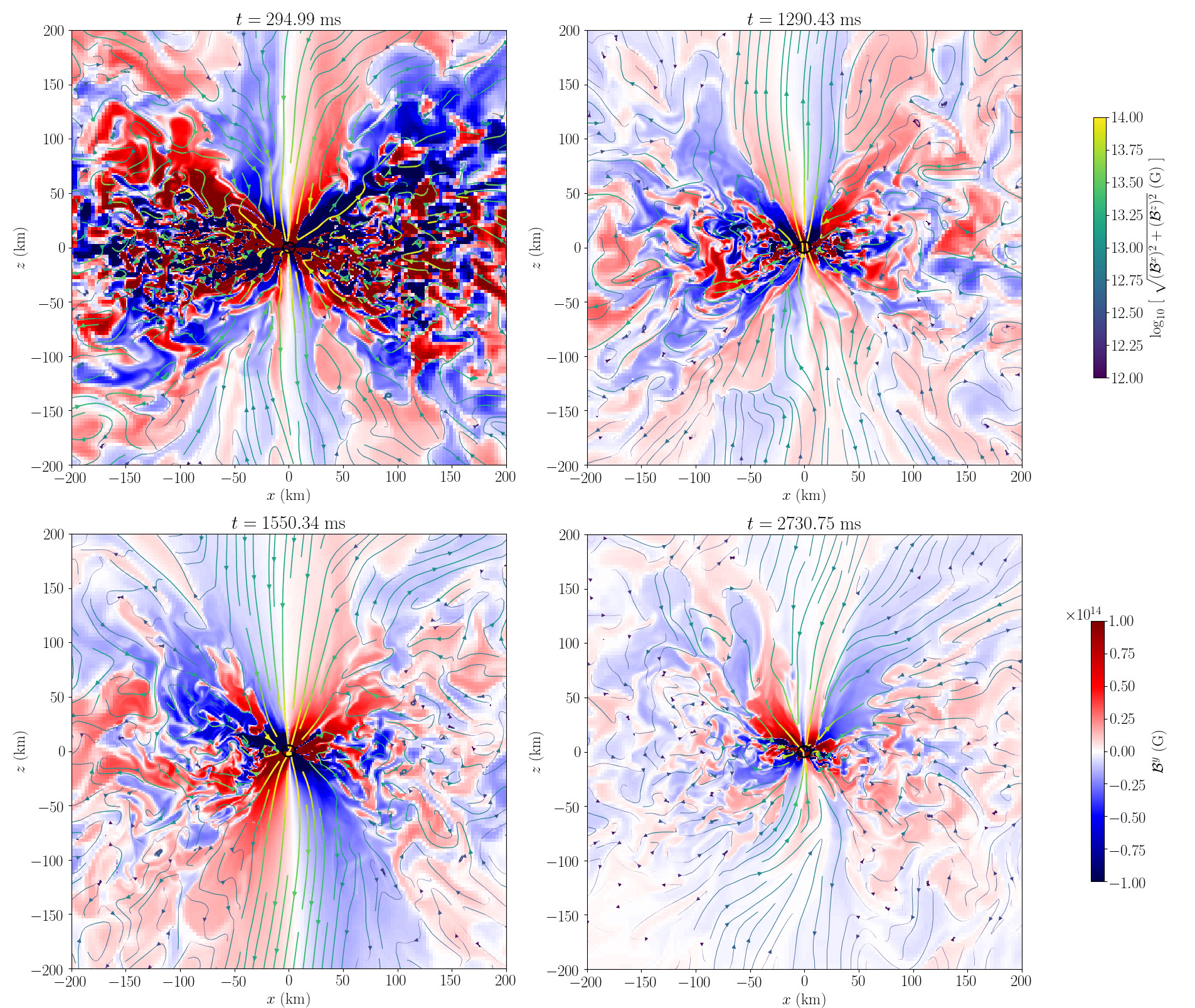} \\
        \caption{The snapshots of the toroidal magnetic field (color profile) together with the poloidal magnetic-field lines (curves) on the $x$-$z$ plane at selected time slices for model Q4B5n. 
        See also the following link for an animation: \url{https://www2.yukawa.kyoto-u.ac.jp/~kota.hayashi/Q4B5n-mf.mp4}.
	}
        \label{fig:snap_xz_mag_Q4B5n}
      \end{center}
\end{figure*}

\addkh{As in our previous simulations~\cite{hayashi2022jul}, in addition to the dynamical and post-merger ejecta, we find a launch of the collimated outflow of matter and Poynting flux in the magnetosphere, which is established after the magnetic tower effect associated with the black-hole spin develops a large-scale helical magnetic fields around the spin axis of the black hole.}

Figure~\ref{fig:plum} shows the time evolution of $L_{\rm iso}$: 
an isotropic-equivalent Poynting luminosity, which we define using the Poynting luminosity for $\theta < 10^{\circ}$ and $r \approx 1500\,{\rm km}$ as
\begin{eqnarray}
  L_{\rm iso} := \frac{2}{1-\cos(10^{\circ})} L_{\theta < 10^{\circ}, r\approx 1500\,{\rm km}}, 
\end{eqnarray}
where
\begin{eqnarray}
&&  L_{\theta < 10^{\circ}, r\approx 1500\,{\rm km}} \nonumber \\
&&:= -\int_{\theta < 10^{\circ},r\approx 1500\,{\rm km}}{T^{\mathrm{(EM)}}}^{~r}_t \sqrt{-g} dS_r.
\end{eqnarray}
$T^{\mathrm{(EM)}}_{~\mu\nu}$ denotes the electromagnetic part of the energy-momentum tensor, \addms{i.e., the second line of Eq.~(\ref{tmunu}). } 
The polar angle is defined by $\theta=\tan^{-1}(\sqrt{x^2+y^2}/z)$.
A particular value ($10^\circ$) is chosen for the surface integral because the opening angle of the funnel region is initially as narrow as $\sim 10^\circ$.
We always assume an observer located along the $z$-axis for evaluating $L_{\mathrm{iso}}$.

For models Q4B5n and Q4B5tn, the typical maximum value of $L_{\rm iso}$ is   $O(10^{50})\,{\rm erg/s}$.
The high-Poynting luminosity stage, which is designated by $L_{\mathrm{iso}}\agt3\times10^{49}$\,erg/s in this paper, is identified for $t\approx  300$--2500\,ms for model Q4B5n.
For model Q4B5tn, the high-Poynting luminosity stage for the upper and lower hemispheres is identified for $t\sim500$--1500\,ms and $\sim1000$--1600\,ms, respectively.
During the high-Poynting luminosity stage, $L_{\mathrm{iso}}$ varies with time by more than an order of magnitude for these models, reflecting the variation of the magnetic-field strength and configuration. 
The reason for this variability is that the magnetic fields with high field strengths are often provided from the disk to the black hole \addms{and the polar region} by the MRI dynamo activity. 
Figure~\ref{fig:BUT} shows the average value of the toroidal field $b_{\bar\varphi \mathrm{,ave}}$ as a function of time and polar angle $\theta$ for models Q4B3e15, Q4B5n, and Q4B5tn. 
Here, $x$, $y$, and $z$ are defined with respect to the black-hole center (the location of the puncture). 
The toroidal field is defined by $b_{\bar\varphi}=(xb_y-yb_x)/\sqrt{x^2+y^2}$.
The average is taken with respect to the azimuthal angle $\varphi=\tan^{-1}(y/x)$ at the selected radius of $r:=\sqrt{x^2+y^2+z^2} \approx 150$\,km. 
Figure~\ref{fig:BUT} displays the so-called butterfly structure~\cite{Brandenburg2005}; the polarity of the toroidal magnetic field flips in a quasi-periodic manner with the period of $\sim 20$ local orbital periods. 
Also, the strong magnetic-field fluxes continuously ascend from the equatorial plane to the surface of the accretion disk.

Although this butterfly structure induced by the MRI dynamo was already found in our previous models~\cite{hayashi2022jul}, we find an interesting new feature in our new computation; the MRI dynamo activity in the accretion disk determines the magnetic-field structure in the magnetosphere. In the dynamo activity, the magnetic fields often ascend from the disk to the vertical and polar regions. For most of the cases, they do not cancel out or alternate the fields originally stayed in the polar region, and thus, the polarity of the magnetosphere is unchanged.
However, for exceptional cases, the inversion of the polarity is achieved. For model Q4B5n, this occurs at $t\sim1.1$\,s and $1.4$\,s (see the middle panel of Fig.~\ref{fig:BUT}).
For $t \alt 1.1$\,s, the polarity of $b_{\bar\varphi \mathrm{,ave}}$ is positive in the polar region of the upper hemisphere. 
Then at $t\sim1.1$\,s, the polarity flips to negative, following the polarity flip at an inner region of the accretion disk. Subsequently, at $t\sim1.4$\,s, the polarity flips back to positive. 
For $L_{\mathrm{iso}}$ of this model, there are three characteristic peaks at $t\sim0.3$\,s, $1.3$\,s, and $1.5$\,s. These peaks reflect the variation of the butterfly structure at the polar region: During the polarity flips in progress, the intensity of the outgoing Poynting flux and $L_{\mathrm{iso}}$ naturally drop, because the magnetic field in the polar region is not aligned and the magnetosphere loses a coherency with respect to the magnetic-fields lines.

This polarity flip in the magnetosphere is also found in the snapshots of the magnetic-field structure. Figure \ref{fig:snap_xz_mag_Q4B5n} shows the toroidal and poloidal magnetic field structures on the $x$-$z$ plane at $t\sim0.3$\,s, $1.3$\,s, $1.5$\,s, and $2.7$\,s for model Q4B5n. The first three panels correspond to the snapshots at which $L_{\mathrm{iso}}$ is at local peaks. The figure clearly shows the polarity flip of both poloidal and toroidal magnetic fields in the magnetosphere. Around these time ranges, 
the magnetic fields ascending from the disk reconnect the originally-existing fields in the magnetosphere, and subsequently, the polarity is changed.

The middle and bottom panels of Fig.~\ref{fig:BUT} also show the magnetic-field polarity flip near the polar region for $t>2.5$\,s of model Q4B5n and for $t\sim1.0$\,s on the upper hemisphere of model Q4B5tn.
However, for these stages, no peak in $L_{\mathrm{iso}}$ is found and its typical value is lower than $\sim10^{49}$\,erg/s, which is an order of magnitude lower than the peak luminosity. 
Our interpretation for this is that the magnetic fields ascending from the disk due to the MRI dynamo activity disturb or deform the magnetosphere, but are not strong enough or aligned enough to replace the polarity of the field completely and reform the magnetosphere that can launch a high-intensity Poyting flux with $L_{\mathrm{iso}}\sim10^{50}$\,erg/s.

We note that the polarity flip was already reported in magnetohydrodynamics simulations for the accretion disks around a spinning black hole~\cite{christie2019sep,liska2020}. For these simulations, the authors also found a turbulent state of the accretion disks. 
Thus, the polarity flip is likely to occur often, if magnetic fluxes with high field strengths are ejected from the inner region of the disks in a turbulent state. 

For both models Q4B3e15 and SFHoQ4B5, the maximum value of $L_{\mathrm{iso}}$ is $\sim 5\times10^{49}$\,erg/s, which is slightly lower than those of other models for which typically $L_\mathrm{iso}\sim10^{50}$\,erg/s in the bright stages. Our interpretation is that this is due to the lower magnetic-field strength in the magnetosphere. For these models, the rest-mass density in the disk at the time when the magnetosphere is formed is lower. 
The reason for model Q4B3e15 is that it takes a longer time to form the magnetosphere than for the other DD2 models and the reason for model SFHoQ4B5 is that the disk mass is smaller than for the DD2 models. 
As we already mentioned, the field strength in the magnetosphere is determined by the field strength of the disk at which the equipartition state is achieved. 
For model Q4B3e15, the equipartition is achieved in the relatively late stage, at which the rest-mass density and internal energy of the disk are relatively low. 
This leads to a lower magnetic-field strength in the magnetosphere for these two models. 
As a result of the lower field strength, the Poynting luminosity, which is powered by the Blandford-Znajek mechanism, becomes lower. 
The lower maximum value of $L_{\mathrm{iso}}$ for model SFHoQ4B5 is understood as the physical result, while that for model Q4B3e15 could be the numerical artifact due to the insufficient grid resolution.

The high-Poynting luminosity stage for model Q4B3e15 starts at $\sim2600$\,ms and lasts for $\sim400$\,ms, entering the fading stage at $\sim3000$\,ms. For model SFHoQ4B5, the high-Poynting luminosity stage starts at $\sim1100$\,ms. We do not find a clear fading stage for this model, but $L_{\mathrm{iso}}$ appears to gradually decrease to $\sim10^{49}$\,erg/s at the termination of the simulation. We indeed find for this model that the opening angle increases with a timescale of a few seconds, and thus, we expect that $L_{\mathrm{iso}}$ will eventually drop in this timescale. 

The isotropic-equivalent Poynting luminosity of $L_{\mathrm{iso}}\sim 10^{50}$\,erg/s together with the opening angle of $\theta \sim 10^{\circ}$  (cf.~Fig.~\ref{fig:3D4_Q4B5tn}) fairly agrees with those for short-hard gamma-ray bursts (or at least for low-luminosity short-hard gamma-ray bursts) in the assumption that the conversion efficiency of the Poynting flux to the gamma-ray radiation is sufficiently high (i.e., close to unity)~\cite{nakar2007apr,berger2014jun}. 

\addkh{
The jet efficiency, which is defined by the ratio of the Poynting luminosity to mass accretion rate on the apparent horizon is $\alt 10^{-2}$ in the present simulation. 
Although the value is increasing, it is still much smaller than that for Ref.~\cite{christie2019sep}, which increases up to $\sim 1$.
This difference is mostly because the infalling mass flux near the equatorial plane is still high in our simulation. 
Around the polar region, however, the ratio of the Poynting flux to the infalling mass flux is $\sim 1$ locally, and through such a region, the black hole is capable of powering outgoing Poynting luminosity into the magnetosphere (see also the discussion of the next subsection).
}

\subsubsection{MADness parameter}  \label{sec:madness}

\begin{figure}[!th]
      \begin{center}
        \includegraphics[scale=0.4]{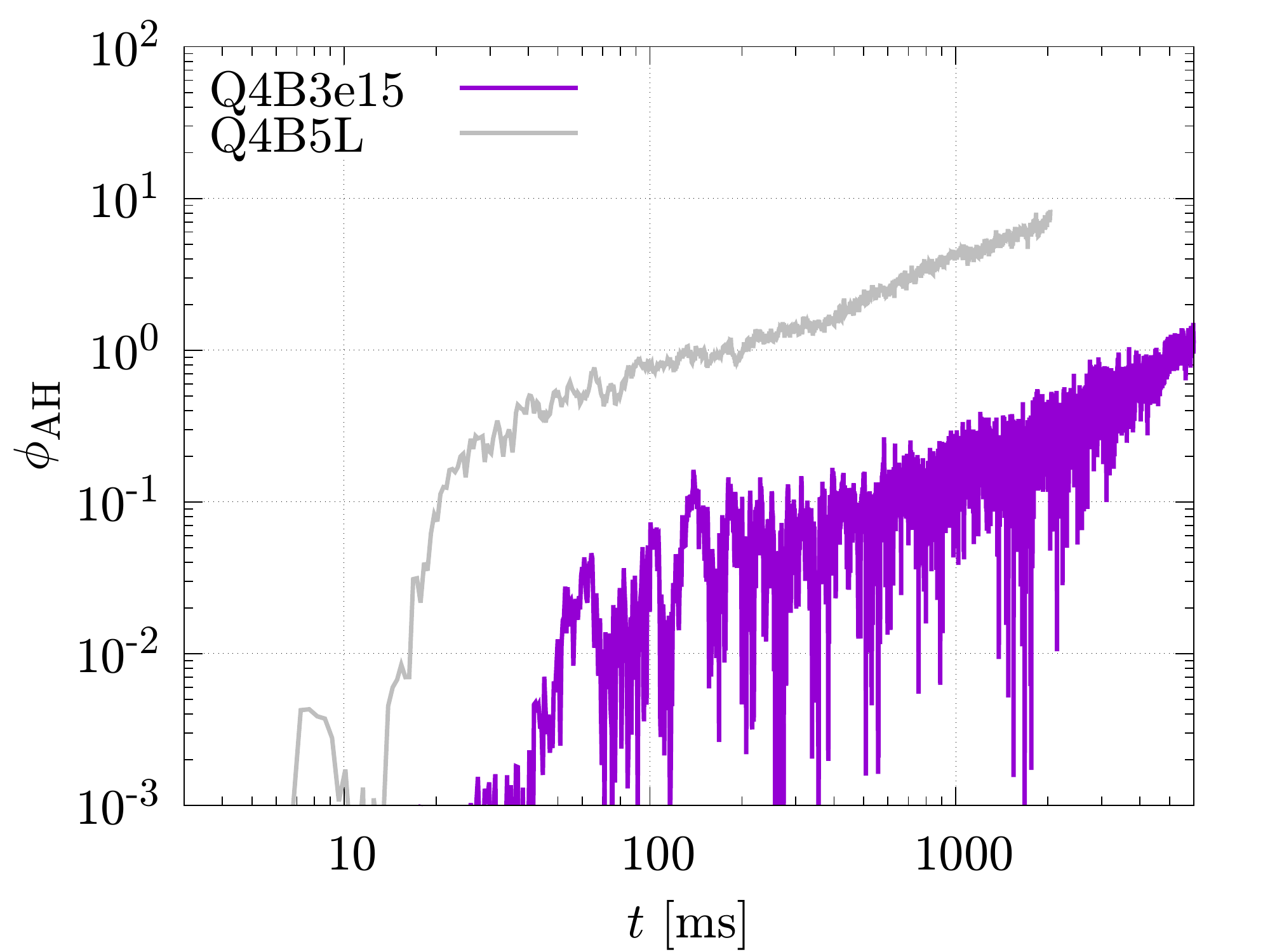} \\
        \includegraphics[scale=0.4]{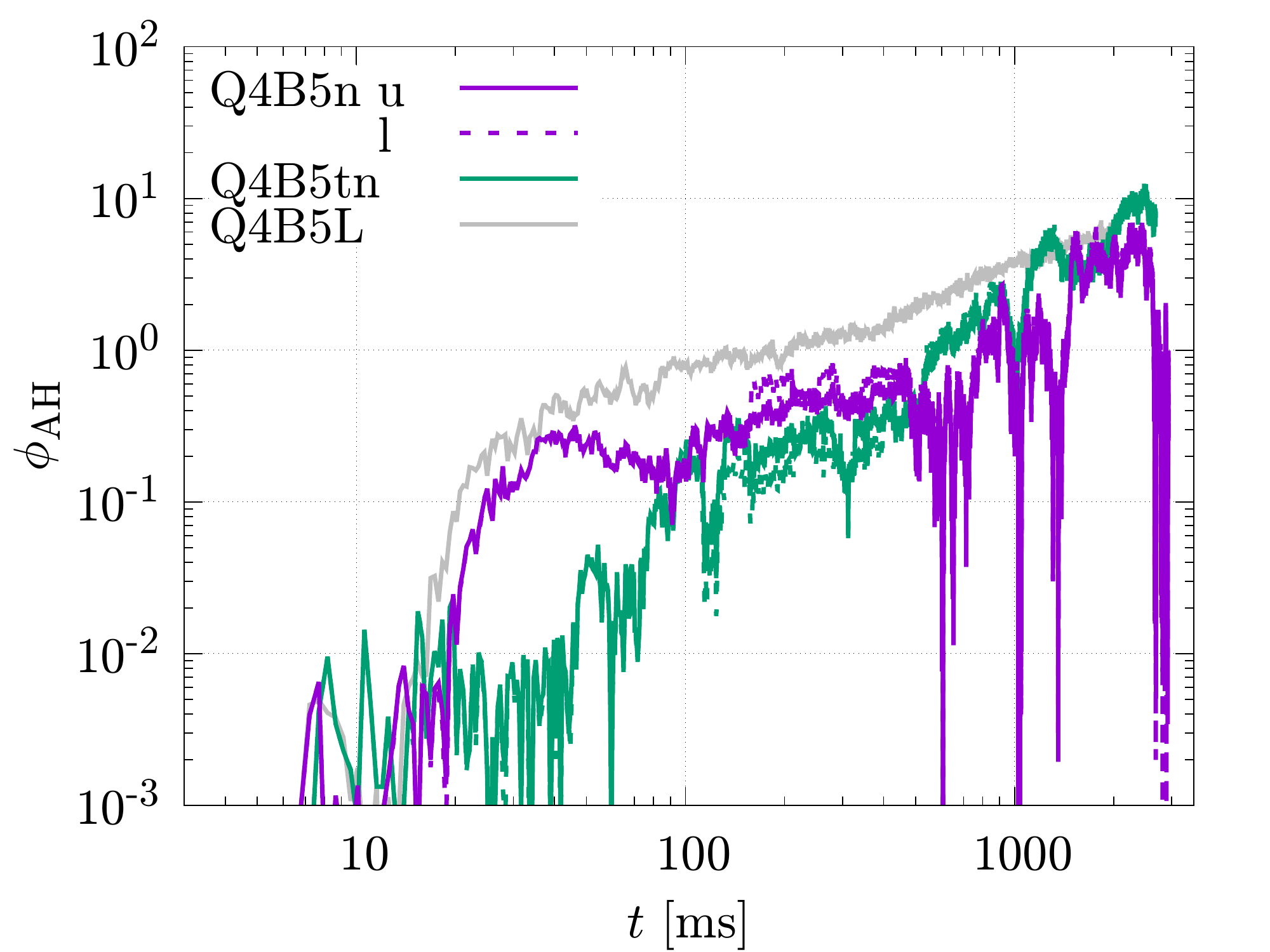} \\
        \includegraphics[scale=0.4]{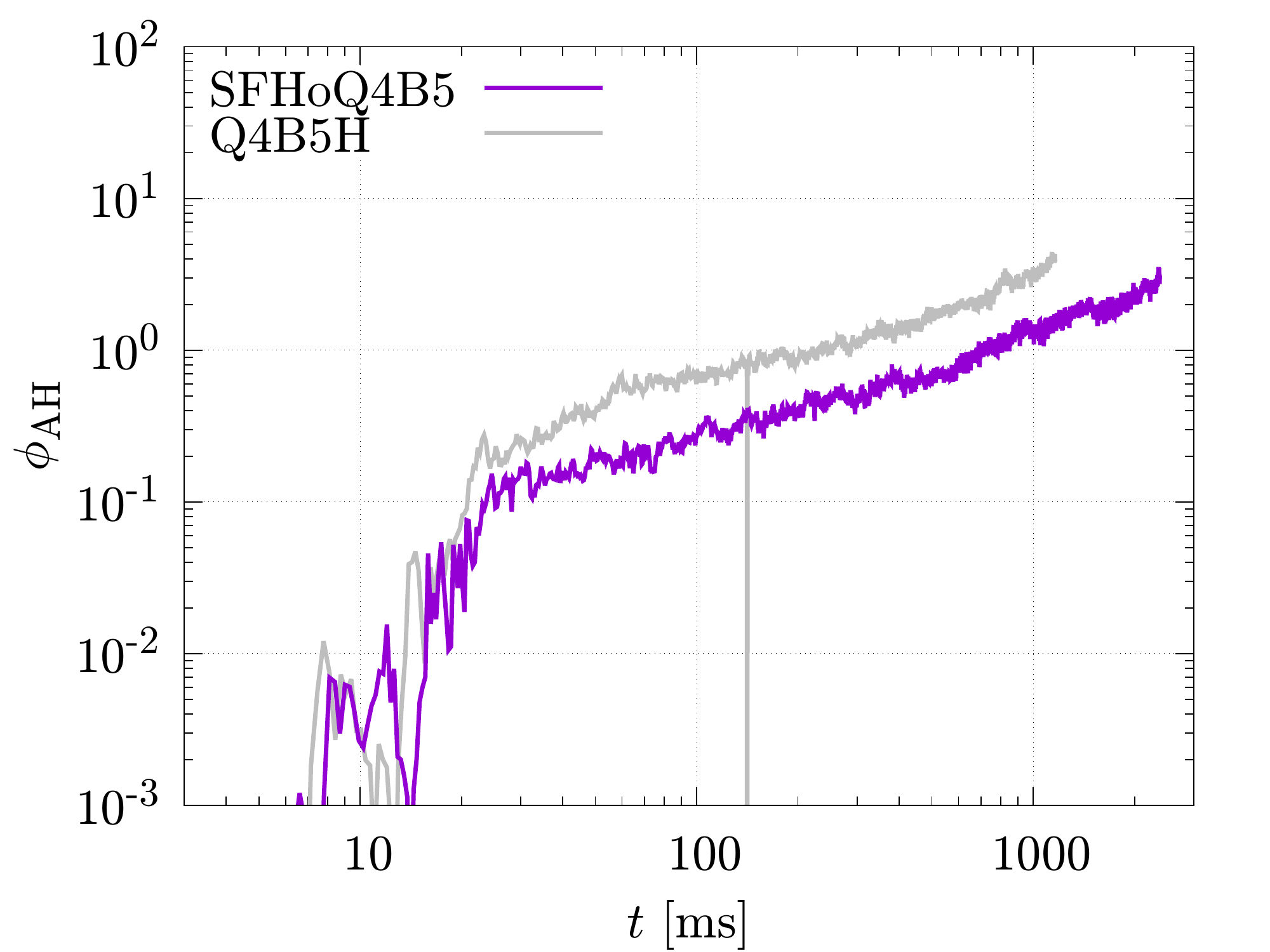} 
        \caption{The time evolution of the MADness parameter 
        for models Q4B3e15 (top panel), Q4B5n and Q4B5tn (middle panel), and SFHoQ4B5 (bottom panel).
        The results for models Q4B5L and Q4B5H of our previous paper \cite{hayashi2022jul} are also shown for comparison (in grey color).
        The characters ``u'' and ``l'' denote the upper and lower hemispheres, respectively.
	}
        \label{fig:mad}
      \end{center}
\end{figure}

\begin{figure}[!th]
      \begin{center}
        \includegraphics[scale=0.4]{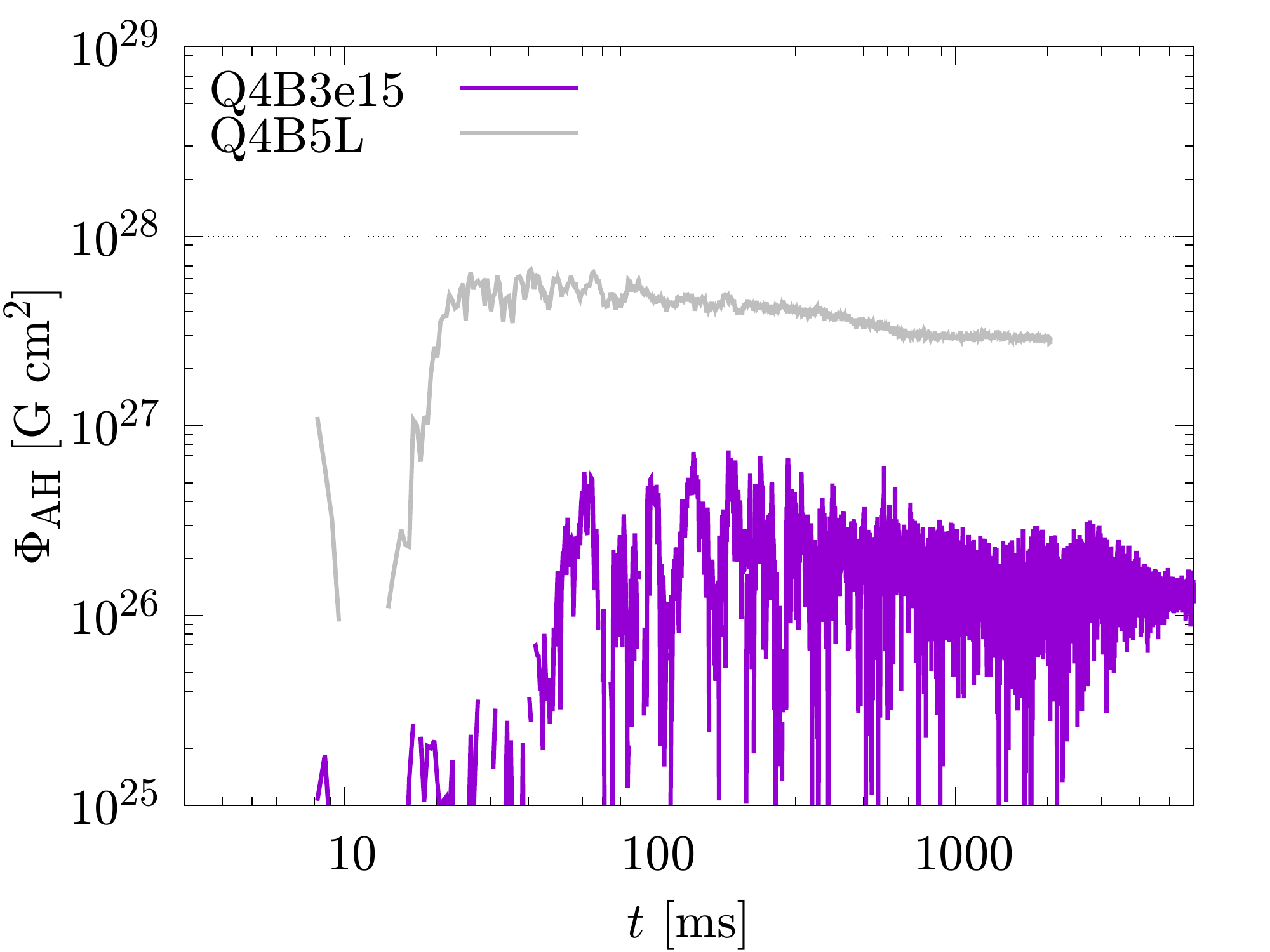} \\
        \includegraphics[scale=0.4]{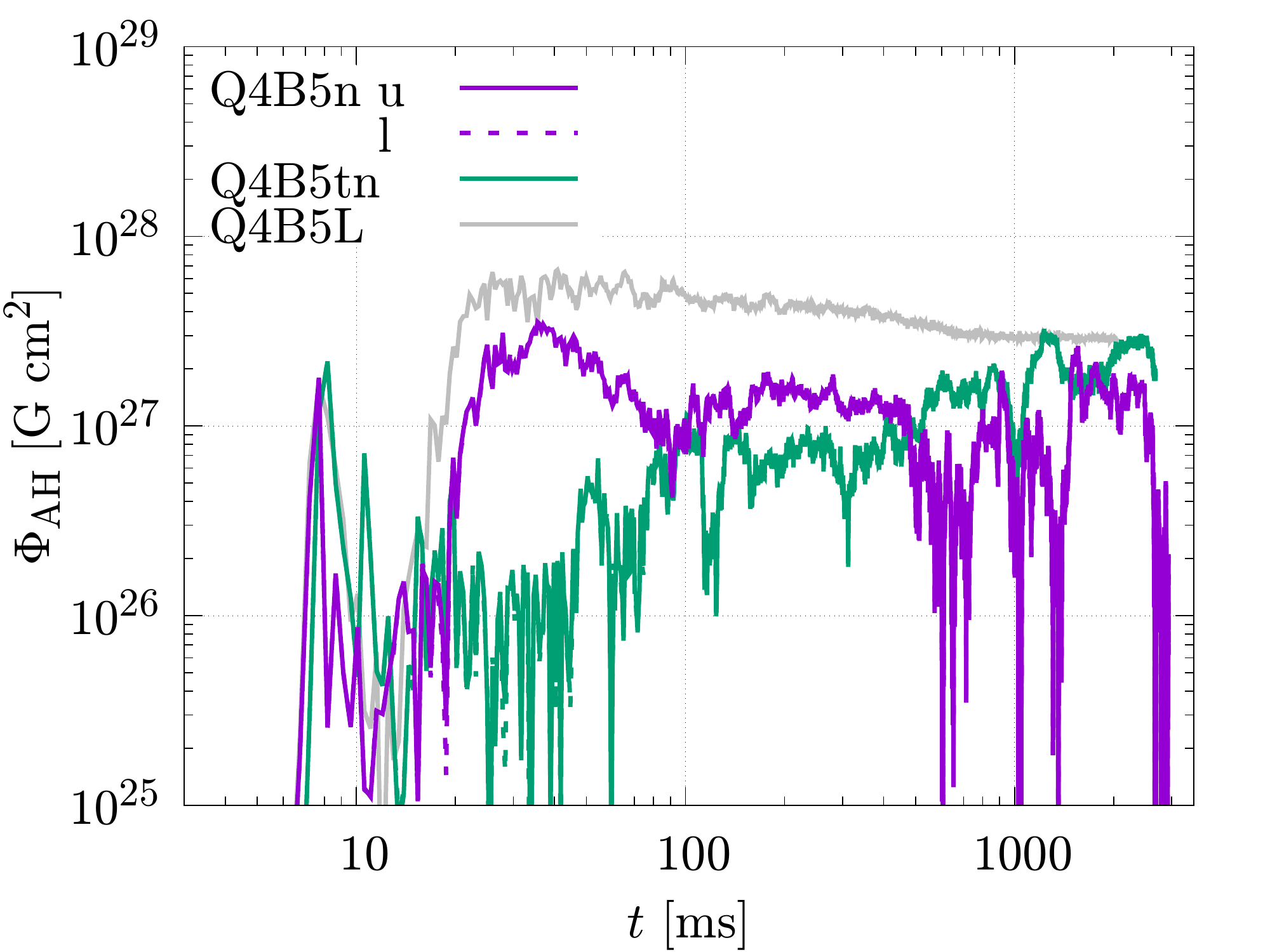} \\
        \includegraphics[scale=0.4]{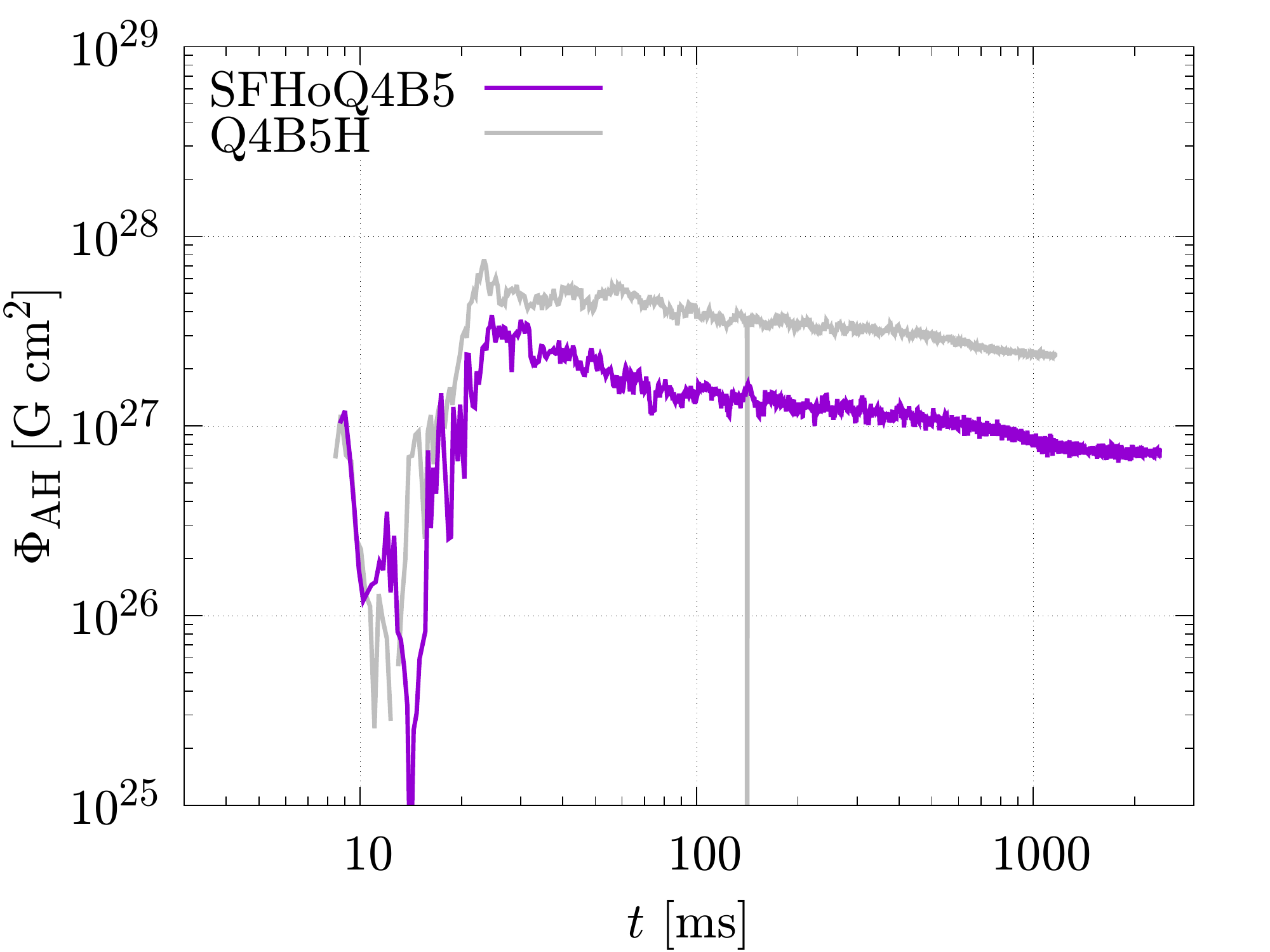} 
        \caption{The same as Fig.~\ref{fig:mad} but for the 
        time evolution of the magnetic flux evaluated on the apparent horizon. 
	}
        \label{fig:phiB}
      \end{center}
\end{figure}

\begin{figure*}[!th]
      \begin{center}
        \includegraphics[scale=0.345]{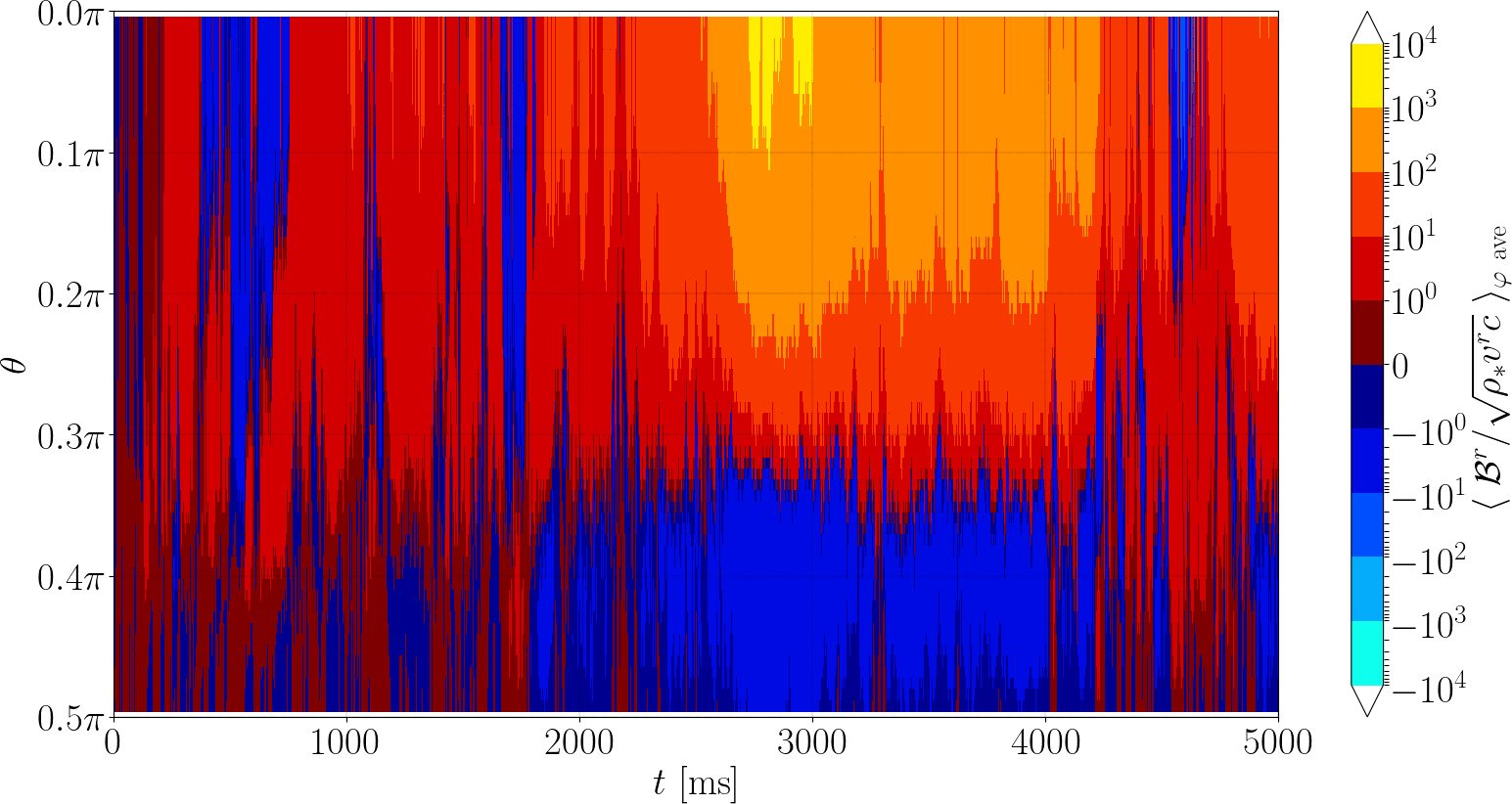} \\ 
        \vspace{2mm}
        \includegraphics[scale=0.345]{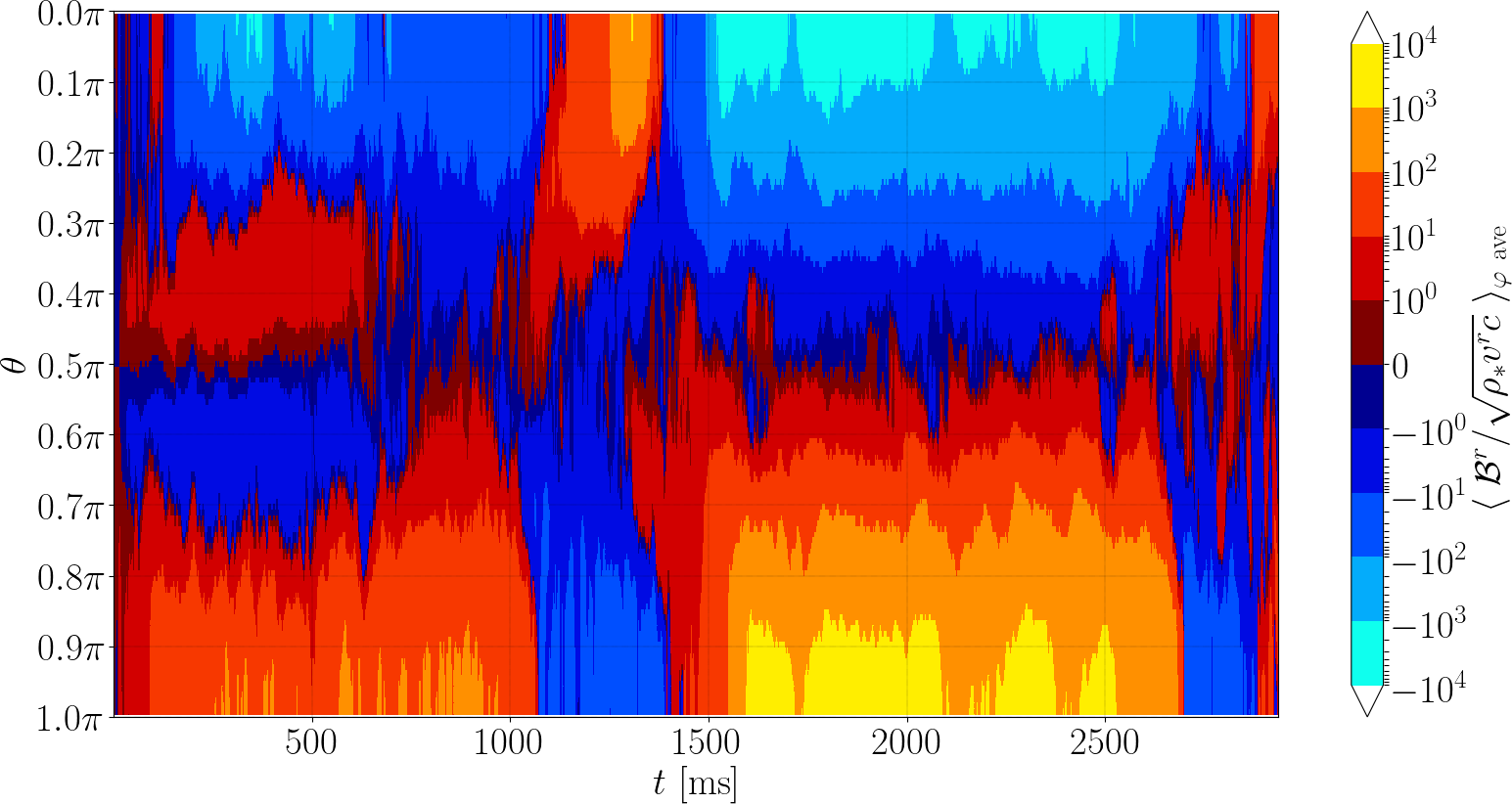} \\
        \vspace{2mm}
        \includegraphics[scale=0.345]{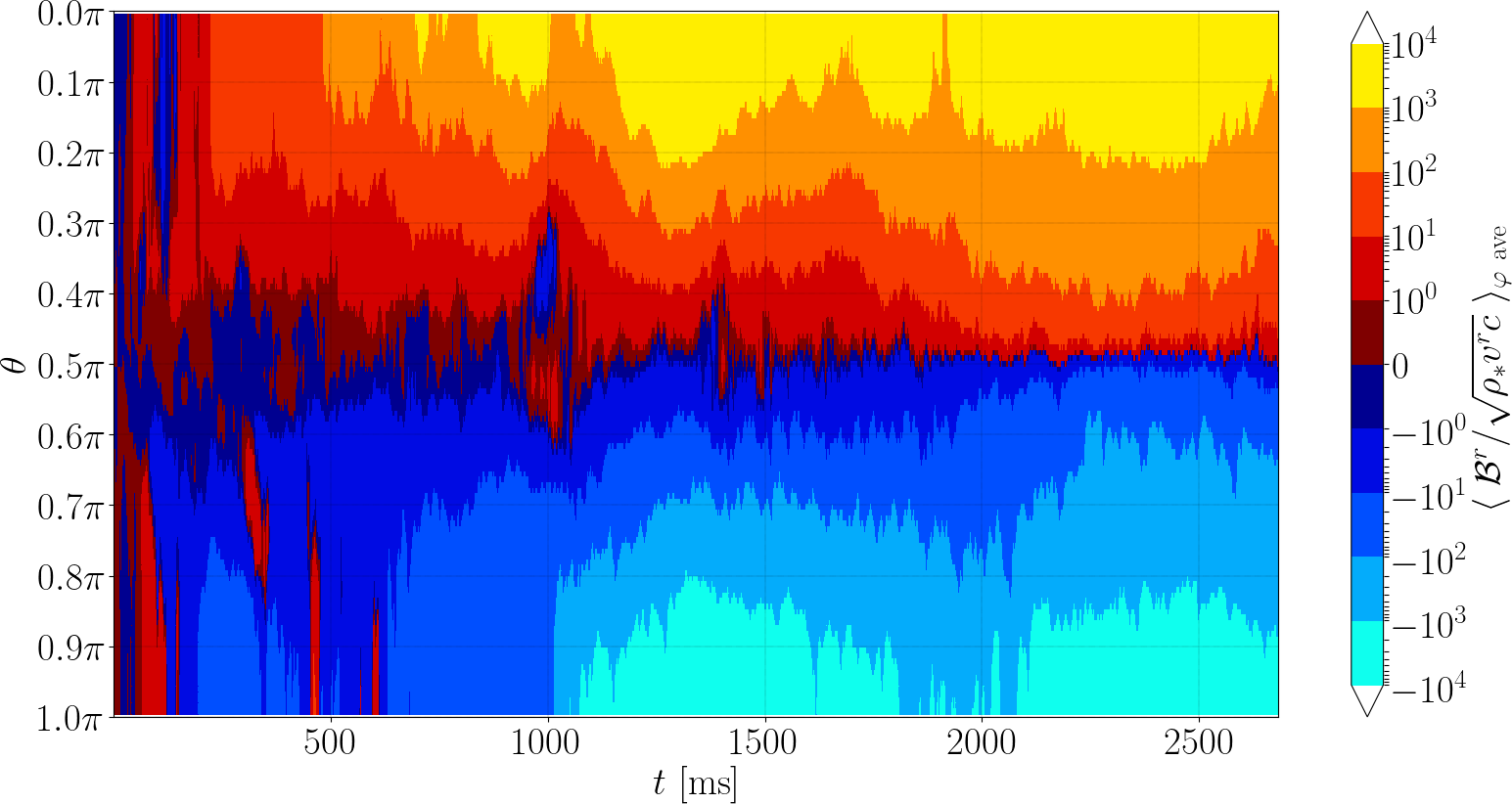} \\
        \caption{The profile of the local MADness parameter $\phi_{\mathrm{AH,local}}$ along the polar direction ($\theta$) as a function of time for models Q4B3e15 (top panel), Q4B5n (middle), and Q4B5tn (bottom).
        }
        \label{fig:lmad}
      \end{center}
\end{figure*}

\begin{figure*}[!th]
      \begin{center}
        \includegraphics[scale=0.28]{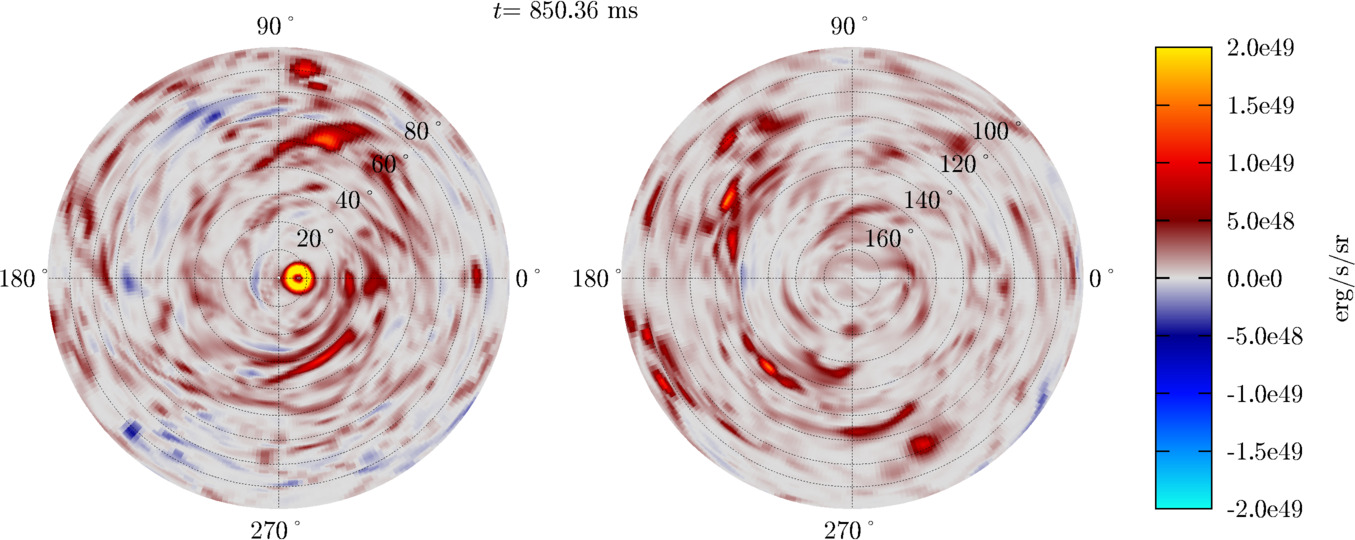} \\ 
        \vspace{2mm}
        \includegraphics[scale=0.28]{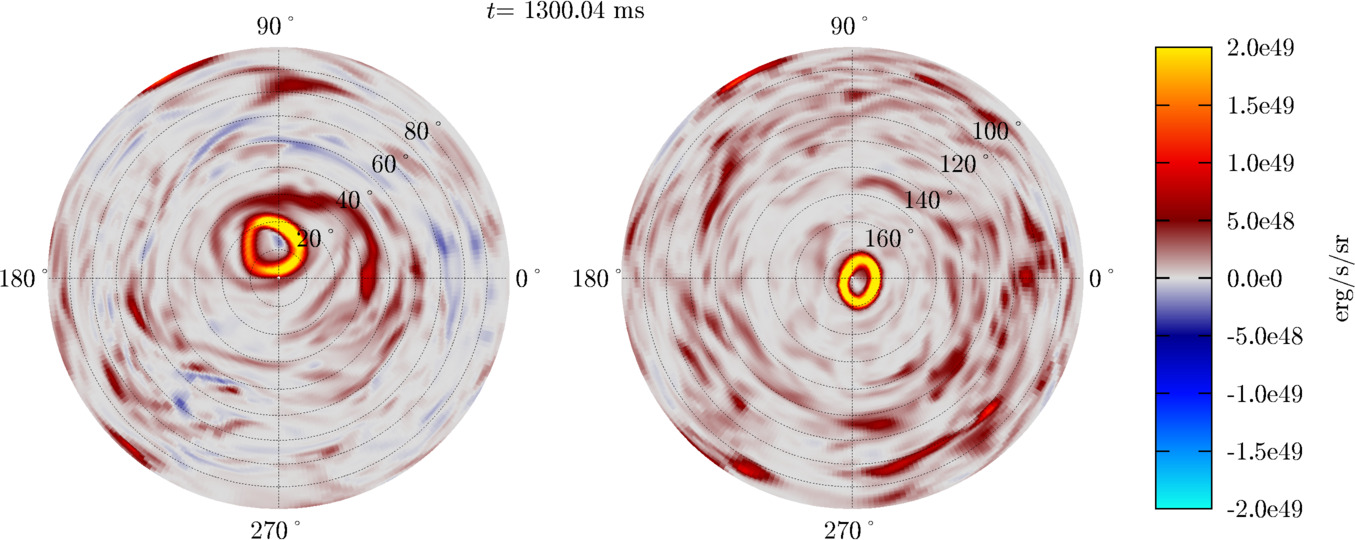} \\
        \vspace{2mm}
        \includegraphics[scale=0.28]{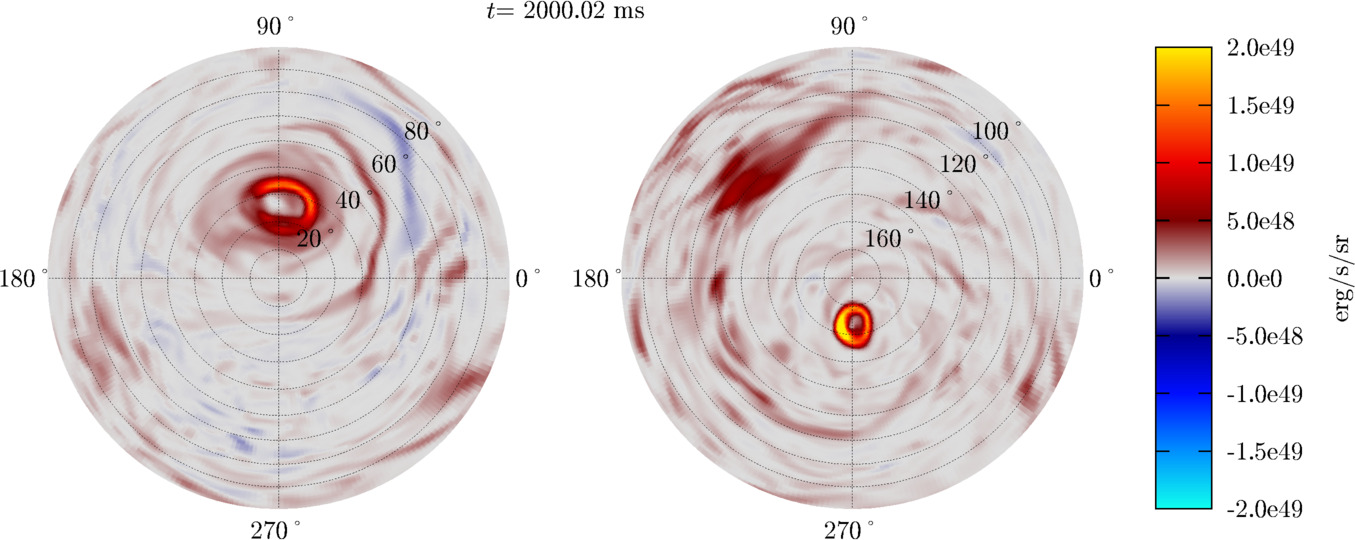} \\
        \vspace{2mm}
        \includegraphics[scale=0.28]{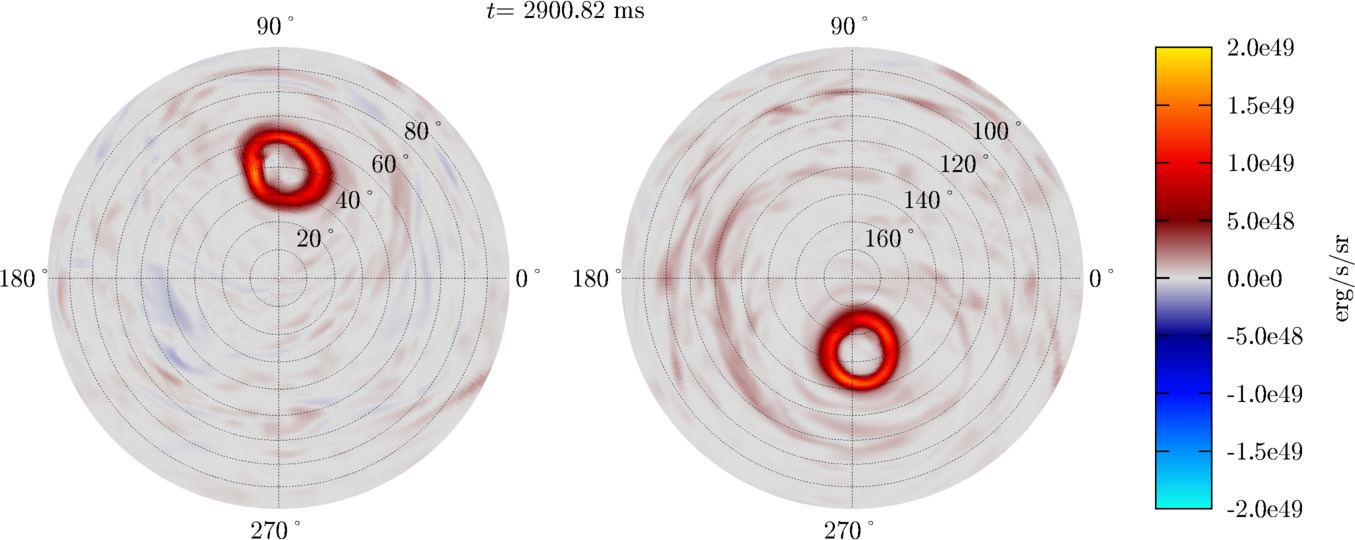} \\
        \caption{The angular distribution of the Poynting flux per steradian on a sphere of $r\approx1500$\,km for model Q4B5tn at selected time slices. 
        The left and right panels display the upper and lower hemispheres, respectively.
        The bright color displayed in the polar region stems from the Blandford-Znajek effect, while for other regions, the magnetic field accompanied by the outflowing matter contributes mainly to the Poynting flux. 
        The region for which the Poynting flux is intense moves in the direction of the $y$-axis from the vicinity of the pole.
        At the same time, the opening angle of the Poynting flux gradually increases.
        See also the following link for the animation: \url{https://www2.yukawa.kyoto-u.ac.jp/~kota.hayashi/Q4B5tn-pf.mp4}.
        }
        \label{fig:ang_dist_pflux_Q4B5tn}
      \end{center}
\end{figure*}

Figure~\ref{fig:mad} shows the time evolution of the so-called MADness parameter which is defined by~\cite{tchekhovskoy2011}
\begin{eqnarray}
  \phi_{\mathrm{AH}} := \frac{ \Phi_{\mathrm{AH}} }{ \sqrt{ \dot{M}_{\mathrm{AH}} 4\pi r_{\mathrm{AH}}^2 c } },
\end{eqnarray}
where
\begin{eqnarray}
  \dot{M}_{\mathrm{AH}}&:=&\oint_{\mathrm{AH}} \rho \sqrt{-g} u^{i} d S_{i} \nonumber \\
  &\approx&\oint_{r=\max{(r_{\mathrm{AH}}})} \rho_{*} v^r r^2\sin{\theta} d\theta d\varphi, 
  \end{eqnarray}
and
  \begin{eqnarray}
  \Phi_{\mathrm{AH}}&:=&\oint_{\mathrm{AH}} B^i \sqrt{\gamma} d S_{i} \nonumber \\
  &\approx&\oint_{r=\max{(r_{\mathrm{AH}}})} \mathcal{B}^r r^2 \sin{\theta} d\theta d\varphi,  
\end{eqnarray}
which denotes the magnetic flux penetrating the apparent horizon. It is found that for all the models $\phi_\mathrm{AH}$ remains smaller than 10 within the simulation time.
Thus, the accretion disks in our simulations do not satisfy the often-referred condition for the magnetically-arrested disk (MAD), $\phi_{\mathrm{AH}}\agt 50$~\cite{tchekhovskoy2011}.\footnote{\addkh{The present results are quantitatively similar to the model BT in Ref.~\cite{christie2019sep} as we discussed in our previous work~\cite{hayashi2022jul}.}} 
However, as we have already described, a high-intensity Poynting flux is generated even if the condition for the MAD is not satisfied. Thus, 
the MADness parameter $\phi_{\mathrm{AH}}$ may not be suitable for assessing whether the jet is launched in the context of neutron-star mergers. 

Figure~\ref{fig:phiB} shows the time evolution of $\Phi_{\mathrm{AH}}$. For models Q4B5n and Q4B5tn, the magnetic fluxes penetrating the upper and lower hemisphere of the apparent horizon are shown separately, although these two components approximately agree with each other.
For most of the models, the magnetic flux on the apparent horizon reaches its peak in the timescale similar to that for achieving the equipartition in the accretion disk. 
Since $\Phi_{\mathrm{AH}}$ does not increase after the peak is reached, the decrease of the accretion rate is the only path for the increase of the MADness parameter for such models.
For model Q4B5tn, which initially has the toroidal magnetic field in the neutron star, $\Phi_{\mathrm{AH}}$ is low for the early post-merger stage ($t\lesssim50$\,ms) and it increases gradually in the entire simulation time. 
For this case, field lines starting from a point in each hemisphere always end in the same hemisphere for the early stage. 
However, due to the MRI turbulence and dynamo, the poloidal field is developed and subsequently penetrates the black-hole horizon, resulting in the increase of  $\Phi_{\mathrm{AH}}$. By contrast, for the pure poloidal initial field, the black hole horizon is penetrated by the poloidal field from the early stage. 

Because the MADness parameter defined on the entire horizon surface might not be a good indicator for assessing the launch of the strong Poynting flux, instead of it, we propose another parameter based on the local quantities.
The point is that the magnetic-field lines that generate the strong Poynting flux do not penetrate the black-hole horizon in the vicinity of the equator, at which dense matter infalling from the accretion disk is always present and the (low-beta) magnetosphere is not formed. 
This suggests that focusing on the polar region on the apparent horizon for evaluating the MADness-like quantity would be a better strategy. 
Thus, we introduce a local MADness parameter, $\phi_{\mathrm{AH, local}}$, which is defined by
\begin{eqnarray}
  \phi_{\mathrm{AH,local}} := \frac{ \mathcal{B}^{r} }{ \sqrt{ \rho_{*} v^{r} c } }.
\end{eqnarray}
Figure~\ref{fig:lmad} shows the azimuthal-average value of the local MADness parameter $\phi_{\mathrm{AH, local}}$ as a function of time and polar angle $\theta$ for models Q4B3e15, Q4B5n, and Q4B5tn. 
For all the models, we find time intervals with $L_{\mathrm{iso}}\gtrsim 3\times10^{49}$\,erg/s, and for such time intervals we always find $\phi_{\mathrm{AH,local}} \gtrsim 100$ at polar region of the apparent horizon. 
This suggests that the black hole has the ability to form a magnetosphere and launch a jet if the value of $\phi_{\mathrm{AH,local}}$ at the polar region exceeds $100$, even if the value of $\phi_\mathrm{AH}$ is smaller than 50. 

We note here that $\phi_{\mathrm{AH,local}}$ only gives us the necessary condition for the launch of a jet with high-Poynting luminosity.
The disturbance or the deformation of the magnetosphere far from the horizon associated with the evolution of the accretion disk could result in a low value of $L_{\mathrm{iso}}$.
For example the local MADness parameter exceeds 100 at $t\gtrsim1700$\,ms but $L_{\mathrm{iso}}$ falls below $10^{49}$\,erg/s for Q4B5tn model.
This is due to the deformation (tilt) of the magnetosphere induced by the post-merger mass ejection. The details of this behavior are given in the next subsection.

Focusing on the polarity, the local MADness parameter for model Q4B5n shows interesting behavior. 
As we already pointed out in this subsection, this model shows a clear butterfly structure of $b_{\bar\varphi \mathrm{,ave}}$ extending to the polar region due to the flip of the magnetic-field polarity, and this flip occurs at the peaks of $L_{\mathrm{iso}}$.
A similar flip is also observed for the local MADness parameter for $t=1000$--$1500$\,ms (see Fig.~\ref{fig:lmad}). 
Thus, we conclude that the complete flip of the magnetic-field polarity in the magnetosphere is the result of this polarity flip on the apparent horizon. 
Just like in the MRI dynamo and butterfly structure of $b_{\bar\varphi \mathrm{,ave}}$ in the accretion disk, the polarity flip starts in the vicinity of the equatorial plane and propagates to the polar region. 
This is because the value of $\phi_{\mathrm{AH,local}}$ is lower than $10$ near the equatorial plane and the fluid dynamics dominates over the magnetic-field dynamics.
This feature enables the matter accretion from the disk to occasionally carry the magnetic field with opposite polarity. 
Then, the magnetic tower effect enhances the magnetic-field strength along the polar direction.
In this process, the preexisting magnetic field near the pole is dissipated away due to the reconnection by the magnetic field with opposite polarity ascending from the equatorial region, which replaces the polarity of the field penetrating the polar region of the horizon.
Once the magnetic tower effect is in action, the magnetic field is amplified by the winding (associated with the black-hole spin) and the matter is pushed outward.
As a result, a high-$\phi_{\mathrm{AH,local}}$ region is realized near the pole, where the magnetic-field strength is high and the rest-mass accretion rate is low. 
In the high-$\phi_{\mathrm{AH,local}}$ region, the magnetic-field dynamics dominates the fluid dynamics, and hence, the polarity flip cannot start in the polar region. 

\subsubsection{Time duration for high Poynting luminosity} \label{sec:time_duration_pflux}

The stage with a high value of $L_{\mathrm{iso}}$ continues for $\sim400$--$2200$\,ms and subsequently starts decreasing. 
This is particularly clear for models Q4B3e15, Q4B5n, and Q4B5tn. 
For these models, we confirmed that the value of $L_{\mathrm{iso}}$ decreases by nearly two orders of magnitude in the fading stage. 
This is consistent with the duration of the short-hard gamma-ray bursts, whose typical duration is $\sim1$\,s~\cite{nakar2007apr}.

In our previous paper~\cite{hayashi2022jul}, we discussed that the decrease of $L_{\mathrm{iso}}$ is due to the increase in the opening angle of the funnel region and the resulting decrease of the magnetic-flux density in the magnetosphere.
The opening angle of the strong Poynting-flux region increases from $\lesssim10^{\circ}$ to $\sim20^{\circ}$.
This is directly related to the accretion disk evolution.
The location of the funnel wall is determined by the balance between the gas pressure of the thick torus and the magnetic pressure at the funnel wall.
In the seconds-long evolution of the torus (disk), the rest-mass density and the gas pressure at the funnel wall gradually decrease due to the post-merger mass ejection and matter accretion onto the black hole.
On the other hand, the magnetic pressure in the magnetosphere and at the funnel wall does not decrease significantly, in particular for the late stage of the evolution.
Thus the magnetic pressure can eventually exceed the gas pressure at the original position of the funnel wall, resulting in the gradual expansion of the funnel region.
As discussed previously, this mechanism could be one of the ingredients that determine the time duration of short-hard gamma-ray bursts.

For the SFHoQ4B5 model, the opening angle also increases with time but with a longer timescale. As a result, we do not find the clear fading stage of $L_{\mathrm{iso}}$ for this model in our simulation time. 
Due to the high computational cost, we terminated the simulation at $t\sim2500$\,ms, but if we evolve the system longer, the fading stage is likely to be present.

From the results for models Q4B3e15, Q4B5n, and Q4B5tn, we find two additionally possible mechanisms for the fade-away of the Poynting luminosity, which could also explain the time duration of short-hard gamma-ray bursts. 
For both mechanisms, the non-trivial evolution of the magnetosphere associated with the evolution of the accretion disk is essential. 

For models Q4B3e15 and Q4B5n, in the very late stage of our simulation, the aligned magnetic field is dissipated away, and as a result, the magnetosphere with aligned magnetic fields disappears. 
Figure~\ref{fig:snap_xz_mag_Q4B5n} shows that for $t\alt 1550$\,ms the poloidal magnetic-field lines in the polar region are aligned approximately with the black-hole spin axis ($z$-axis). In the magnetosphere along the $z$-axis, the magnetic-field lines are clearly helical and maintain the high-intensity Poynting flux. 
However, it is found that at $t\approx2730$\,ms the magnetic field in the polar region is not aligned anymore. 
Moreover, the magnetic-field dynamics cannot govern the fluid dynamics and the clear magnetosphere disappears. 
In the absence of the well-ordered magnetic field, the system cannot maintain the high-intensity Poynting flux.

Figure~\ref{fig:BUT} shows that for model Q4B5n with $t\sim1500$--$2500$\,ms for which the high Poynting luminosity is maintained (see Fig.~\ref{fig:plum}), the polarity of the poloidal magnetic field at the polar region remains to be preserved and does not reverse. 
However, for $t\gtrsim2500$\,ms, the Poynting luminosity decreases. For this late stage, the polarity in the polar region frequently reverses in response to the polarity reversal in the disk near the equatorial plane. The local MADness parameter in Fig.~\ref{fig:lmad} also shows similar behavior.
We interpret that the disappearance of the magnetosphere and the decrease of the Poynting flux stem from the MRI dynamo activity in the accretion disk.
Due to the MRI dynamo activity, the polarity of the magnetic field in the accretion disk is reversed quasi-periodically and the magnetic flux continuously ascends toward the polar region from the equatorial region.  
If the magnetic field ascending from the disk has the polarity different from that in the magnetosphere and its field strength is high enough, the magnetic field that is  originally located in the magnetosphere could be dissipated away by the magnetic-field reconnection. 
This is what is observed for $t\gtrsim2500$\,ms of model Q4B5n, and for $t\gtrsim3000$\,ms of Q4B3e15 model. 
If the strength of the ascending magnetic field is even stronger, then it could replace the polarity completely and reform the magnetosphere again.
This is what is observed for $t\approx1000$--$1500$\,ms of model Q4B5n.
However, we expect that the magnetosphere formation will not occur again after the disappearance of the strong magnetic fields for models Q4B3e15 and Q4B5n because as already discussed, the magnetic-field strength in the disk is determined by the rest-mass density (and thus internal energy density) achieved for the equipartition relation. Since the rest-mass density decreases as a result of the disk expansion, the post-merger mass ejection, and the mass accretion onto the black hole, 
the magnetic-field strength in the disk also decreases. Therefore, the revival of a magnetosphere with strong magnetic fields is not possible in the late stage of the disk. 
In the absence of strong magnetic fields, the Poynting luminosity is low because the Blandford-Znajek Poynting luminosity is proportional to the square of the magnetic-field strength~\cite{blandford1977}. 

Model Q4B5tn shows the second mechanism. In this mechanism, the magnetosphere still exists, but the Poynting luminosity becomes apparently low if we measure along the direction of the black-hole spin. The mechanism of this is the tilt of the accretion disk and magnetosphere.
Figure~\ref{fig:ang_dist_pflux_Q4B5tn} shows the angular distribution of the Poynting flux per steradian defined by $-T^{~r}_t\sqrt{-g}/\sin{\theta}$ on a sphere of $r\approx1500$\,km for model Q4B5tn.
For $t\lesssim1300$\,ms the strong Poynting-flux region is approximately aligned with the polar direction ($z$-axis direction and black-hole spin direction).
However, at $t\sim2000$\,ms the strong Poynting-flux region starts deviating from the polar region clearly, and at $t\approx2900$\,ms it is tilted by $25^{\circ}$--$40^{\circ}$ from the $z$-axis direction in the $y$-$z$ plane.
The isotropic-equivalent luminosity is defined by the integration for $0^{\circ}<\theta<10^{\circ}$ and $170^{\circ}<\theta<180^{\circ}$ on the upper and lower hemispheres, respectively, and thus, it decreases significantly by the tilt.
We note that the black-hole spin axis is aligned with the $z$-axis during the entire simulation time.
As Fig.~\ref{fig:2D4_Q4B5tn} shows, the system approximately has the equatorial-plane symmetry in the early stage of the post-merger evolution. However, later, the system loses the symmetry; the accretion disk misaligns with the equatorial plane and the major axis of the magnetosphere also deviates from the $z$-axis direction.
This is due to the asymmetric nature of the post-merger mass ejection resulting from the turbulent state of the accretion disk.\footnote{\addkh{This asymmetric mass ejection is always present in the absence of equatorial symmetry, and the origin is a small turbulence in the disk. The asymmetry is enhanced subsequently. It is difficult to mention what happens in the long-term evolution process, but one important point is that there is no mechanism that suppresses the enhancement of the asymmetry. Thus, once the equatorial asymmetry is induced in the matter distribution, the asymmetry is always enhanced, and as a result, the asymmetric mass ejection takes place.}}
Specifically, in this model, the post-merger ejecta carries a large amount of the $y$-component of the angular momentum and this breaks the symmetry of the accretion disk. 
Indeed, the remnant accretion disk has the $y$-component of the angular momentum, and as a result, the system tilts in the $y$-$z$ plane. 
\footnote{We note that the system is still capable of launching a jet in an off-axis direction. Indeed we still find the strong Poynting-flux region even after  $L_{\mathrm{iso}}$ (in our definition) drops.}

We find not only the tilt of the high Poynting-flux region, but also the widening of the opening angle in Fig.~\ref{fig:ang_dist_pflux_Q4B5tn} like in the models of our previous simulations~\cite{hayashi2022jul}. For model Q4B5tn in Fig.~\ref{fig:ang_dist_pflux_Q4B5tn}, at $t\approx850$\,ms, the opening angle is $\lesssim10^{\circ}$ but it increases to $\sim30^{\circ}$ at $t\approx2900$\,ms. Thus, the intensity of the Poynting flux decreases simultaneously with tilting. This widening of the opening angle results from the decrease of the gas pressure at the funnel wall due to the post-merger mass ejection and matter accretion onto the black hole, as we already described.

We have found the two new possible mechanisms for the decrease of the Poynting luminosity in addition to one mechanism that we already found in our previous paper~\cite{hayashi2022jul}. For all the three mechanisms, the evolution of the accretion disk and the post-merger mass ejection are critical processes. 
Our results show that irrespective of the mechanisms, the timescale of $\sim1$\,s for the high Poynting-luminosity stage is determined by the evolution timescale of the accretion disk, which is determined by the neutrino cooling and magnetohydrodynamics turbulence that control the post-merger mass ejection. 

A word of caution is appropriate here. The system, specifically the accretion disk, is in a turbulent state by the MRI. That is, the evolution of the disk and MRI dynamo activity are determined by a stochastic process. 
This implies that we cannot precisely predict the strength of the magnetic field that penetrates the black hole and forms the magnetosphere and the angular momentum that is carried away by the post-merger ejecta. 
It is also not easy to predict whether the black hole will always be penetrated by the magnetic field strong enough to form a magnetosphere and launch the Poynting flux that can explain typical short-hard gamma-ray bursts. 
It is also not easy to predict by which mechanisms the high Poynting-luminosity stage is terminated. However, our simulation results show that once the magnetosphere is formed, its subsequent evolution is determined by the global properties of the post-merger ejecta and the accretion disk.

\section{Conclusion} \label{sec:conclusion}
We have reported the new results of general-relativistic neutrino-radiation magnetohydrodynamics simulation for seconds-long black hole-neutron star mergers. 
We employed a variety of setups that are different from those in our previous paper~\cite{hayashi2022jul}, while the mass of the black hole and neutron star and the black-hole spin are identical with those of the previous setting so that the neutron star is disrupted by the tidal force of the black hole. 
The difference in the setup for the present work from the previous work is summarized as follows: 
(1) The initial maximum magnetic-field strength in the neutron star is assumed to be $3 \times 10^{15}$\,G, which is by a factor 17 weaker than in the previous simulations.
(2) The toroidal magnetic field is assumed for the initial magnetic-field configuration in the neutron star.
(3) The equatorial-plane symmetry is removed from the simulation.
(4) The SFHo EOS in addition to the DD2 EOS is employed to model the neutron star.
We performed four new simulations for $\sim 2.5$--$6$\,s in order to self-consistently follow the dynamical mass ejection, remnant disk evolution, post-merger mass ejection, and generation of collimated Poynting flux in the magnetosphere which may drive a short-hard gamma-ray burst. 

We found, irrespective of the difference in the setups listed above, that the essential part of the merger and post-merger processes is unchanged.
First, the matter with its mass $\approx 0.046M_{\odot}$ (DD2 models) or $\approx 0.014M_{\odot}$ (SFHo model) is dynamically ejected right after the tidal disruption of the neutron star in the timescale of a few\,ms.
Then the accretion disk with the initial rest mass $M_{\mathrm{disk},0}\approx 0.28M_{\odot}$ (DD2 models) or $\approx 0.22M_{\odot}$ (SFHo model) is formed around the remnant black hole.
In the accretion disk, the magnetic field is amplified by magnetohydrodynamics effects such as the MRI, winding, and Kelvin-Helmholtz instability.
The MRI turbulence and dynamo action induce effective viscosity that enhances the angular-momentum transport. As a result, the mass accretion and the expansion of the disk are induced. 
In addition, a portion of the matter with the strong magnetic fields is outflowed  vertically from the disk. 
This plays a role in the disk expansion to the vertical direction. In the turbulent process, thermal energy is generated, but in the first several hundred ms, the generated heat is dissipated primarily by the neutrino emission and it does not have a significant effect on the post-merger mass ejection. 

The disk expands gradually due to the angular momentum transport effects. 
As a result, the maximum temperature of the disk drops below $\sim 3$\,MeV, and the neutrino luminosity drops below $\sim 10^{51}$--$10^{51.5}$\,erg/s.
Then the neutrino emission cannot carry away an appreciable fraction of the thermal energy generated by the turbulent process from the disk, and the generated thermal energy induces the convective motion in the disk, which carries the thermal energy generated around the inner edge of the disk to the outer region. 
This convective motion contributes to the heating for the outer part of the disk, and eventually, induces the post-merger mass ejection.
In our present setups, the mass of the post-merger ejecta is $\agt 0.030M_{\odot}$ (DD2 model), or $\agt 0.018M_{\odot}$ (SFHo model), which are $\sim 10\%$ of $M_{\mathrm{disk},0}$. 

As in our previous paper~\cite{hayashi2022jul}, we found, irrespective of the models, that there are two components in the electron fraction distribution for the ejected matter. One is a low-electron fraction component ($Y_\mathrm{e} < 0.1$) produced by the dynamical ejecta and the other is a mildly neutron-rich component ($0.1 \alt Y_\mathrm{e} \alt 0.4$) produced by the post-merger ejecta. \addms{We found that the electron fraction distribution depends weakly on the EOS, and the dependence is reflected in the elemental pattern of the $r$-process nucleosynthesis~\cite{Wanajo:2022jgw}.}
Also, there are two components in the velocity distribution. 
One is a fast component (up to $v \sim 0.4c$) produced by the dynamical ejecta and the other is a relatively slow component ($v < 0.1c$) produced by the post-merger ejecta. 
These distributions are suitable for reproducing an elemental abundance pattern similar to the solar abundance and those of the metal-poor stars~\cite{Wanajo:2022jgw}.  

We analyzed the ratio of the anisotropic stress to the pressure to evaluate the alpha viscous tensor, $\alpha_{ij}$. 
It is found that all the components of $\alpha_{ij}$ have a value between $10^{-2}$ and $10^{-1}$. 
However, the value of the $r\varphi$ or $xy$ components of this tensor is larger than the other components. 
This reflects that not only the magnetically-induced viscous effect but also intrinsic magnetohydrodynamics effects play a role in the momentum transport. 
The large value of $\alpha_{r\varphi}$ suggests that a magnetohydrodynamics effect associated with global magnetic fields such as the magneto-centrifugal effects~\cite{blandford1982} play an important role in the angular momentum transport.  

We found quantitative differences between the results for the models with low and high initial magnetic-field strengths. 
Because it takes a longer time to amplify the magnetic field up to saturation and to achieve an equipartition state in the disk from the low initial field strength (with a limited grid resolution), the evolution of the accretion disk and post-merger mass ejection are delayed. However, essentially no differences are found in the properties of the ejected matter. 
\addkh{We note that even with the low initial magnetic-field strength employed here, the initial field strength is still several orders higher than the realistic value. 
The simulation starting from a neutron star endowed with $\sim10^{12}$\,G is ultimately necessary.}

Accompanying the turbulent disk formation, a funnel-shaped magnetosphere with the low rest-mass density and the aligned helical magnetic-field lines are formed near the spin axis of the black hole. 
This magnetosphere is a magnetically dominated region and is in an approximate force-free state. 
The helical magnetic field lines that form the magnetosphere penetrate the black hole and it extracts the rotational kinetic energy of the rapidly spinning black hole by the Blandford-Znajeck mechanism~\cite{blandford1977}.
Then, the collimated outgoing Poynting flux is generated with the opening angle of $\sim 10^{\circ}$, and its isotropic-equivalent luminosity is $\sim 10^{50}$\,erg/s.
The high Poynting luminosity stage continues for $\sim 0.5$--2\,s, and the luminosity subsequently decreases. These properties are consistent with typical short-hard gamma-ray bursts~\cite{nakar2007apr,berger2014jun}. 

For the model with the SFHo EOS as well as for previous models~\cite{hayashi2022jul}, the Poynting luminosity is likely to drop due to the spreading of the funnel wall and the decrease of the magnetic-field strength.  
The spreading of the funnel wall is caused by the decrease of the gas pressure from the torus at the funnel wall. 
However, we also found other two processes that result in the decrease of the Poynting luminosity in the late stage of the post-merger evolution.
First, for the model with a low initial magnetic-field strength and the model with no equatorial-plane symmetry, the Poynting luminosity for a given observer drops due to the disappearance of the magnetosphere stemming from the reconnection of the magnetic-field lines.  
This is caused by the MRI dynamo activity in the accretion disk, which enforces the magnetic flux with a variety of the polarity to be ejected quasi-periodically from the disk to the polar region. 
When strong magnetic fields with the polarity opposite to that in the magnetosphere emerge from the disk, they pair-annihilate by the reconnection, and the magnetosphere temporarily disappears.
If the magnetic field emerging from the disk is strong enough by any chance, it replaces the magnetic field in the magnetosphere, \addms{resulting in the polarity flip}. 
However, this replacement is only found in the model with no equatorial-plane symmetry, and the disappearance occurs for the models with low initial magnetic-field strength and the model with no equatorial plane symmetry.
Second, for the model with the initially toroidal magnetic field in the neutron star, the Poynting luminosity drops due to the tilt of the magnetosphere. 
Because the post-merger ejecta occasionally carries the angular momentum component not parallel to the black hole spin axis, the accretion disk is enforced to tilt, in particular in the late-time evolution.
Then, it results in the tilt of the magnetosphere because the funnel structure of the magnetosphere is determined by the gas pressure from the disk (torus).
Irrespective of these mechanisms, the evolution process of the accretion disk does determine the evolution process of the magnetosphere. 
These three mechanisms all include stochastic processes, and it is not feasible to precisely predict which mechanisms determine the evolution process of the magnetosphere. However, these three could be plausible mechanisms to make short-hard gamma-ray bursts as short as $\sim 0.5$--$2$\,s.

\begin{acknowledgments}
  We thank Sho Fujibayashi, Kunihito Ioka, and Shinya Wanajo for useful discussions. 
  Numerical simulations were performed on
  Sakura, Cobra, and Raven clusters at Max Planck Computing and Data Facility,
  Yukawa-21 at Yukawa Institute for Theoretical Physics of Kyoto University,
  and Cray XC50 at CfCA of National Astronomical Observatory of Japan.
  This work was in part supported by Grant-in-Aid for Scientific Research 
  (grant Nos.~19K14720, 20H00158, 22K03617, and 23H04900) of Japanese MEXT/JSPS.
  KH was supported by JST SPRING (grant No. JPMJSP2110).
\end{acknowledgments}

\bibliography{paper}

\end{document}